# A Multi-Core Processor Platform for Open Embedded Systems

A DISSERTATION
PRESENTED TO THE GRADUATE SCHOOL
OF SCIENCE AND TECHNOLOGY
OF KEIO UNIVERSITY
IN CANDIDACY FOR THE DEGREE
OF DOCTOR OF PHILOSOPHY IN ENGINEERING

RECOMMENDED FOR ACCEPTANCE
BY THE CENTER FOR COMPUTER SCIENCE

**September 2009**

**INOUE, Hiroaki**

# ABSTRACT

Recent proliferation of embedded systems has generated a bold new paradigm, known as open embedded systems. While traditional embedded systems provide only closed base applications (natively-installed software) to users, open embedded systems allow the users to freely execute open applications (additionally-installed software) in order to meet various user requirements, such as user personalization and device coordination.

Key to the success of platforms required for open embedded systems is the achievement of both the scalable extension of base applications and the secure execution of open applications. Most existing platforms, however, have focused on either scalable or secure execution, limiting their applicability.

This dissertation presents a new secure platform using multi-core processors, which achieves both scalability and security. Four techniques feature the new platform: (1) seamless communication, by which legacy applications designed for a single processor make it possible to be executed on multiple processors without any software modifications; (2) secure processor partitioning with hardware support, by which Operating Systems (OSs) required for base and open applications are securely executed on separate processors; (3) asymmetric virtualization, by which many OSs over the number of processors are securely executed under secure processor partitioning; and (4) secure dynamic partitioning, by which the number of processors allocated to individual OSs makes it possible to be dynamically changed under secure processor partitioning.

# ACKNOWLEDGEMENTS

Many thanks go to Prof. Hideharu Amano for his grateful support and encouragement of my Ph. D. degree. I was fortune to have been one of his students at Keio University. Ten years ago, I, as a master student, engaged in a national research project, called the JUMP-1 project, with him. I remember that his leadership was instrumental in my research on the memory-based processor of the project. He has been my excellent role model as a professional researcher.

I also would like to thank Prof. Kenji Kono, Prof. Takahiro Yakoh, and Prof. Nobuyuki Yamasaki for graciously serving as the reviewers of this dissertation. Their insightful comments helped me to improve this dissertation.

My research has been enriched through collaborations with Tsuyoshi Abe, Yoshiharu Asakura, Hiroshi Chishima, Masato Edahiro, Toshitaka Fujioka, Yoshimi Fukagawa, Masao Fukuma, Masajiro Fukunaga, Satoshi Hieda, Hirofumi Higuchi, Akihisa Ikeno, Kazuhisa Ishizaka, Yoshiyuki Ito, Masayoshi Kai, Masaki Kondo, Kouji Maeda, Toshiya Matsui, Satoshi Matsushita, Hideaki Nagata, Yukikazu Nakamoto, Naoki Nishi, Yoshinori Saida, Junji Sakai, Kazue Sako, Naoki Sato, Kenichi Sawada, Naotaka Sumihiro, Kenji Suzuki, Satoshi Suzuki, Yasuaki Tadokoro, Mikiya Tani, Sunao Torii, Masaki Uekubo, Kazutoshi Usui, Yuki Utsuhara, Kazutoshi Wakabayashi, Mitsuhiro Watanabe, Masakazu Yamashina, Koji Yoshida, and all others involved in the MP98 project of NEC Corporation. I would like to express my deep appreciation for their great contributions. In addition, I would like to thank Muneo Fukaishi, Akira Funahashi, Yuji Hamada, Hiroshi Kodama, Masayuki Mizuno, Michitaka Okuno, and Noriaki Suzuki for their kind help in writing this dissertation.

Finally, I wish to thank my family for their immeasurable support.

# CONTENTS

# LIST OF FIGURES

# LIST OF TABLES

# CHAPTER 1
# INTRODUCTION

This chapter introduces the concept of open embedded systems, and clarifies our research contributions.

## 1.1 OPEN EMBEDDED SYSTEMS

Recent proliferation of embedded systems has generated a bold new paradigm, known as open embedded systems [Intel 06] [NTT 04a] [NTT 04b]. While traditional embedded systems provide only closed base applications (*i.e.*, natively-installed software, such as mailer and browser in mobile phones) to users, open embedded systems allow the users to freely execute open applications (*i.e.*, additionally-installed software that includes user-level programs, libraries, and device drivers) as well as base applications. Open applications may be downloaded from any web sites in order to add various functionalities to embedded systems. They also may communicate with other embedded systems in order to offer device coordination to users.

FIGURE 1.1 shows three useful service examples of open embedded systems. The first service example is a driver-assist service, in which a drive recorder equipped with a car stores a lot of driving information in coordination with a notification event sent from a car navigation system when the car approaches an accident-prone area. The second service example is a virtual-terminal service, in which a user makes it possible to virtually use multiple mobile terminals for private and business scenes on a physical terminal by means of the free install of carrier software packages. The last service example is an anti-crime service, in which a child's mobile phone automatically calls an emergency contact number (*e.g,* a home number) in coordination with a notification event sent from town's monitoring cameras when town's monitoring cameras detect that a child is moving out of town. Leveraging open applications, open embedded systems make it possible to meet various user requirements, such as user personalization and device coordination, unlike traditional embedded systems.

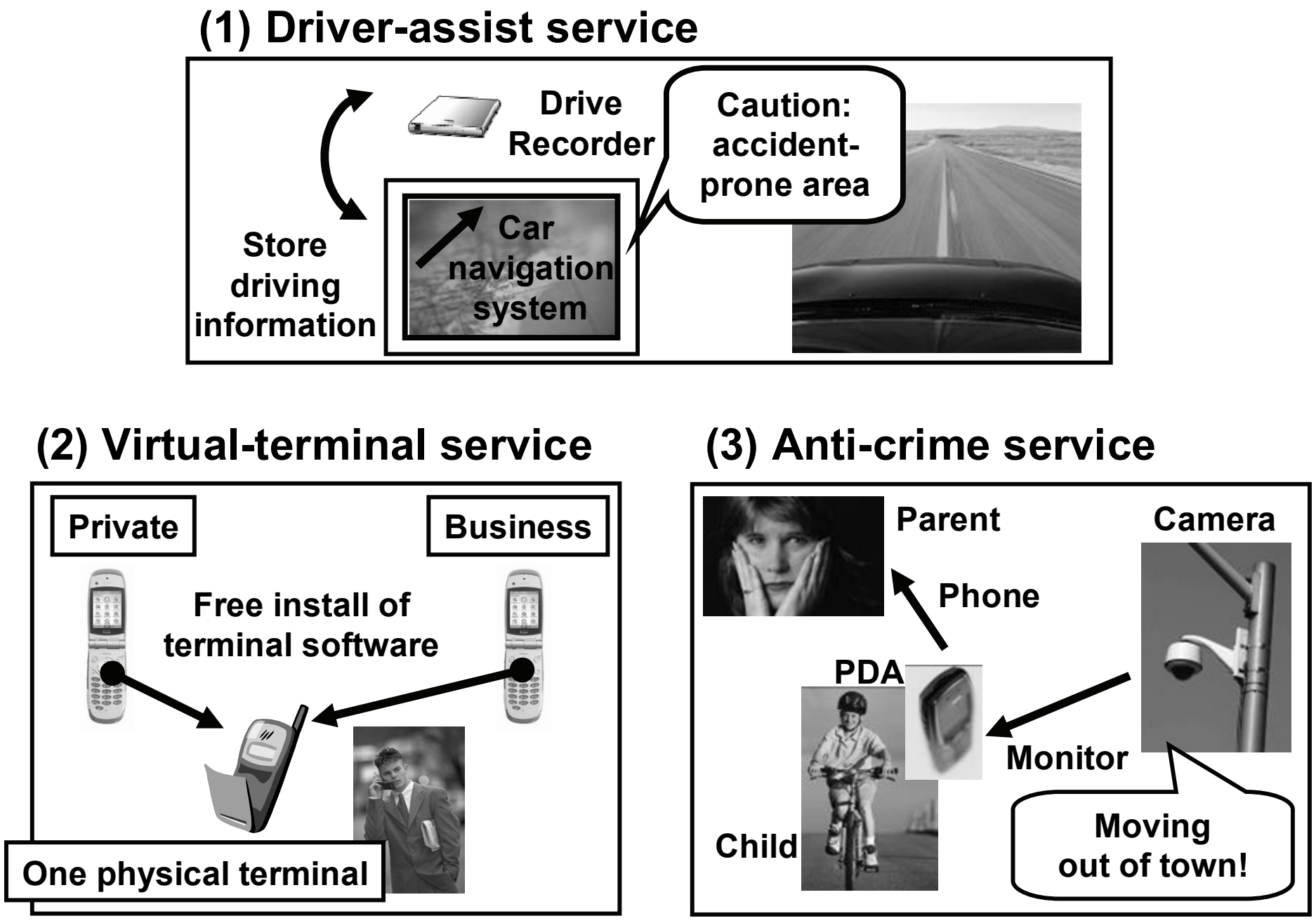

**FIGURE 1.1: EXAMPLES OF OPEN EMBEDDED SYSTEMS**

Open applications need to be executed on the same platform as base applications since the open applications that we target include device drivers as well as user-level programs by definition. This requirement of open embedded systems indicate that open embedded systems need to support at least two isolated execution environments (*i.e.*, domains) for the separate execution of base and open applications. A domain is here defined as an execution environment formed on a native OS. While a base domain executes base applications, an open domain executes a group of open applications. Further, additional open domains may be required in order to isolate many groups of open applications themselves. FIGURE 1.2 summarizes the features of open embedded systems.

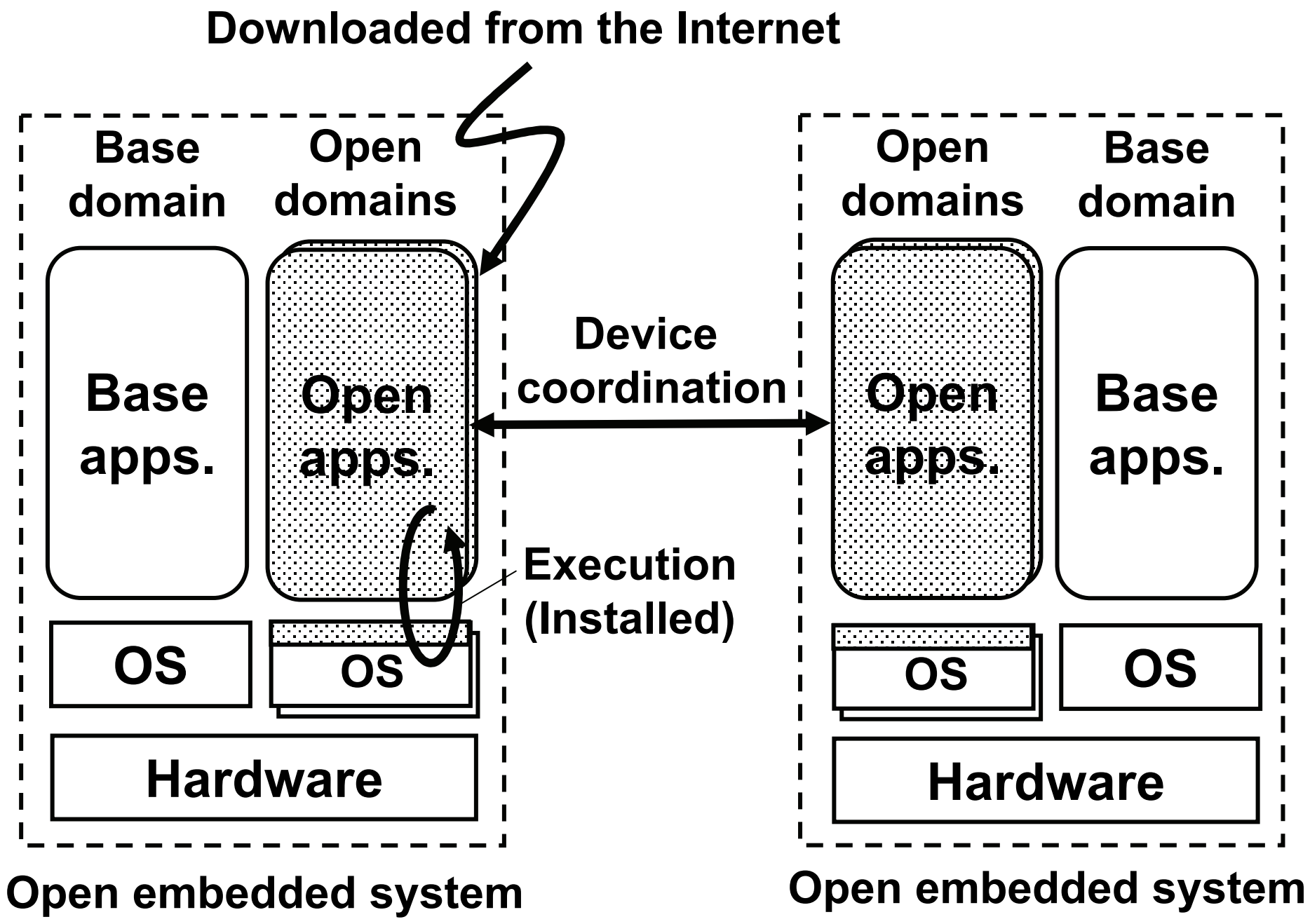

**FIGURE 1.2: FEATURES OF OPEN EMBEDDED SYSTEMS**

In order to deploy this new paradigm on traditional embedded systems, platforms used for open embedded systems require the achievement of the following design objectives:

- **Scalable[1] functionality for base applications:** The major driving force of embedded systems still enriches the functionality of base applications in order to maximize user experience. The development costs of base applications, however, would seem to reach an extraordinarily-high value since the number of lines of source codes required for base applications rapidly increase [Morgan 04]. The platforms used for open embedded systems also need to support the scalable extension of base applications in a cost-effective way.

- **Hardened security for open applications:** The flexibility of open embedded systems would seem to result in a two-edged blade because new groups of open applications might contain bugs or viruses. FIGURE 1.3 shows the recent trends in viruses for mobile terminals. As shown in this figure, the number of mobile viruses rapidly increases [Gostev 06] [Gostev 07]. This means that base applications must be clearly protected from malicious open applications in order to maintain the minimum functionality of embedded systems. In addition, open embedded systems need to securely isolate many groups of open applications in order to prevent mutual interference among the application groups.

- **Base features for embedded systems:** Unlike traditional computing systems, embedded systems need to be able to operate with limited resources. Open embedded systems also require the careful consideration for base features, such as performance overhead and memory footprint.

[1] The word "scalability" means the extensibility to various technical attributes, such as the number of processors, functionalities, and the number of clients.

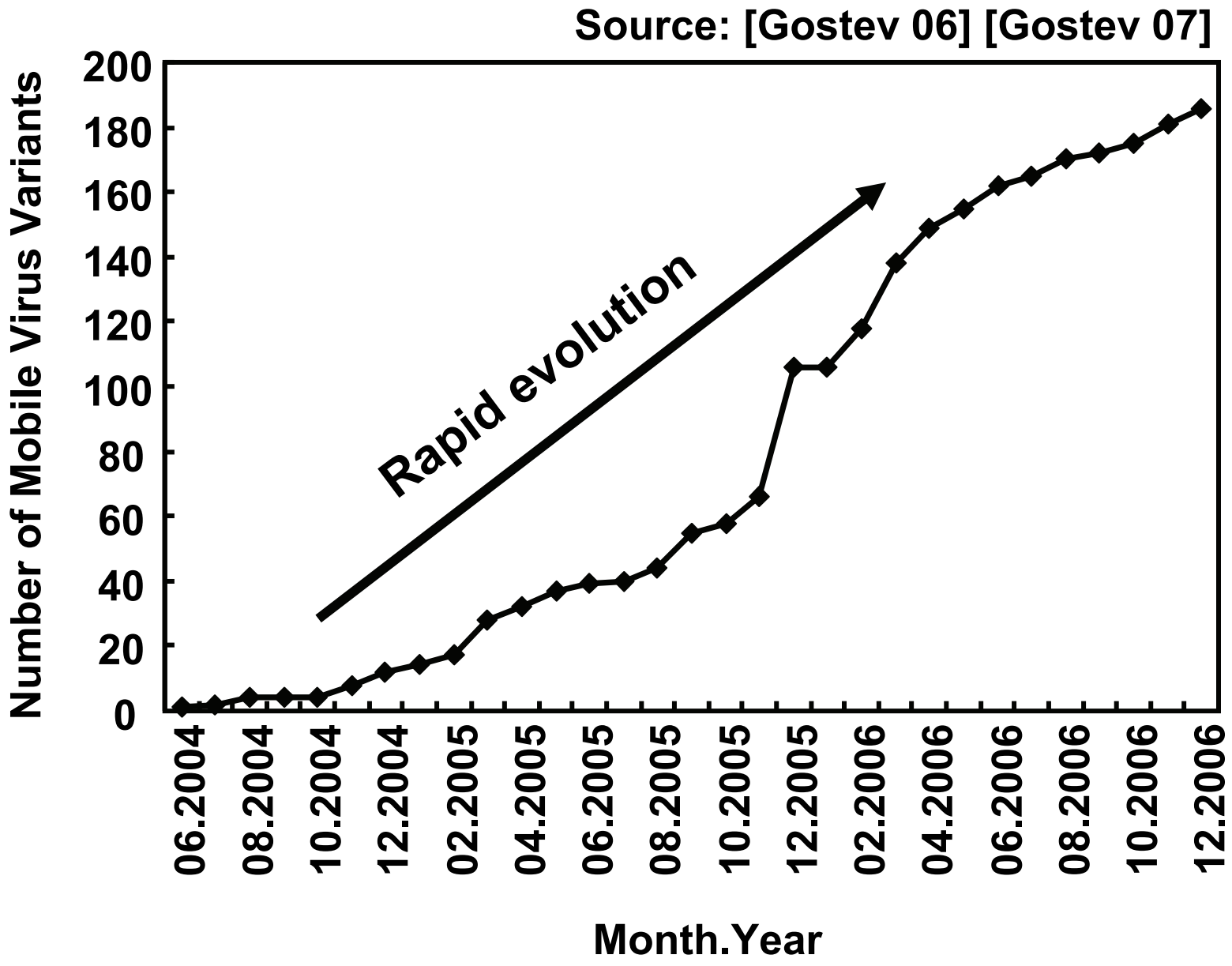

FIGURE 1.3: TRENDS IN MOBILE VIRUSES

## 1.2 SECURITY MODEL

Two aspects help classify security required for embedded systems: data security and program security. The goal of data security is the protection of the integrity and privacy of confidential data. Much work on data security, such as XOM [Lie 00], AEGIS [Suh 05], TPM [TCG 06], and SENSS [Zhang 05] helps prevent untrusted software executed on a processor from stealing private keys or modifying applications and OSs.

The goal of program security is the protection of the correct execution of programs. We classify attacks against program security into two directions: vertical attacks and horizontal attacks. Vertical attacks are ones that try to take control of programs on other domains by exploiting the vulnerability of the underlying platform. For example, the vulnerability of the *ptrace* system call allowed processes to obtain root privileges on Linux OS (version 2.4.18). Horizontal attacks are ones that try to change control flows of programs on other domains by means of inter-program communication. For example, the vulnerability of Apache web servers (version 1.3.24) allowed web clients to modify web contents because the web servers had a software flow that misinterpreted invalid requests encoded using chunked encoding.

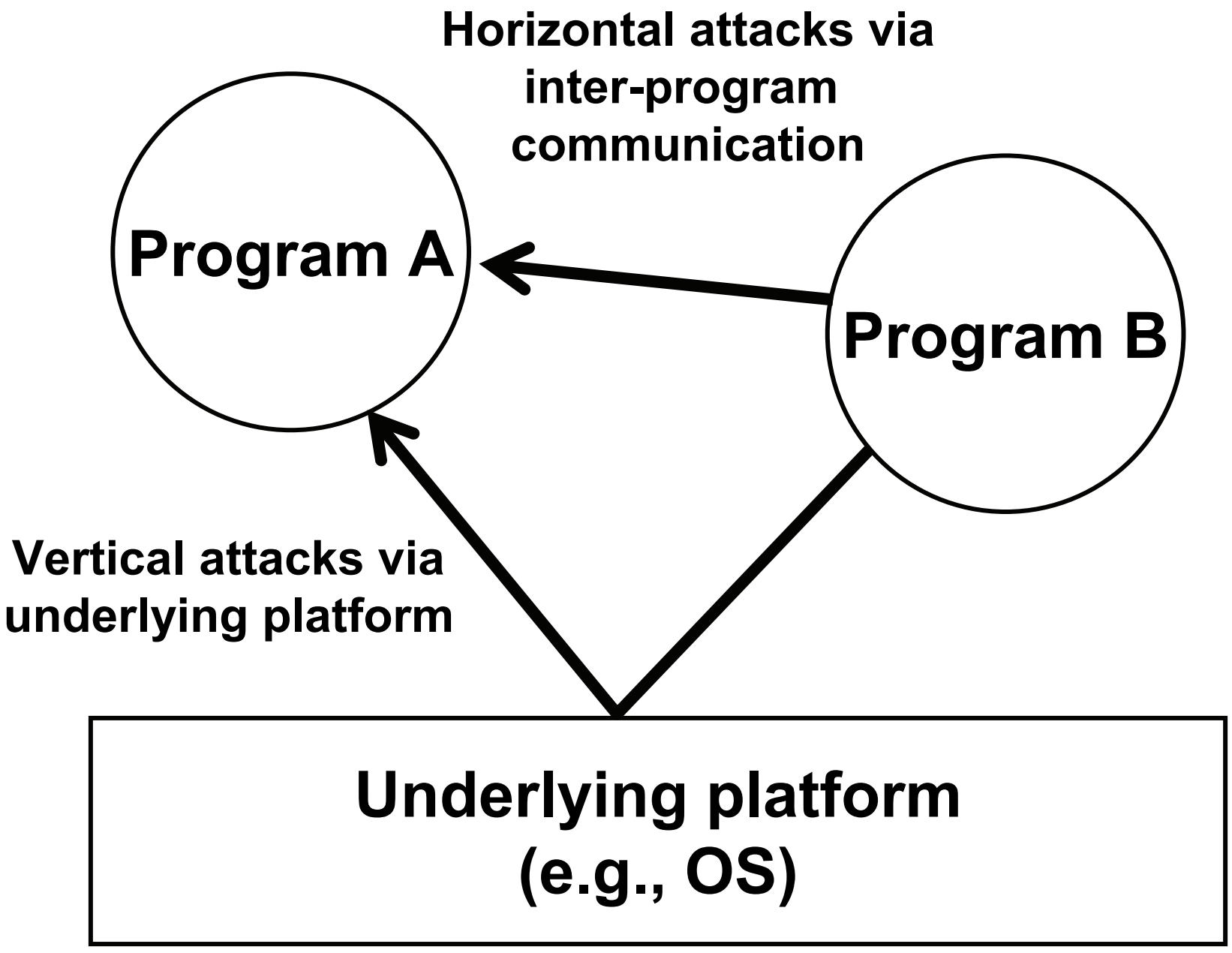

**FIGURE 1.4: SECURITY MODEL**

This dissertation focuses on the program security which blocks vertical attacks. A security capability which makes underlying platform more secure is most important for the execution of open applications. Without the security capability, open embedded systems would seem to fail to execute any class of open applications (*e.g.*, device drivers) on native OSs since vertical attacks launched from malicious open applications make it possible to compromise the native OSs. In addition, other security capabilities which help enhance both data security and program security against horizontal attacks would seem to fail to be trustworthily implemented without reliable underlying platform.

## 1.3 CHALLENGES

Various electric hardware components, such as processors, memories and I/Os, form the basis of platforms for open embedded systems. In particular, processor architecture becomes an important factor in meeting the above requirements of open embedded systems since the architecture directly involves with the execution of both base and open applications.

In recent trends of processor architecture, multi-core processors would seem to be one promising technology direction. A multi-core processor is defined as a processor which contains multiple cores (processors) in a chip. Conventional single-core processors need to operate at a high clock frequency in order to provide sufficient performance to both base and open applications, which makes it difficult to reduce power dissipation. By way of contrast, multi-core processors enable the desired level of performance to be achieved with a number of processors that operate at moderate clock frequencies, which helps to keep power dissipation low [Torii 05].

*From the software point of view*, processing on multi-core processors is classified into two types (see FIGURE 1.5): (a) Asymmetric Multi-Processing (AMP) and (b) Symmetric Multi-Processing (SMP) [Sakai 07] [Sakai 08].

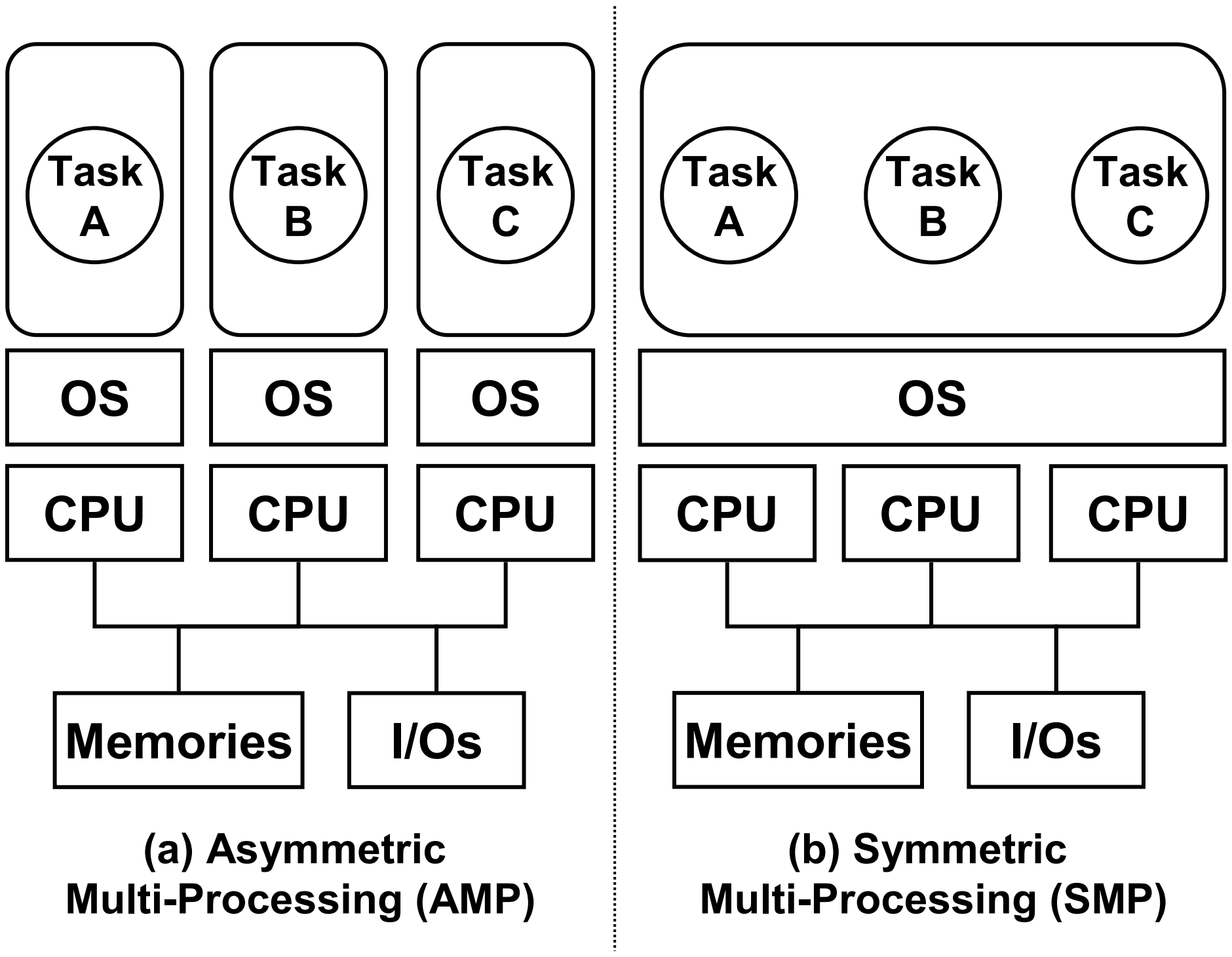

**FIGURE 1.5: TWO TYPES OF MULTI-PROCESSING**

With AMP, multiple OSs are executed on different processors. Various tasks are fixedly assigned to each processor. While multiple OSs help separate the execution of base applications from that of open applications, the OSs make it difficult for legacy base applications designed for a single-core processor to be executed over multiple processors without any software modifications. This means that AMP improves the secure execution of open applications, sacrificing the scalable extension of base applications. It should be noted that AMP still provides vulnerable protection among OSs since malicious open applications make it possible to exploit the security holes of OSs.

With SMP, a single OS manages multiple processors. The OS enables tasks to be transparently executed over multiple processors. While legacy base applications designed for a single-core processor are executed over multiple processors without any software modifications, a single OS results in causing mutual interference among base and open applications. This means that SMP improves the scalable extension of base applications, limiting the secure execution of open applications.

Moreover, both AMP and SMP have a common concern about the support of many domains (*i.e.*, OSs) used for various groups of open applications. While AMP needs to increase the number of processors for the support of many OSs, SMP supports only a single OS in a system. In order to cope with this issue, virtualization would seem to be a good solution. Conventional virtualization technologies, however, have a degree of security vulnerability [Hacker 07]. In addition, the technologies unfit for embedded systems in terms of base features, such as performance overhead and memory footprint, since traditional virtualization technologies have been originally developed for computing systems.

From the above discussion, use of traditional multi-core processors poses major challenges to the achievement of open embedded systems since neither AMP nor SMP is in itself satisfactory.

## 1.4 STATE OF THE ART

Existing research on multi-core processors, however, has satisfied the requirements on neither scalable nor secure execution. In terms of scalable execution of base applications, the techniques used for traditional platforms [Accetta 86] [Fleisch 86] [Maloy 04] [MPI 97] [Mullender 90] [OMG 04] [Paulin 06] [Rozier 88] [Sharif 99] [Steen 99] [Tan 02] require a wide range of software modifications for either OSs or applications. This software incompatibility prevents the scalable extension of base applications especially on AMP. Moreover, in terms of secure execution of open applications, the techniques used for traditional platforms [Armstrong 05] [Barham 03] [Baratloo 00] [Chen 08] [Cowan 98] [Dike 00] [Evans 02] [Fortify 09] [Gondo 07] [Gong 03] [Johnson 07] [Loscocco 01] [Neiger 06] [Openwall 01] [Qualcomm 04] [Seshadri 07] [Shinagawa 09] [Sugerman 01] [Whitaker 02] still have potential vulnerability on both AMP and SMP because the platforms provide only software-based protection.

**TABLE 1.1: STATE OF THE ART**

| Requirements | Items | AMP | SMP |
|---|---|---|---|
| Scalable extension of base applications | Problem | Unsolved | Solved |
| | Reason | A wide range of software modifications | |
| Secure execution of open applications | Problem | Unsolved | Unsolved |
| | Reason | Only software-based protection | |

## 1.5 CONTRIBUTIONS

The primary contributions of this dissertation are the attainment of a multi-core processor platform for open embedded systems. Our multi-core processor platform addresses the challenges of both AMP and SMP in order to achieve both the scalable extension of base applications and the secure execution of open applications. FIGURE 1.6 summarizes the qualitative advantage of our multi-core processor platform, compared with existing work. Four innovative techniques feature our platform:

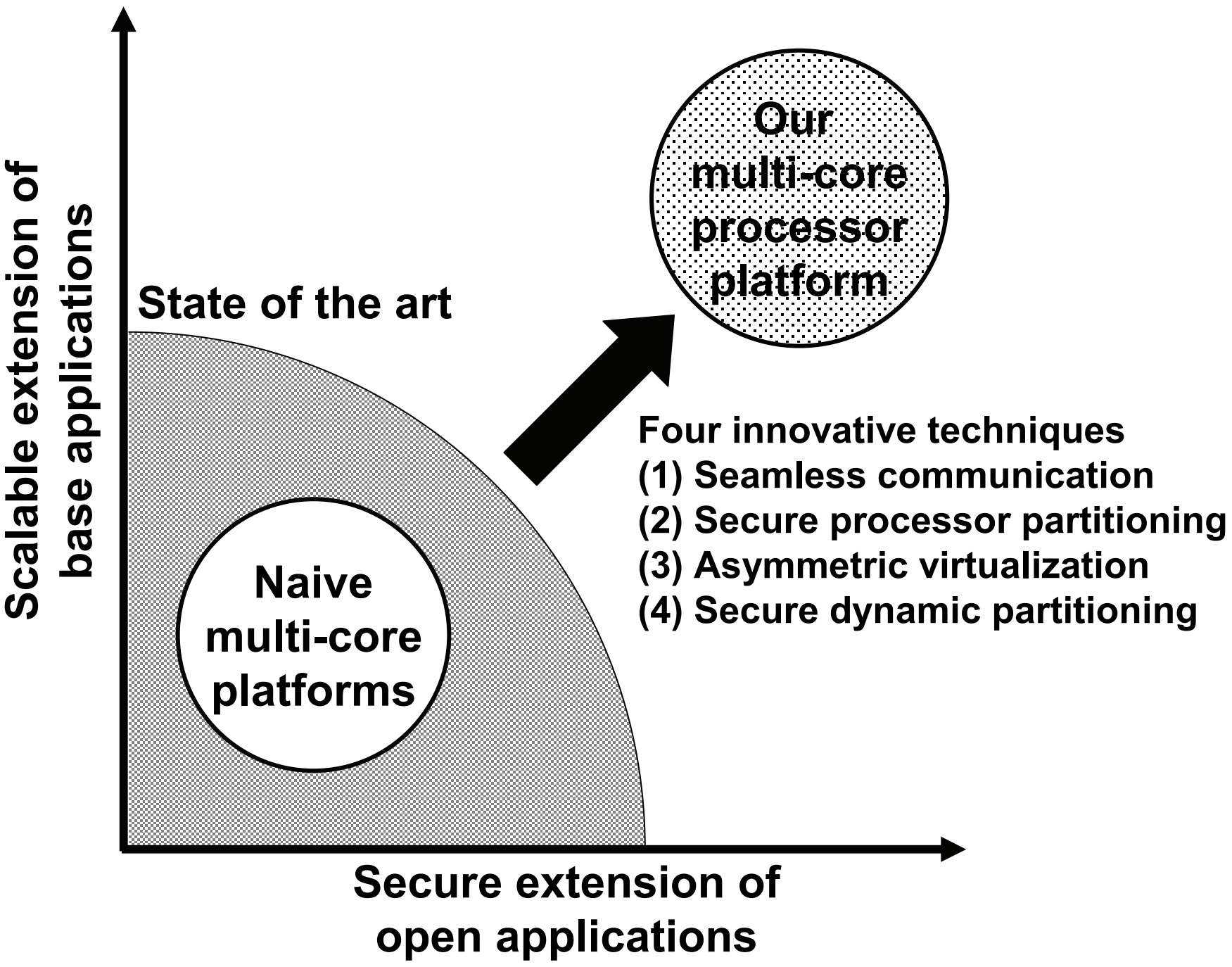

**FIGURE 1.6: CONTRIBUTIONS OF OUR MULTI-CORE PROCESSOR PLATFORM**

- We propose a software approach for seamless communication, by which legacy base applications designed for a single-core processor make it possible to be executed over multiple processors without any software modifications [Inoue 09b]. In this way, our platform achieves the scalable extension of base applications even on AMP.

- We present a hardware-supported approach for secure processor partitioning, by which OSs are mutually protected on separate processors [Inoue 06a] [Inoue 08b]. For the secure execution of open applications, this processor partitioning bases our platform formed on SMP as well as AMP.

- We provide a new type of virtualization, known as asymmetric virtualization, by

which many OSs over the number of processors are securely executed under secure processor partitioning [Inoue 06b] [Inoue 08a]. This virtualization helps provide many secure domains on AMP and SMP for the secure execution of many groups of open applications since it fits for embedded systems.

- We propose secure dynamic partitioning as an extension of secure processor partitioning, by which the number of processors allocated to individual OSs makes it possible to be dynamically changed on SMP under secure processor partitioning [Inoue 07] [Inoue 09a]. In this way, our platform achieves the secure execution of open applications even on SMP.

## 1.6 ORGANIZATION

The remainder of this dissertation is structured as follows.

Chapter 2 introduces our base platform for open embedded systems. It turns out that our platform enables the achievement of both the scalable execution of base applications and the secure execution of open applications.

Chapters 3 through 6 present the detailed design of four innovative techniques with respect to (1) seamless communication, (2) secure processor partitioning, (3) asymmetric virtualization, and (4) secure dynamic partitioning. Evaluations also demonstrate the effectiveness of the four techniques.

Chapter 7 concludes the research contributions presented in this dissertation.

TABLE 1.2 summarizes the overview of our four new techniques, clarifying the requirements of open embedded systems, the issues that traditional multi-core processor platforms have, and the effects of the innovative techniques.

**TABLE 1.2: SUMMARY OF FOUR NEW TECHNIQUES**

| Requirements | Items | AMP | SMP |
|---|---|---|---|
| Scalable extension of base applications | Issue | A wide range of software modifications | |
| | Technique | Seamless communication | |
| | Effect | No need for software modifications | |
| | Chapter | Chapter 3 | |
| Secure execution of open applications (for new groups) | Issue | Security holes exploited by open applications | |
| | Technique | Secure processor partitioning | Secure dynamic partitioning |
| | Effect | Hardware-level secure protection | multiple domains under hardware-level secure protection |
| | Chapter | Chapter 4 | Chapter 6 |
| Secure execution of open applications (for many groups) | Issue | Failure to support many open domains securely | |
| | Technique | Asymmetric virtualization | |
| | Effect | Secure support of many domains | |
| | Chapter | Chapter 5 | |

# CHAPTER 2
# BASE PLATFORM

This chapter describes the structures of our base platform for open embedded systems, and clarifies how our new techniques work on multi-core processors.

## 2.1 OVERVIEW

FIGURE 2.1 illustrates the structural overview of our base platform featured by both hardware and software. The hardware structure uses a multi-core processor. By definition, while a base domain executes base applications, maintaining at least one processor; an open domain executes a group of open applications, based on the classification of open applications.

Our base platform supports two execution modes: a highly-functional mode and a multi-functional mode. A highly-functional mode executes only a base domain in order to maximize user experience of an embedded system. By way of contrast, a multi-functional mode executes a base domain and open domains in order to add various functionalities to the embedded system. In response to user requests and environmental events, the platform dynamically switches the two execution modes, coordinating with the hardware and software structures. In this way, our base platform achieves the functions required for open embedded systems.

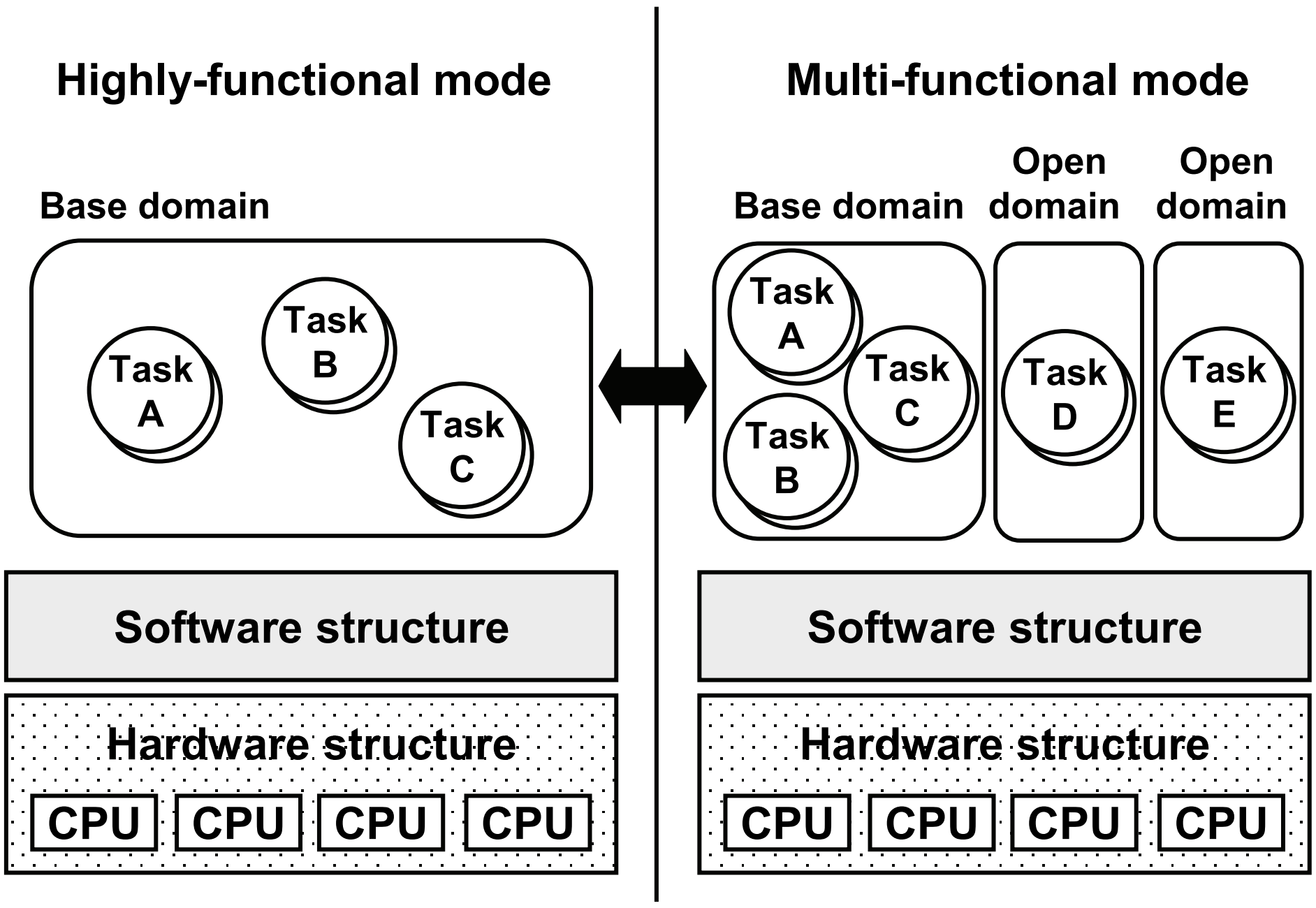

**FIGURE 2.1: BASE PLATFORM**

## 2.2 HARDWARE STRUCTURE

FIGURE 2.2 shows an example of a multi-core processor used for our hardware structure. By definition, a multi-core processor provides multiple processors (*e.g.*, four processors in the figure) in a chip. Most importantly, all processors contained in a chip share the same memories and I/Os due to stern constraints on the number of chip pins. The sharing results in allowing a processor to access the memories and I/Os that software executed on another processor uses. In order to solve the issue of this competitive access, our hardware structure equips with a new unit, called a bus management unit, on a processor bus. This unit allows each processor to access only specified address ranges of memories and I/Os, monitoring bus access at hardware level.

In this way, our bus management unit helps achieve secure processor partitioning for both AMP and SMP. Chapter 4 illustrates the detailed design of the hardware unit.

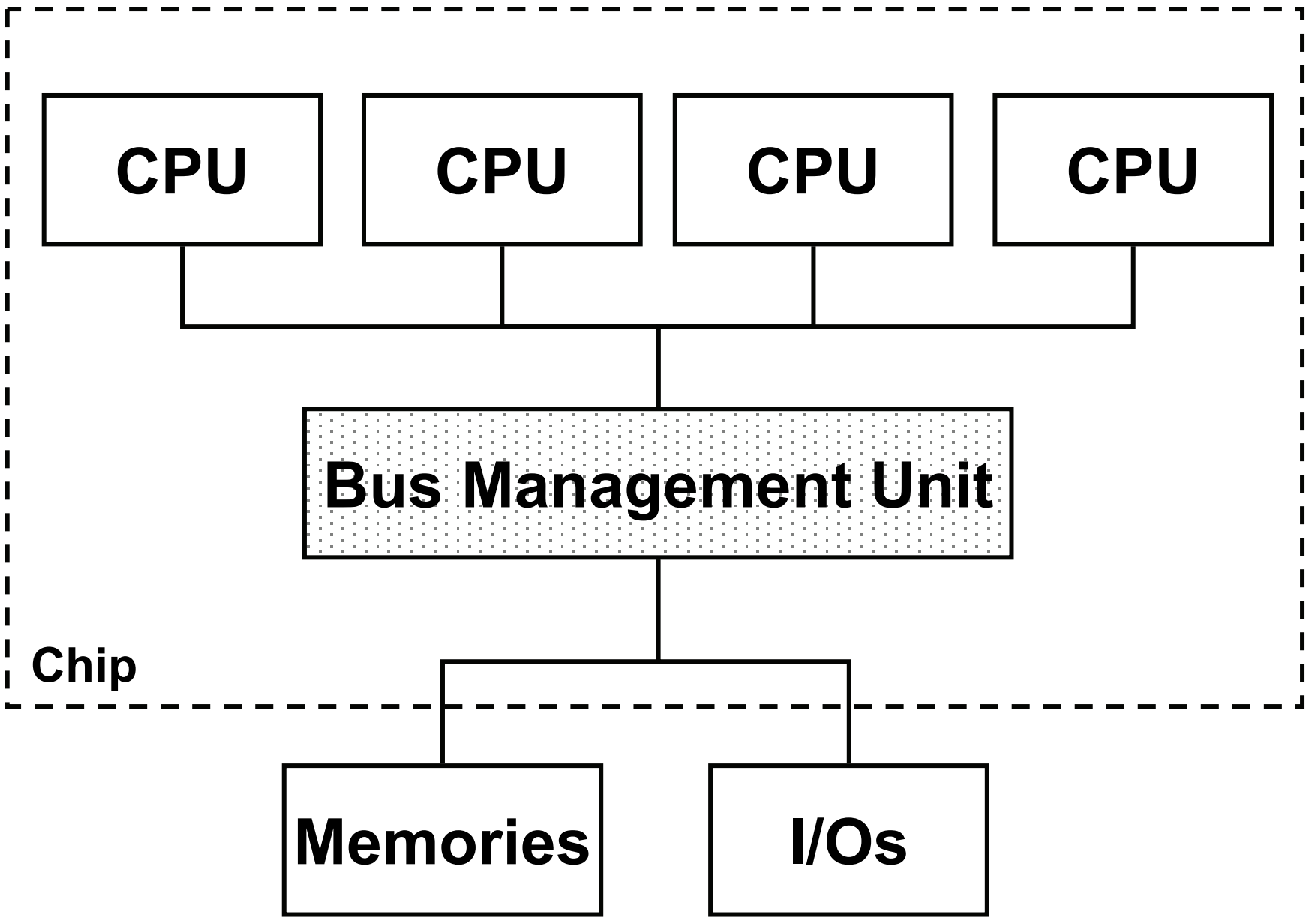

FIGURE 2.2: HARDWARE STRUCTURE

## 2.3 SOFTWARE STRUCTURE

Our software structure supports two types of both AMP and SMP, utilizing the above hardware structure. This software structure may employ heterogeneous OSs as well as homogeneous OSs to support multiple domains.

FIGURE 2.3 describes an example of a software structure on AMP. By definition, a processor executes an OS designed for a single-core processor.

In order to cope with the issue on scalable extension of base applications, we introduce user-level software, referred as to seamless communication software, to the base domain. The communication software enables legacy base applications designed for a single-core processor to be executed on multiple processors without any software modification, by hooking OS system calls and handling them at the user level. This is especially important because, from the viewpoint of base application developers, separate OSs will appear to be a single, uniform OS. Chapter 3 illustrates the detailed design of seamless communication software.

An individual OS forms an open domain on AMP. In this figure, while open domain A and B may execute applications validated by separate software manufacturers, open domain C may execute untrusted applications.

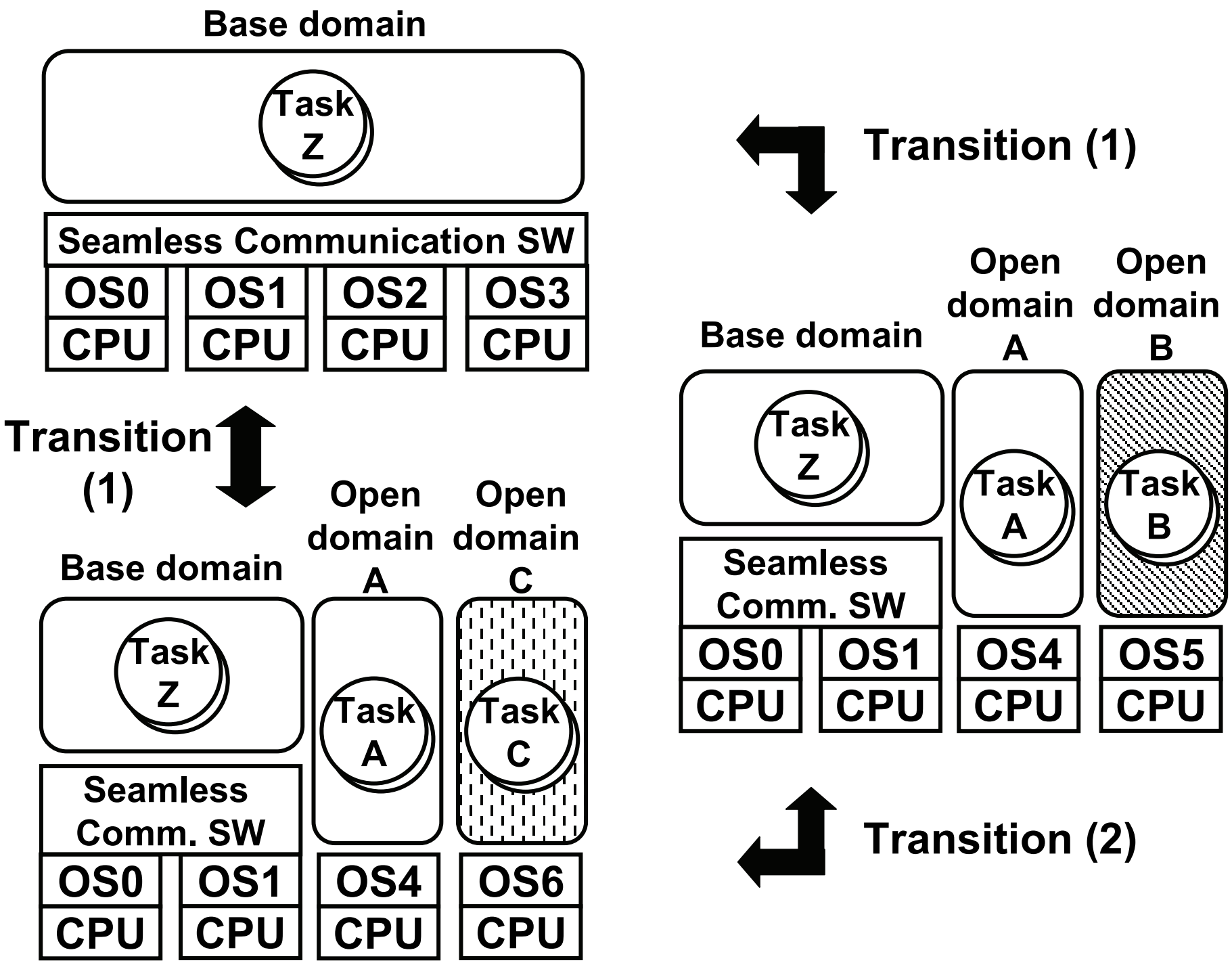

**FIGURE 2.3: SOFTWARE STRUCTURE ON AMP**

Asymmetric virtualization helps support multiple domains for the secure execution of many groups of open applications without any increase of processors in this structure.

The virtualization allocates processors to individual domains, utilizing our bus management unit. In response to user requests, the virtualization invokes transitions between a state with only the base domain and a state with both the base domain and two open domains (*i.e.*, transition (1) in FIGURE 2.3). Moreover, the virtualization invokes transitions between states with different states (*i.e.*, transition (2) in FIGURE 2.3). Chapter 5 illustrates the detailed design of asymmetric virtualization.

The support of many domains (*e.g.*, seven OSs in FIGURE 2.3), however, results in increasing the total memory requirements. In general, embedded systems employ the XIP (eXecute-In-Place) technique [Bird 04], which places read-only data (*e.g.*, instructions) on ROM and copies only other data to RAM in order to reduce the total memory requirements. We apply the XIP technique to multiple homogenous OSs. FIGURE 2.4 shows OS memory images of both ROM and RAM in this structure. While both a boot loader and the kernel text (*i.e.*, OS instructions) are shared among OSs, only kernel data are copied into individual areas of RAM.

FIGURE 2.5 describes an example of a software structure on SMP. Unlike the software structure on AMP, seamless communication software is not employed for the base domain since the OS used for a base domain itself manages multiple processors.

Secure dynamic partitioning helps support multiple domains for the secure execution of many groups of open applications in this structure, coordinating with our bus management unit and asymmetric virtualization. The dynamic partitioning allows processors to be de-allocated from the base domain and allows the processors to be allocated to open domains, and vice versa. The base domain maintains at least one processor.

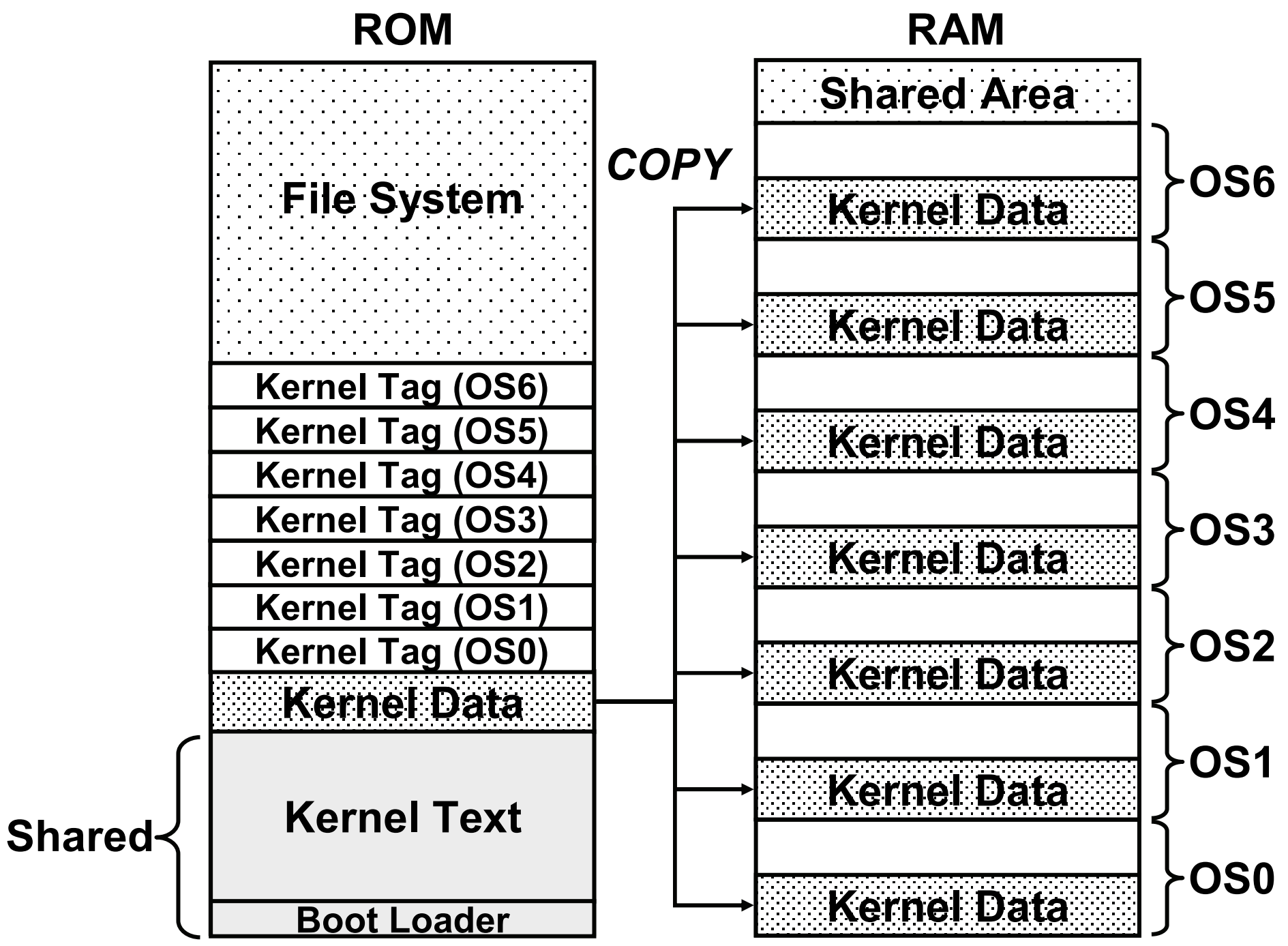

FIGURE 2.4: KERNEL XIP FOR OSS

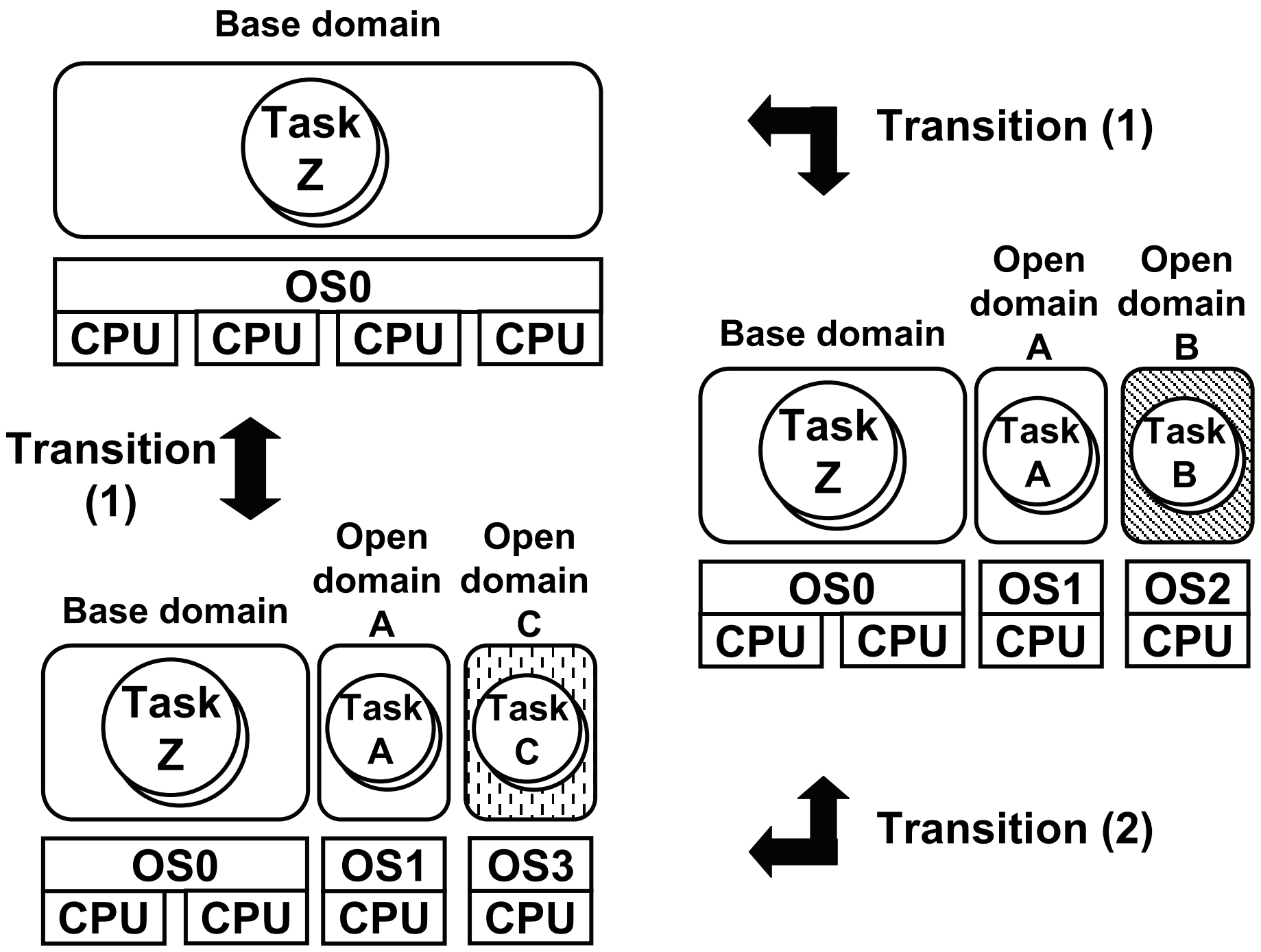

FIGURE 2.5: SOFTWARE STRUCTURE ON SMP

In the same way as the transitions of the software structure on AMP, the dynamic partitioning invokes transitions between a state with only the base domain and a state with both the base domain and two open domains (*i.e.*, transition (1) in FIGURE 2.5). Moreover, the virtualization invokes transitions between states with different states (*i.e.*, transition (2) in FIGURE 2.5). Chapter 6 illustrates the detailed design of secure dynamics partitioning.

It should be noted that our software structure needs software modifications for the secure execution of open applications in spite of no software modifications for the scalable extension of base applications. The software modifications are, however, not a practical problem since the modifications create new additional values, costing for newly added applications (*i.e.*, not for legacy applications).

## 2.4 SUMMARY

Both hardware and software structures feature our platform. The most important feature of the hardware structure is our bus management unit, which supports the secure execution of open applications at the hardware level. Moreover, three important techniques feature the software structure: seamless communication software, asymmetric virtualization, and secure dynamic partitioning. While seamless communication software addresses the issue of AMP in terms of the scalable extension of base applications, secure dynamic partitioning asymmetric addresses the issue of SMP in terms of the secure execution of new groups of open applications. Finally, our asymmetric virtualization helps support many domains on both AMP and SMP for the secure execution of many groups of open applications.

# CHAPTER 3
# SEAMLESS COMMUNICATION

This chapter describes seamless communication software, which enables legacy base applications designed for a single-core processor to be executed on AMP without any software modifications.

## 3.1 MOTIVATION

Current high-end embedded systems, such as mobile terminals, contain millions of lines of source codes for OSs, application platforms (*i.e.*, MOAP [Tsuji 05] by NTT DoCoMo), and base applications. A wide range of software modifications would seem to be required for AMP since the whole software has been designed for single-core processors. This software incompatibility prevents the scalable extension of base applications on AMP.

Conventional distributed OS and distributed middleware approaches help resolve the above software compatibility issue. Distributed OS approaches (partially) support the application compatibility, modifying single-core OSs in order to equip with system calls extended to multi-core processors. By way of contrast, distributed middleware approaches achieve the OS compatibility, modifying applications in order to use middleware APIs designed for multi-core processors. In other words, neither distributed

OS nor distributed middleware approaches are satisfactory since both approaches require software modifications.

The limitation of conventional approaches, however, indicates that one promising solution would seem to be the application of an approach intermediate between distributed OS and distributed middleware approaches, that is, distributed middleware that equips with system calls extended to multi-core processors. As the equipped system calls, this dissertation focuses on system calls associated with two Inter-Process Communications (IPCs): System V IPC and UNIX Domain Sockets (UDSs) [Stevens 98]. This is because the two IPCs were most frequently used in actual Linux-based mobile terminal software.

Thus, we introduce seamless communication software (*i.e.*, middleware), achieving inter-core communication through the same APIs as intra-core communication, hooking OS system calls and executing them at the user level. In this way, legacy base applications designed for a single-core processor make it possible to be executed on AMP without any software modifications. TABLE 3.1 summarizes the qualitative advantages of our approach, compared with others.

**TABLE 3.1: ADVANTAGES OF SEAMLESS COMMUNICATION SOFTWARE**

| **Feature** | **Distributed OS approaches** | **Distributed middleware approaches** | **Our approach** |
|---|---|---|---|
| OS compatibility | No | Yes | Yes |
| Application compatibility | Partially, yes | No | Yes |

## 3.2 RELATED WORK

Our research differs in a number of respects from the current body of research on communication software.

The major distributed OS approaches include Mach [Accetta 86], Amoeba [Mullender 90], Corus [Rozier 88] and TIPC [Maloy 04]. These approaches provide new OS system calls that are used for both intra-core and inter-core communication. This means that the approaches make it necessary to modify base applications as well as single-core OSs. In particular, distributed OS approaches for System V IPC include Distributed System V IPC in Locus [Fleisch 86] and DIPC [Sharif 99], which provide full compatibility with System V IPC. They, however, require a wide range of OS modifications. Moreover, UDSs are originally designed for intra-core communication at the OS level. This means that base applications need to be modified to use network sockets, such as TCP/IP and UDP/IP, for inter-core communication, instead of UDSs.

The major distributed middleware approaches include CORBA [OMG 04], Globe [Steen 99], MPI [MPI 97] and DSOC [Paulin 06]. While these approaches support inter-core communication at the user level, they require extensive modification of base applications because APIs for the inter-core communication are different from OS system calls used for intra-core communication. In particular, distributed middleware approaches for System V IPC include SHOC IPC [Tan 02], which provides the same API as System V IPC at the user level. It, however, fails to support System V IPC message queues. Moreover, no distributed middleware approaches in terms of UDSs are unknown.

By way of contrast, our approach requires no modification of legacy base applications in terms of System V IPCs and UDSs because the approach hooks OS system calls at the user level.

# 3.3 SYSTEM V μ-IPC

System V μ-IPC extends the function of System V IPC to both intra-core and inter-core communication. It provides the same IPC objects: semaphores, messages queues and shared memories, supporting the same APIs as System V IPC. Here, "μ" (*i.e.*, *mu*) stands for the respective initials: "*m*ulti-core" and "*u*ser-level."

## 3.3.1 OVERVIEW

FIGURE 3.1 illustrates two important components of System V μ-IPC: a System V μ-IPC library and a System V μ-IPC process.

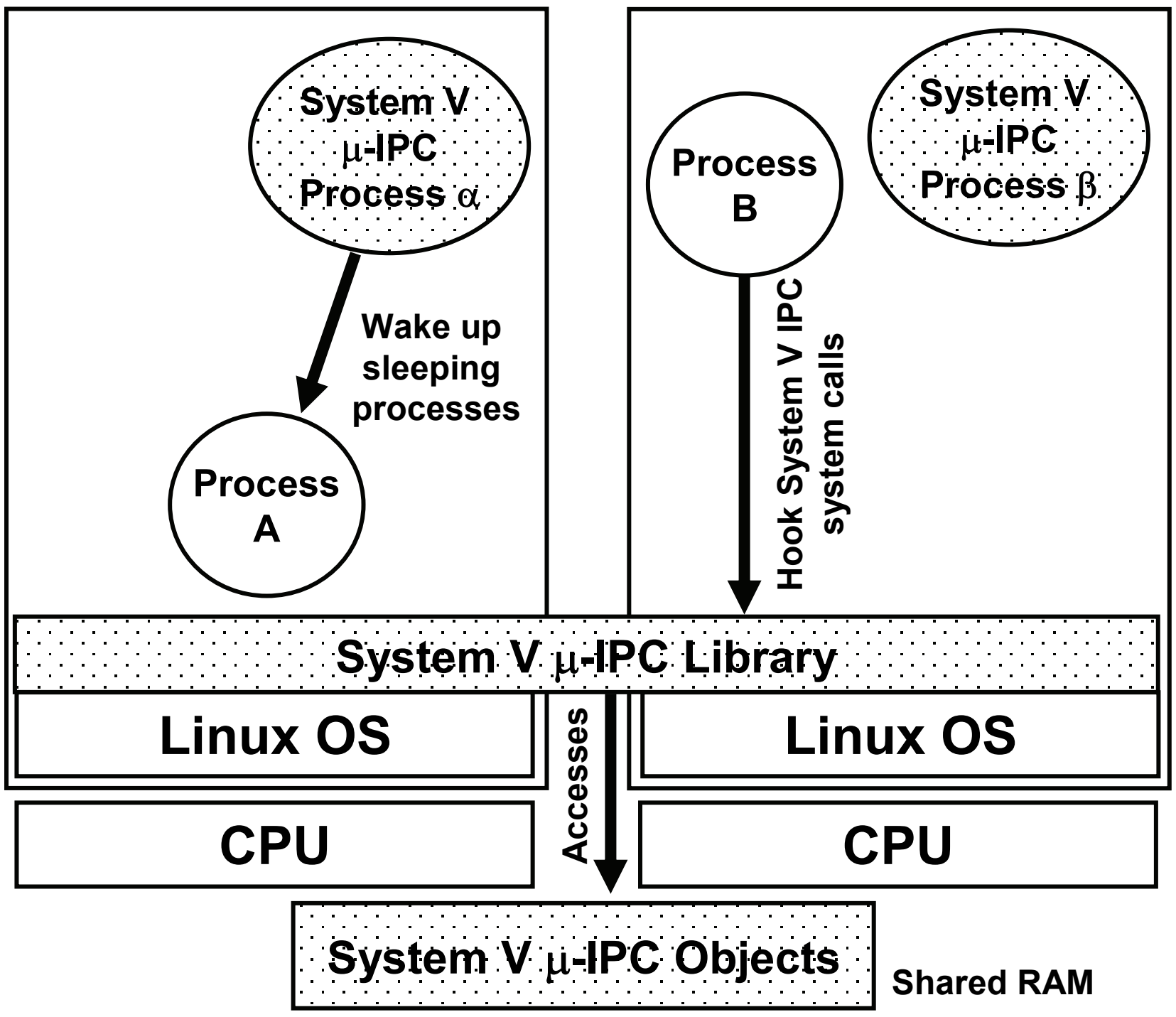

**FIGURE 3.1: SYSTEM μ-IPC COMPONENTS**

A System V μ-IPC library is linked to the base applications which need inter-core communication. The user-level library hooks system calls with respect to System V IPC, executing the system calls within it. Here, the library manages IPC objects in a shared RAM. For example, shared memories of System V IPC are directly mapped into part of the shared RAM. Control blocks required for both semaphores and message queues of System V IPC are stored in another part of the shared RAM. A System V μ-IPC process is a process which helps wake up sleeping processes at the user-level instead of an OS. For example, System V μ-IPC process α wakes up sleeping process A, which waits for a semaphore or a message queue.

FIGURE 3.2 describes how two components of System V μ-IPC work on a multi-core processor.

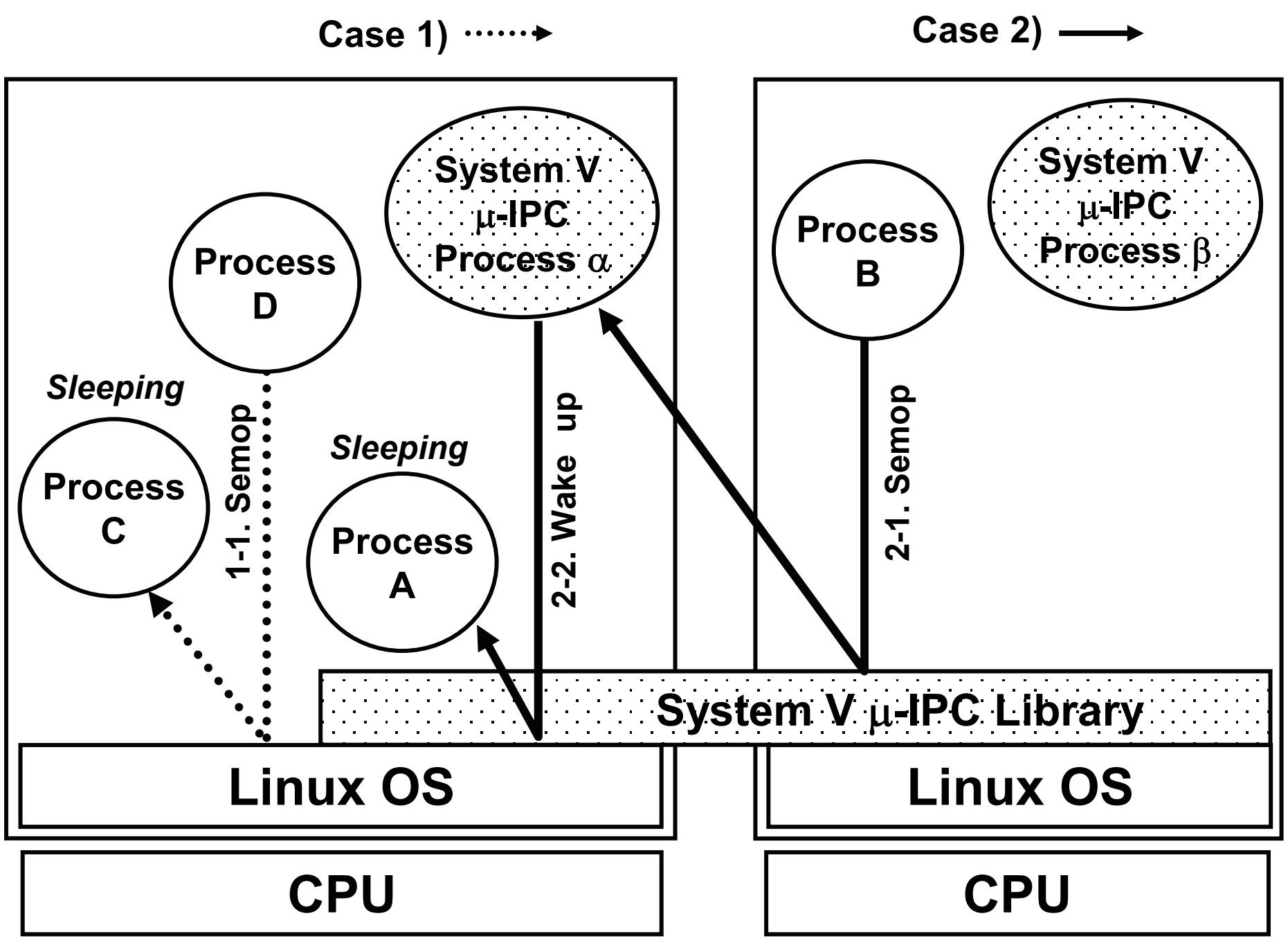

FIGURE 3.2: SYSTEM μ-IPC SEMAPHORE OBJECTS

Two typical cases are shown with semaphore objects: 1) when process D issues a *semop* system call to do an *up* operation to a semaphore, a System V μ-IPC library hooks the system call. Then, the library directly wakes up process C executed on the same processor since the library notices that process C has been waiting for the semaphore; and 2) when process B issues a *semop* system call to do an *up* operation to a semaphore, the System V μ-IPC library hooks the system call in a similar way. Next, the library requests System V μ-IPC process α to wake up process A executed on a different processor since the library notices that process A has been waiting for the semaphore. Finally, System V μ-IPC process α receives the wake-up request, waking up process A.

### 3.3.2 DESIGN

FIGURE 3.3 shows the internal design of our System V μ-IPC library. A System V μ-IPC library consists of four components: an API adaptor, a mutual exclusion component, a memory allocator, and a process controller. An API adaptor provides System V APIs for semaphore, message queue, and shared memory objects. A mutual exclusion component offers the function of a swap-based mutual exclusion mechanism to the API adaptor so that the API adaptor can correctly access shared data. A memory allocator implements a buddy memory allocation technique [Knuth 97], which provides any size of memory from a shared RAM to the API adaptor. A process controller sends a System V μ-IPC process a control message through Inter-Processor Interrupt (IPI) so that the System V μ-IPC can wake up a sleeping process. Here, a control message contains the process ID of a sleeping process.

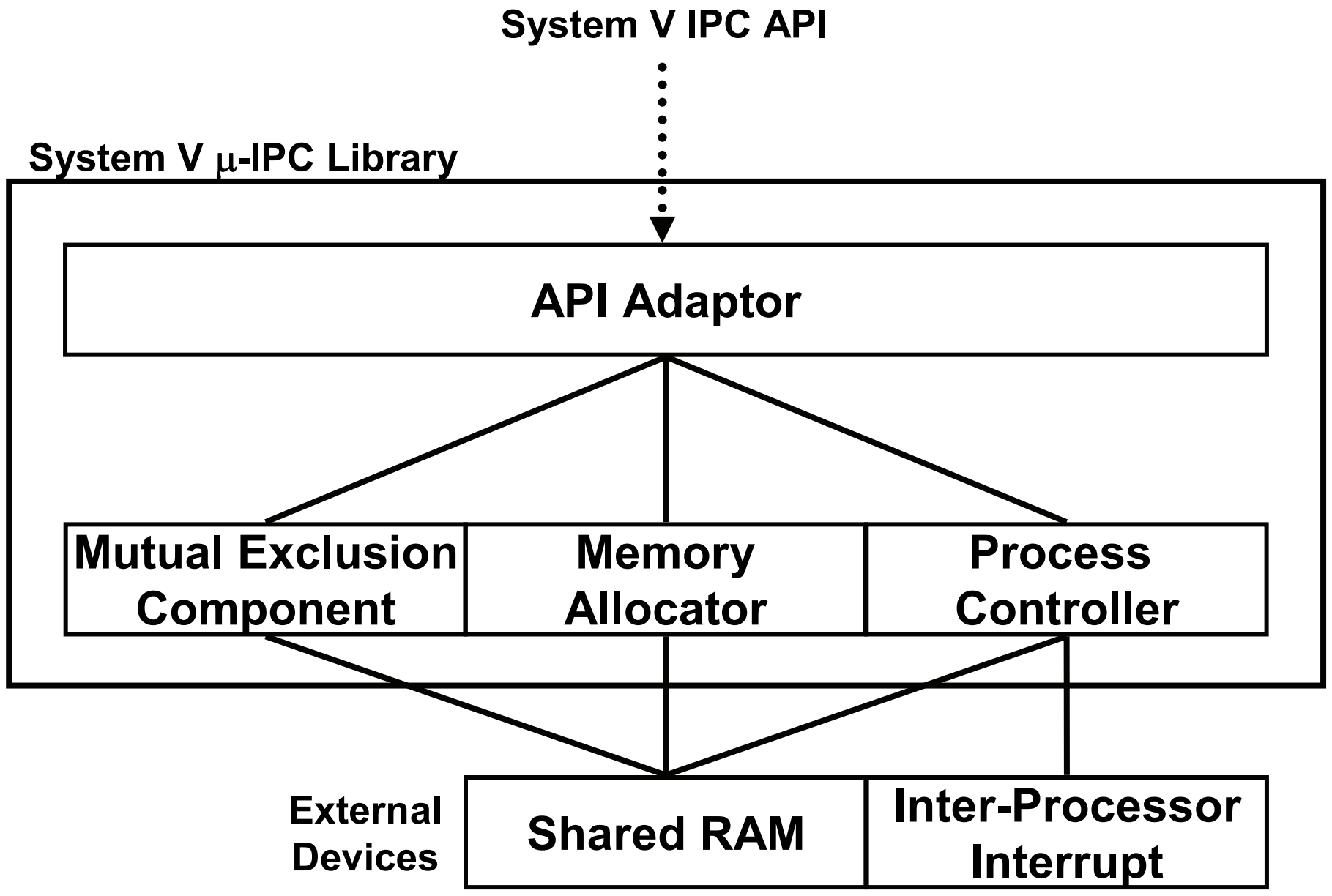

**FIGURE 3.3: INTERNAL DESIGN OF SYSTEM μ-IPC LIBRARY**

FIGURE 3.4 and FIGURE 3.5 show detailed operation flows of a System V μ-IPC Process. In these figures, process 3 wakes up two sleeping processes executed on a different processor, process 1 and 2, through two semaphore objects.

A System V μ-IPC process sleeps through an *ioctl* system call, waiting for an interrupt. Then, process 3 issues a *semop* system call to do an *up* operation to a semaphore. The System V μ-IPC library linked to process 3 hooks the system call. After that, the library puts a control message into a shared RAM with an IPI in order to wake up process 1. Further, process 3 issues a *semop* system call to do an *up* operation to another semaphore. In a similar way, the System V μ-IPC library puts another control message into the shared RAM in order to wake up process 2. The second control message is simply linked to the previous control message without any IPIs. This means that multiple IPIs are not simultaneously sent to the System V μ-IPC process in order to avoid overlapping interrupts. Once an IPI driver receives an IPI, it wakes up the System V μ-IPC process through a signal. In this way, the System V μ-IPC library wakes up a sleeping System V μ-IPC process.

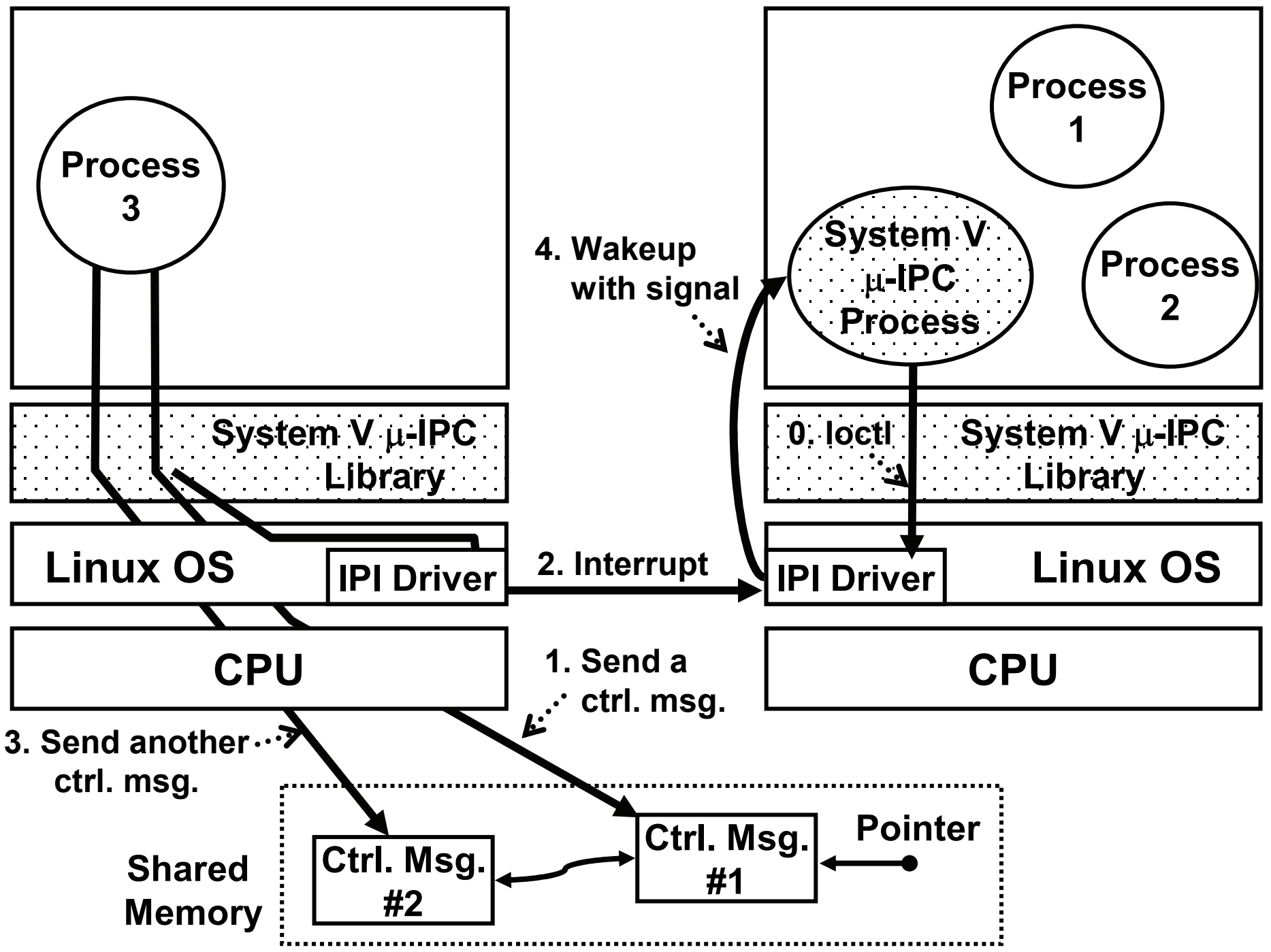

**FIGURE 3.4: OPERATION FLOW OF SYSTEM μ-IPC PROCESS (PART I)**

Second, the awakened System V μ-IPC process retrieves two control messages from a linked list in the shared RAM, waking up two sleeping processes, process 1 and process 2. When the System V μ-IPC process has finished retrieving any control messages, it re-issues an *ioctl* system call in order to wait for an interrupt. In this way, process 3 wakes up two sleeping processes, process 1 and process 2, through a System V μ-IPC process.

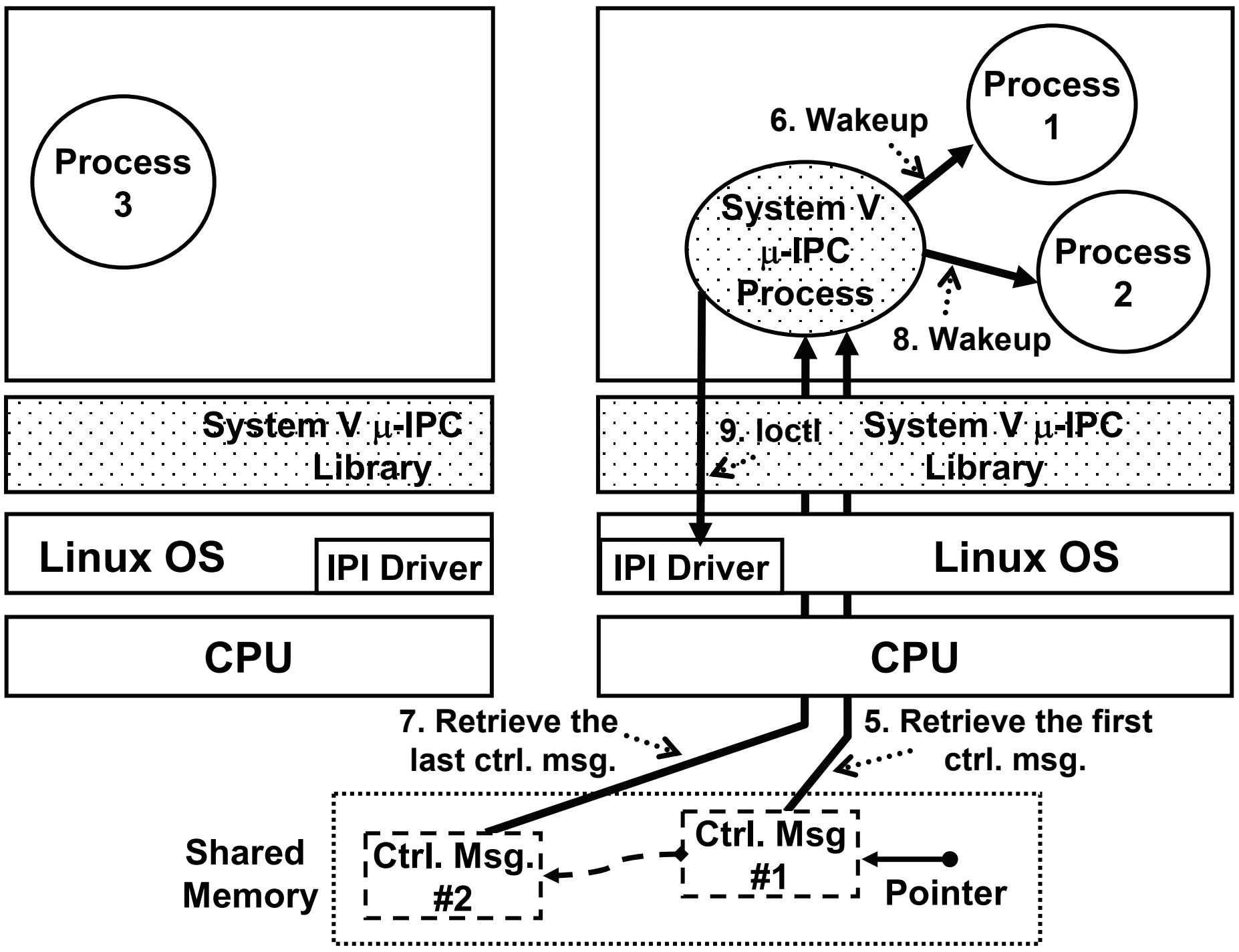

**FIGURE 3.5: OPERATION FLOW OF SYSTEM μ-IPC PROCESS (PART II)**

## 3.4 μ-UDS

μ-UDS extends the function of UDSs to both intra-core and inter-core communication. This user-level software supports the same APIs as UDSs. Here, "μ" (*i.e.*, *mu*) stands for the respective initials: "*m*ulti-core" and "*u*ser-level."

### 3.4.1 OVERVIEW

FIGURE 3.6 illustrates two important components of μ-UDS: a μ-UDS library and a μ-UDS process. A μ-UDS library is linked to the base applications which need inter-core communication (*i.e.*, server applications). The user-level library hooks two system calls, *bind* and *close*, notifying μ-UDS processes executed on different processors of the hooked system calls. A μ-UDS process is a process which helps handle system calls notified from a μ-UDS library at the user level. For example, a μ-UDS process binds a new socket in response to *bind* notifications and closes a socket in response to *close* notifications. In addition, μ-UDS processes support data transfer between different processors.

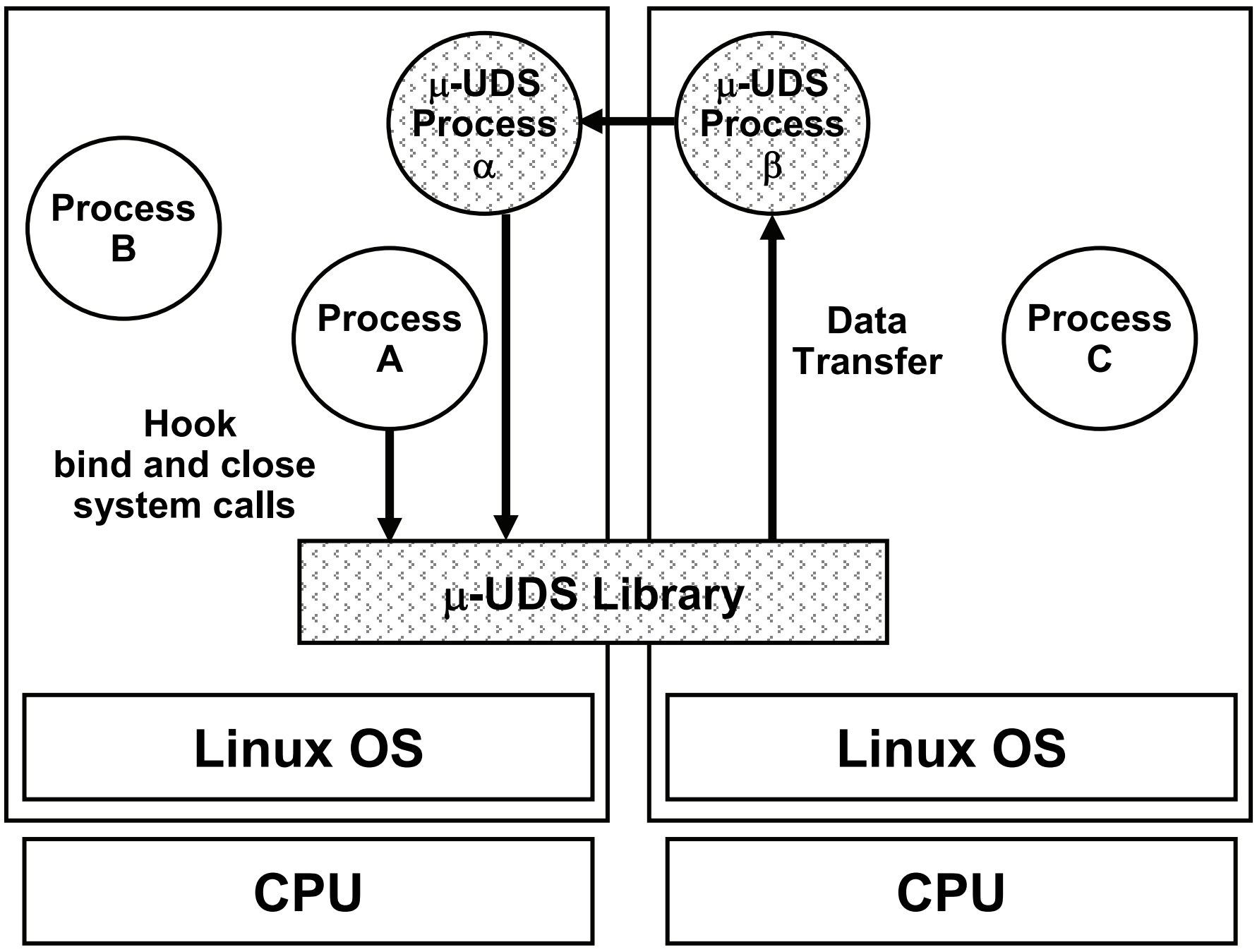

**FIGURE 3.6: μ-UDS COMPONENTS**

FIGURE 3.7 shows how two μ-UDS components work for a *bind* system call. When process A issues a *bind* system call to a UDS, a μ-UDS library hooks the system call. Next, the library notifies μ-UDS processes on different processors of the system call. μ-UDS process β issues a *bind* system call in order to receive data from client processes (*e.g.*, process C). Finally, the library executes a *bind* system call in order to bind a socket to process A.

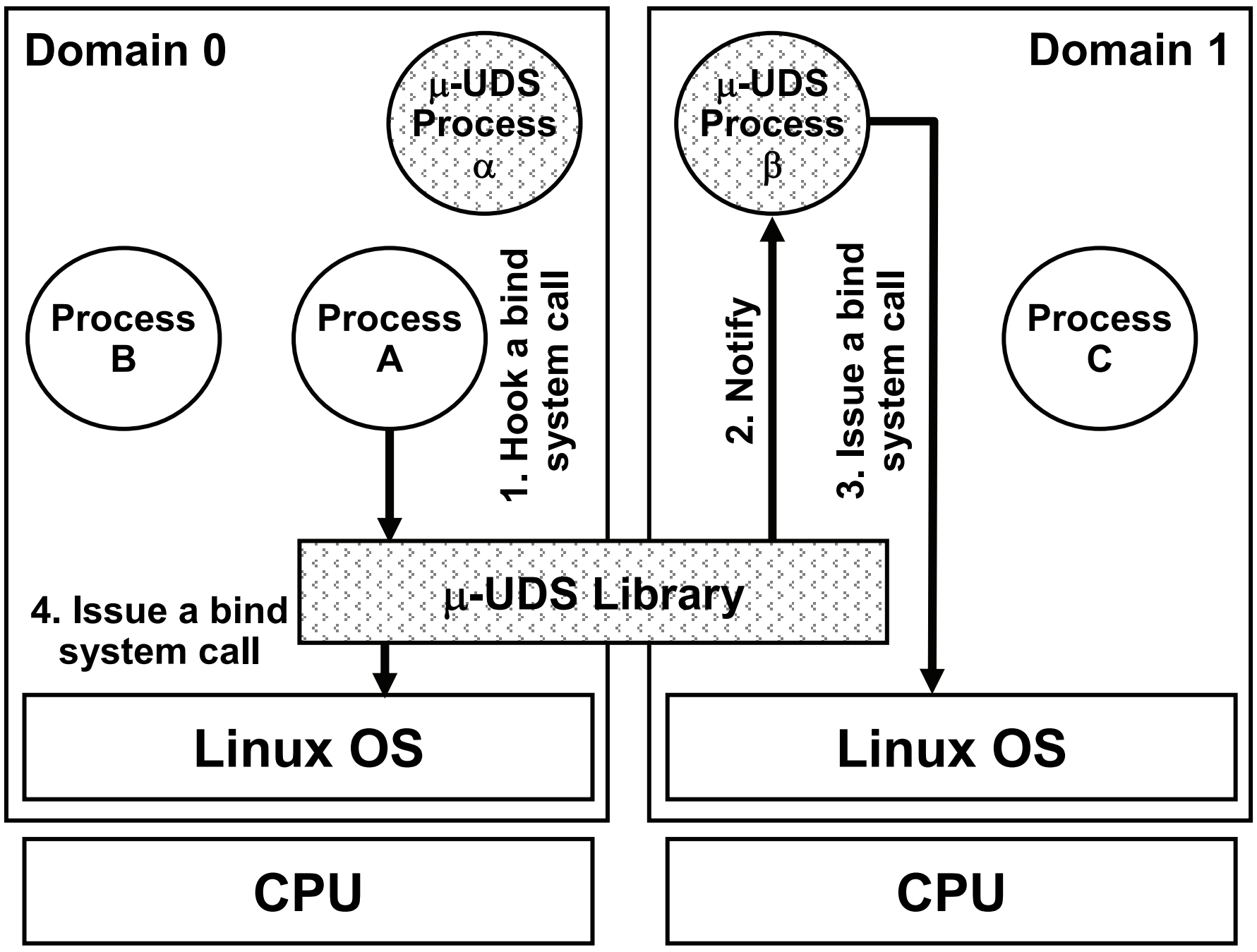

FIGURE 3.7: BIND SYSTEM CALL OF μ-UDS

FIGURE 3.8 shows how two μ-UDS components work for data transfer among processes. Here, two typical cases are shown: 1) when process B sends data to a UDS of an OS, process A receives the data from the OS; and 2) when process C sends data to an OS, μ-UDS process β receives the data from the OS. Next, μ-UDS process β transfers the data to μ-UDS process α. Finally, μ-UDS process α sends the data to process A.

FIGURE 3.9 shows how two μ-UDS components work for a *close* system call. When process A issues a *close* system call to a UDS, a μ-UDS library hooks the system call. Next, the library notifies μ-UDS processes on different processors of the system call. μ-UDS process β issues a *close* system call. Finally, the library executes a *close* system call in order to close a socket bound to process A.

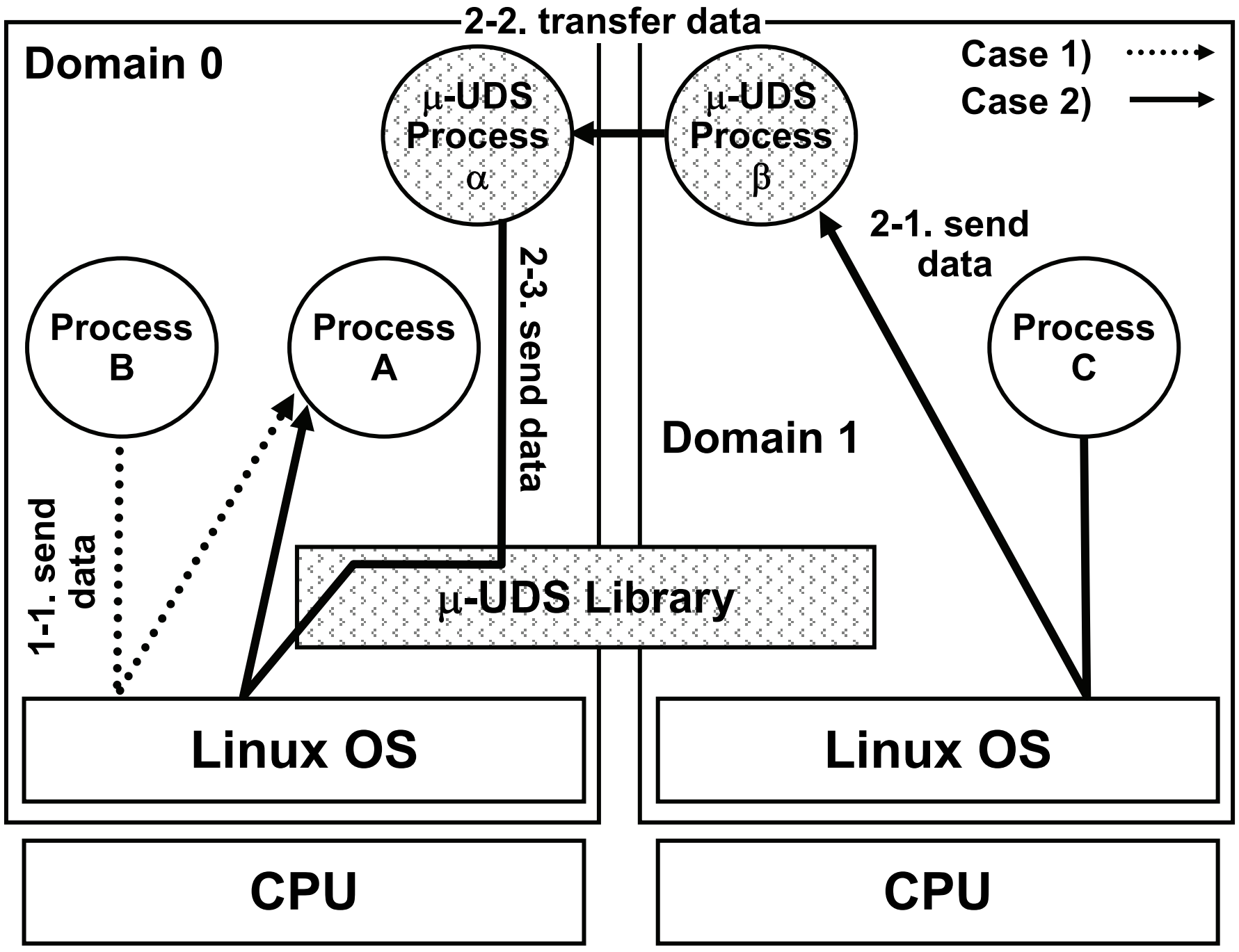

FIGURE 3.8: DATA TRANSFER OF µ-UDS

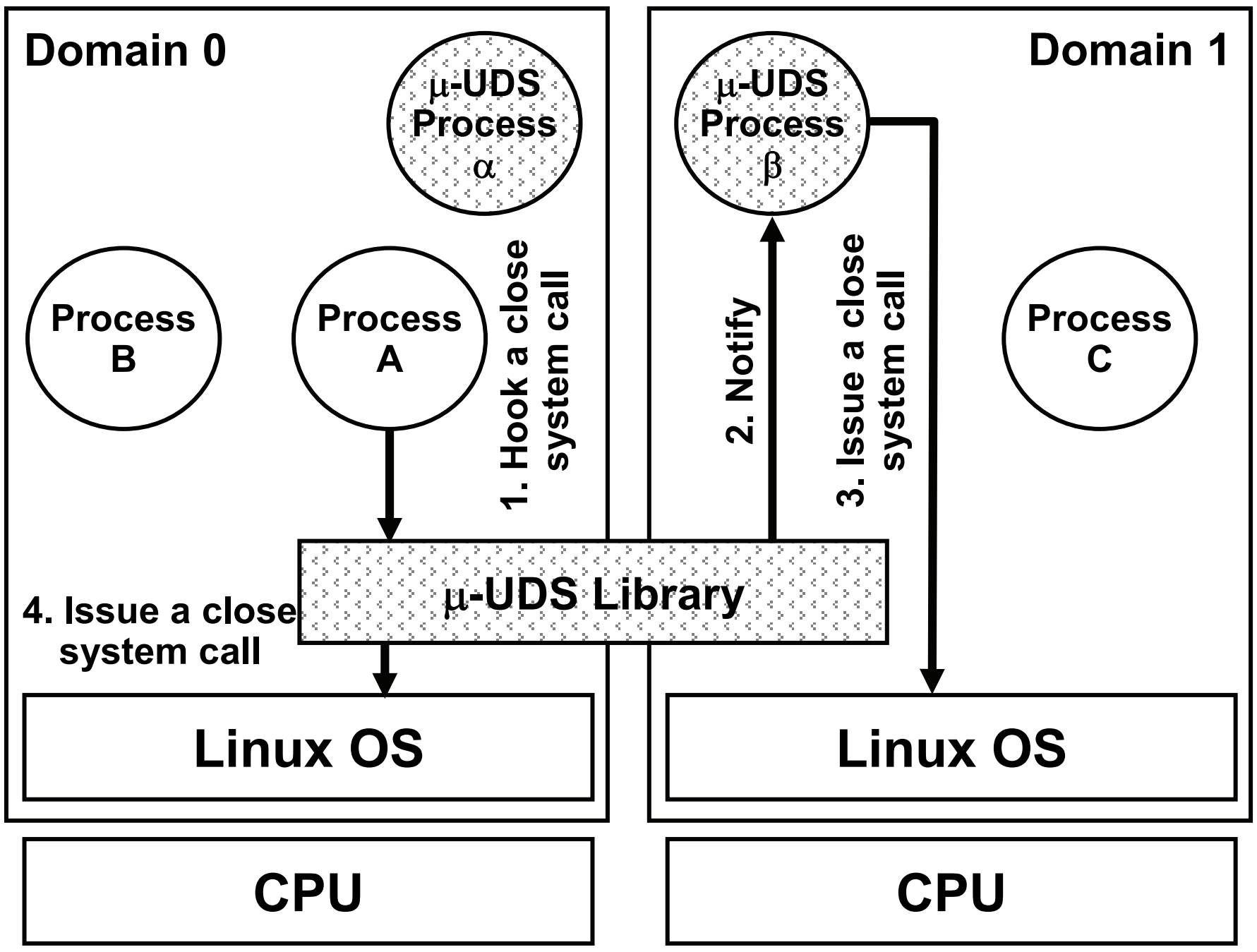

FIGURE 3.9: CLOSE SYSTEM CALL OF µ-UDS

### 3.4.2 DESIGN

FIGURE 3.10 depicts the internal design of our μ-UDS library. A μ-UDS library consists of four components: an API adaptor, a database manager, a μ-UDS communicator, and a system call handler. An API adaptor provides two same APIs as *bind* and *close* system calls. A database manager manages a database based on both a shared memory object and a semaphore object of System V μ-IPC. Each entry in the database contains the ID of a server process using a UDS, the name bound to the UDS, and the socket file descriptor corresponding to the UDS. A μ-UDS communicator sends μ-UDS processes the arguments given from *bind* and *close* system calls via message queue objects of System V μ-IPC. A system call handler executes system calls requested from an API handler.

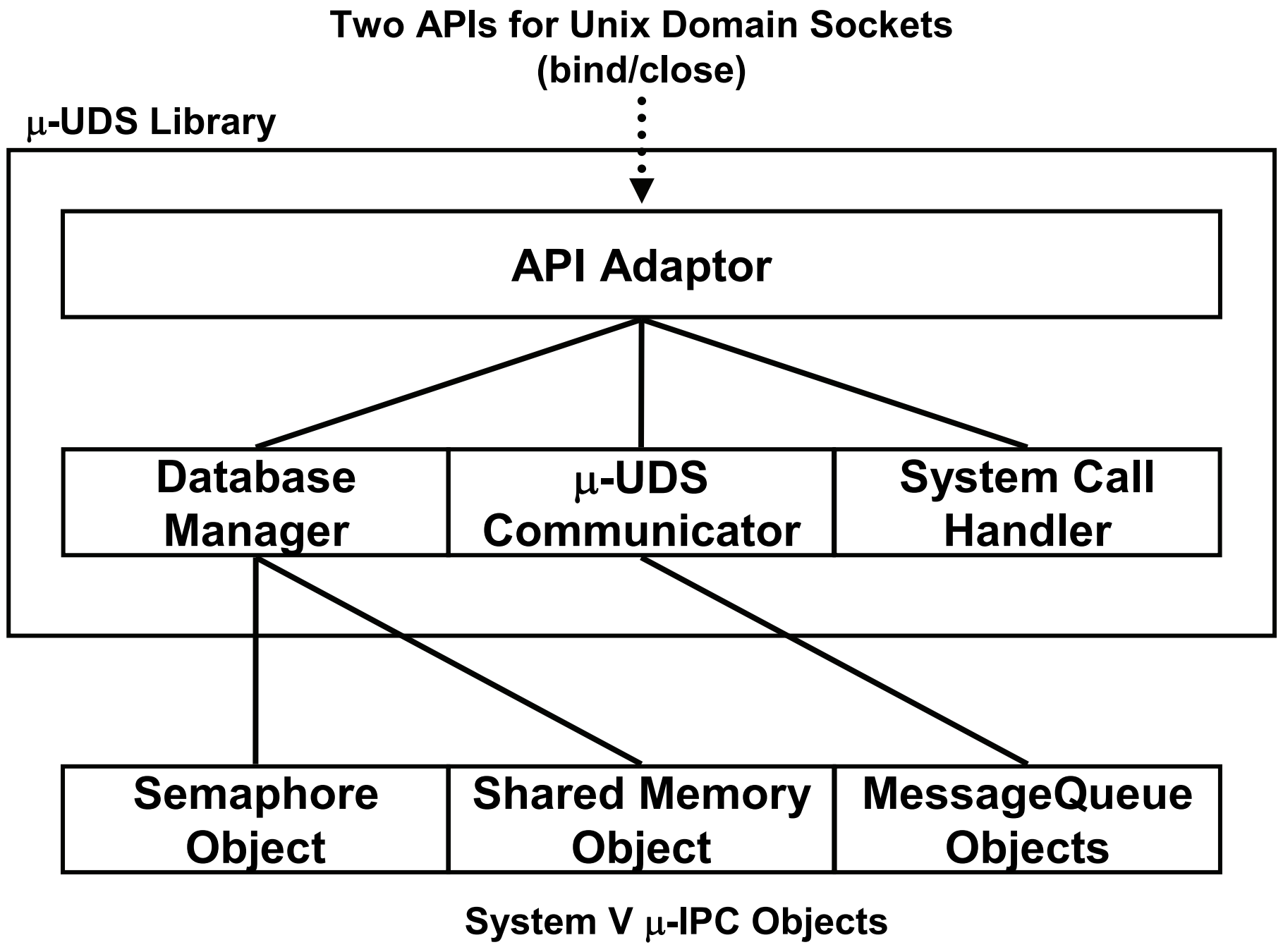

**FIGURE 3.10: INTERNAL DESIGN OF μ-UDS**

FIGURE 3.11, FIGURE 3.12, and FIGURE 3.13 show the detailed operation flows of a μ-UDS process. A μ-UDS process consists of two components: a main controller, and communication threads. Here, the operation flows are corresponding to the examples shown in FIGURE 3.7, FIGURE 3.8, and FIGURE 3.9, respectively.

FIGURE 3.11 illustrates how a μ-UDS process internally handles a *bind* system call (see FIGURE 3.7). First, μ-UDS process β receives the notification of a *bind* system call from a μ-UDS library. The main controller of μ-UDS process β creates a new communication thread. After that, the thread issues a *bind* system call in order to receive data from process C.

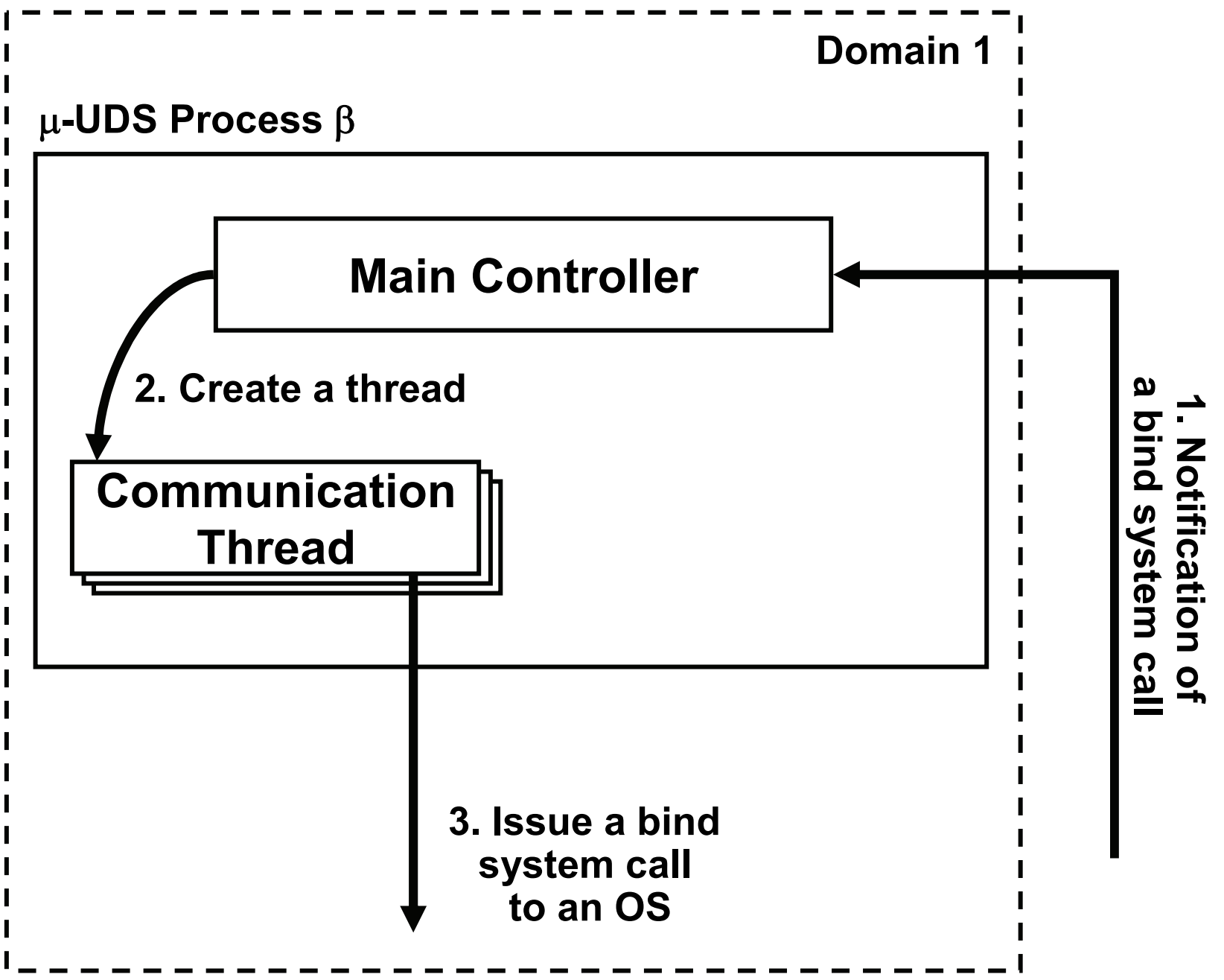

**FIGURE 3.11: OPERATION FLOW OF μ-UDS (BIND SYSTEM CALL)**

FIGURE 3.12 describes how μ-UDS processes internally transfer data between processors (see FIGURE 3.8). A communication thread created by μ-UDS process β receives data from process C, transferring the data to the main controller of μ-UDS process α. Then, the main controller of μ-UDS process α sends the data to process A.

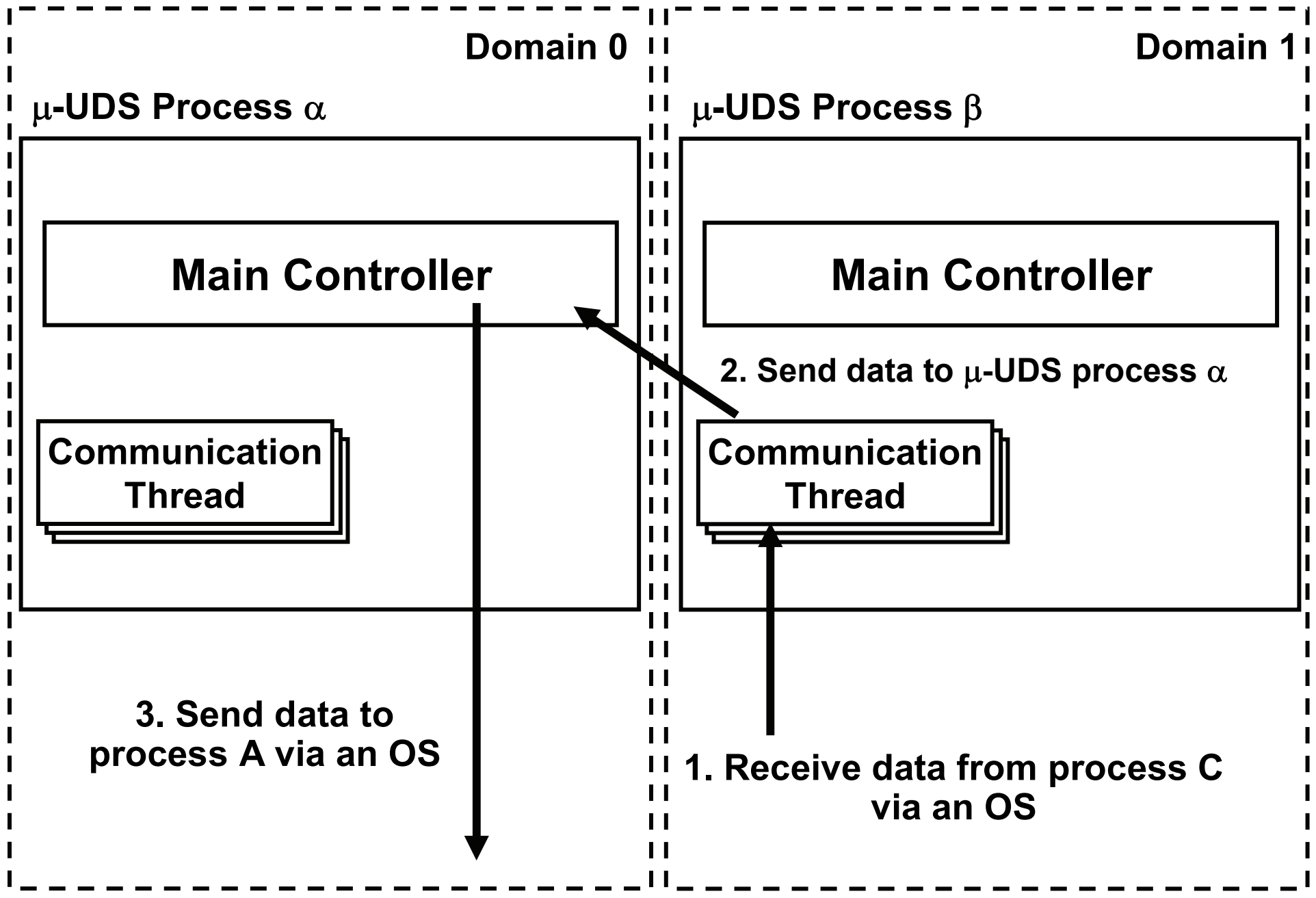

**FIGURE 3.12: OPERATION FLOW OF μ-UDS (DATA TRANSFER)**

FIGURE 3.13 depicts how a μ-UDS process internally handles a *close* system call (see FIGURE 3.9). First, μ-UDS process β receives the notification of a *close* system call from a μ-UDS library. After that, the main controller of μ-UDS process β destroys a communication thread corresponding to a socket bound to process A. Finally, the main controller issues a *close* system call in order to close the socket.

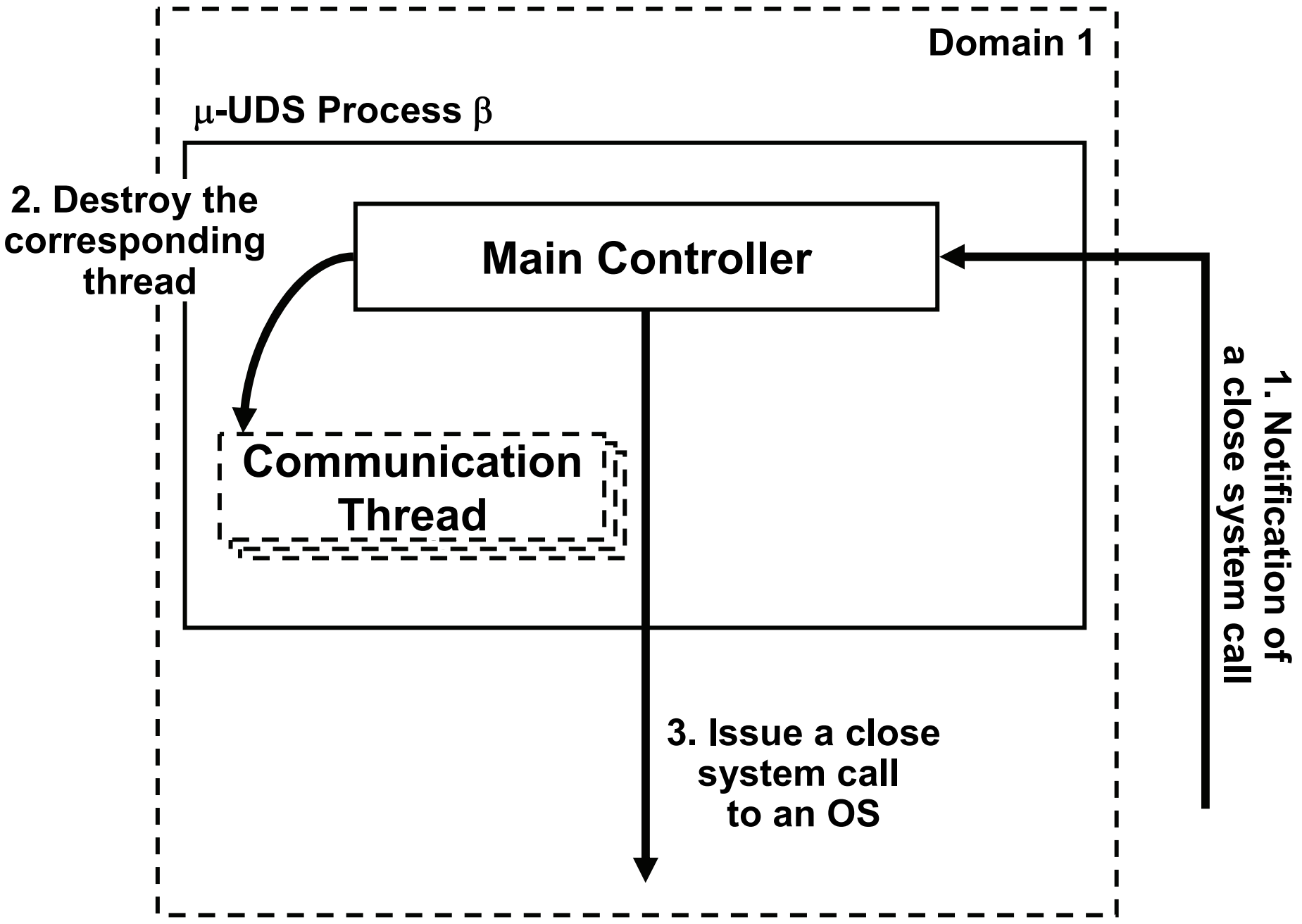

**FIGURE 3.13: OPERATION FLOW OF μ-UDS (CLOSE SYSTEM CALL)**

## 3.5 EVALUATION

TABLE 3.2 summarizes our AMP evaluation environment, called MP211 [Torii 05]. MP211 is a mobile application processor which equips with three ARM processors. Interestingly, the area of three processors occupies only 15% of this whole SoC, as shown in FIGURE 3.14.

**TABLE 3.2: AMP EVALUATION ENVIRONMENT**

| Item | Feature |
|---|---|
| Processors | ARM926EJ-S x 3 |
| Cache | I: 16KB, D: 16KB |
| Frequency | ARM: 200MHz, Bus: 100MHz |
| OS | Linux |

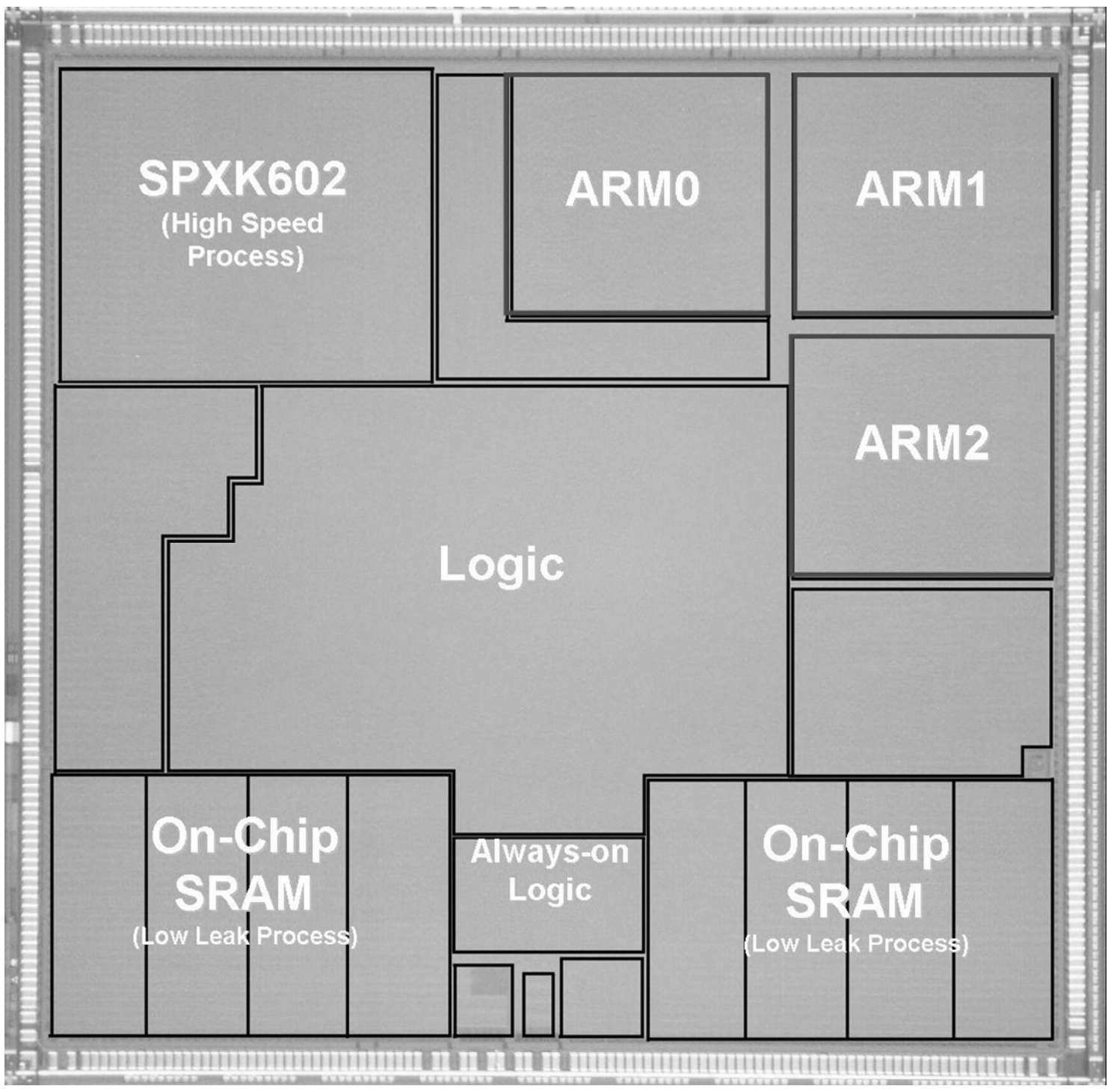

**FIGURE 3.14: MP211 AS AMP EVALUATION ENVIRONMENT**

Evaluations show that our seamless communication software actually worked on AMP (see Section 3.5.1). Further, both System V μ-IPC and μ-UDS achieved higher performance (see Section 3.5.2, and 3.5.3) and lower code size (see Section 3.5.4) than did other approaches.

### 3.5.1 SUCCESSFUL EXAMPLE FOR SEAMLESS IPCS

As a successful example of the application of our seamless communication software, FIGURE 3.15 demonstrates that the menu screen of an in-house Linux-based mobile terminal on MP211.

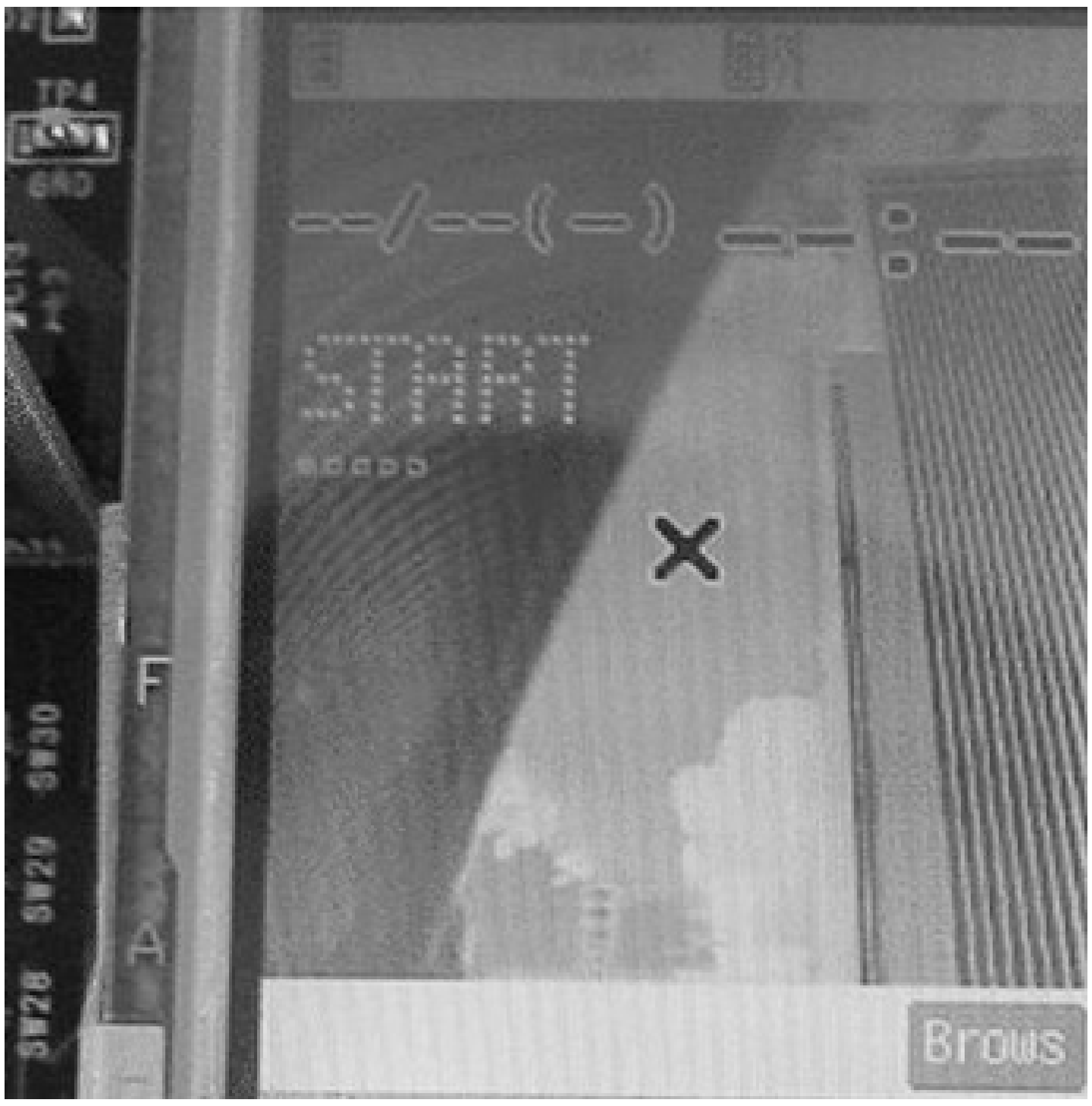

**FIGURE 3.15: LINUX-BASED MOBILE TERMINAL ON AMP**

This means that the seamless communication software enabled actual in-house Linux-based mobile terminal software designed for a single processor to be run on three processors without software modifications. *This result is especially remarkable because it cannot be achieved with other approaches.* Further, the performance overhead of the seamless communication software was small enough because its performance overhead was only 0.1% for the original in-house mobile terminal software.

### 3.5.2 SYSTEM V μ-IPC PERFORMANCE

In order to study the performance effectiveness of our System V μ-IPC, we measured the execution time of System V IPC systems calls. We compared our System V μ-IPC with a distributed System V IPC of SHell Over a Cluster (SHOC) [Tan 02] as an available reference. This comparison must be interpreted very carefully because SHOC is implemented on PC clusters.

FIGURE 3.16 shows the normalized execution time of three system calls: a *semop* system call to a remote semaphore object, a *shmat* system call to a shared memory object, and a *shmdt* system call to a shared memory object. Here, the execution time of a *semop* system call to a local semaphore object is normalized to 1. Values in parenthesis indicate the measured execution time with respect to each system call for the reference. This evaluation demonstrates that our System V μ-IPC achieved roughly 5.0 times faster normalized execution time (13.8 times faster measured execution time) than did SHOC. Thus, our System V μ-IPC is efficiently designed for AMP with shared memory.

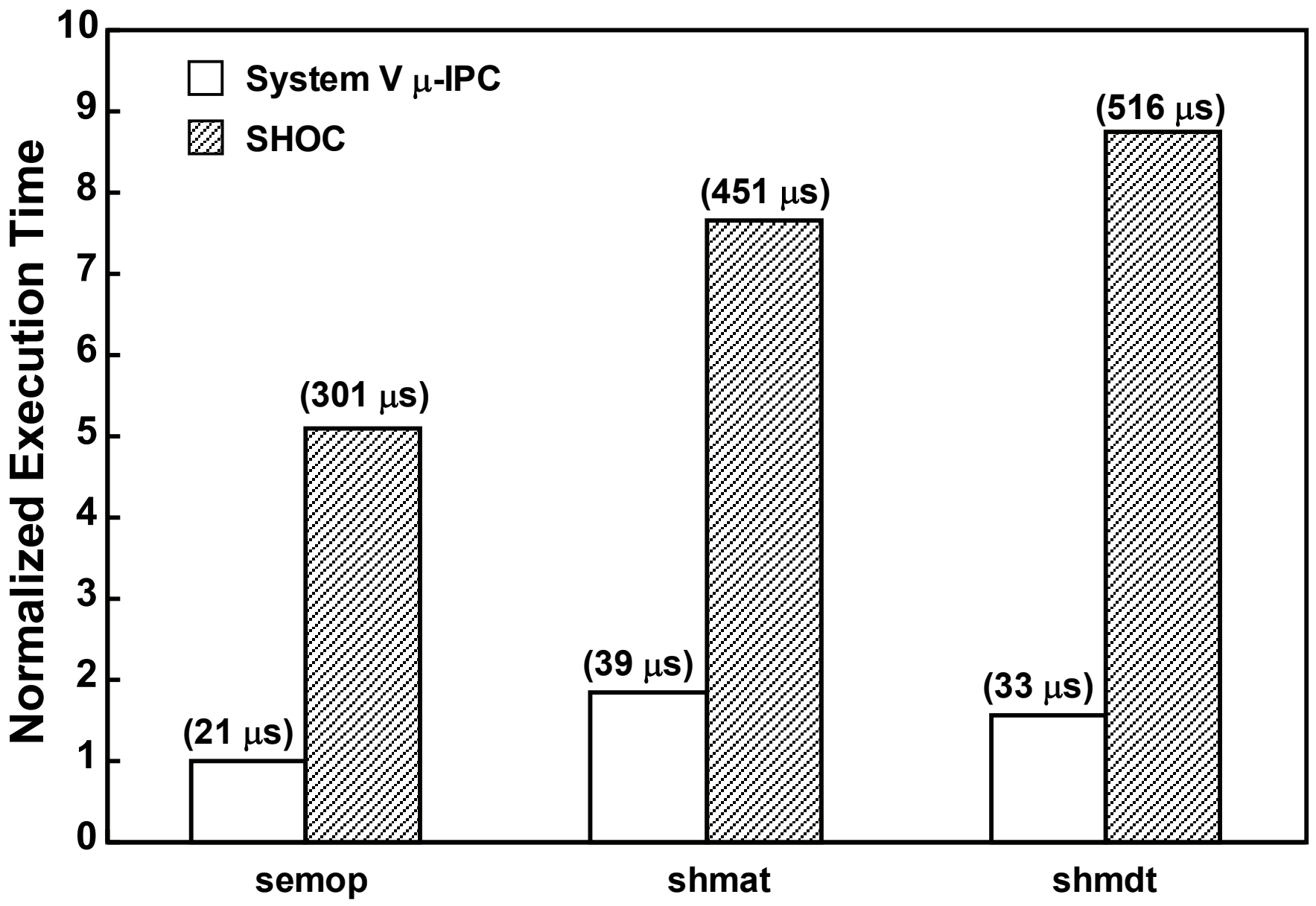

**FIGURE 3.16: SYSTEM V μ-IPC PERFORMANCE**

It should be noted that, in our System V μ-IPC, the absolute execution time of the compared system calls is not necessarily slow although we used the normalized execution time for the comparison of different platforms. With respect to the execution time of a *semop* system call to a local semaphore object (*i.e.*, the base-reference execution time on a single processor), our System V μ-IPC achieves the execution time of 21 μs on a 200-MHz processor while SHOC achieves the execution time of 59 μs on a 450-MHz processor.

### 3.5.3 μ-UDS PERFORMANCE

In order to study the performance effective of our μ-UDS, we measured the bandwidth of both our μ-UDS and Linux network sockets. FIGURE 3.17 shows the bandwidths of connection-less communication on two processors in the transfer of 1 MB of data in fragment sizes of 1KB, 4 KB and 16 KB.

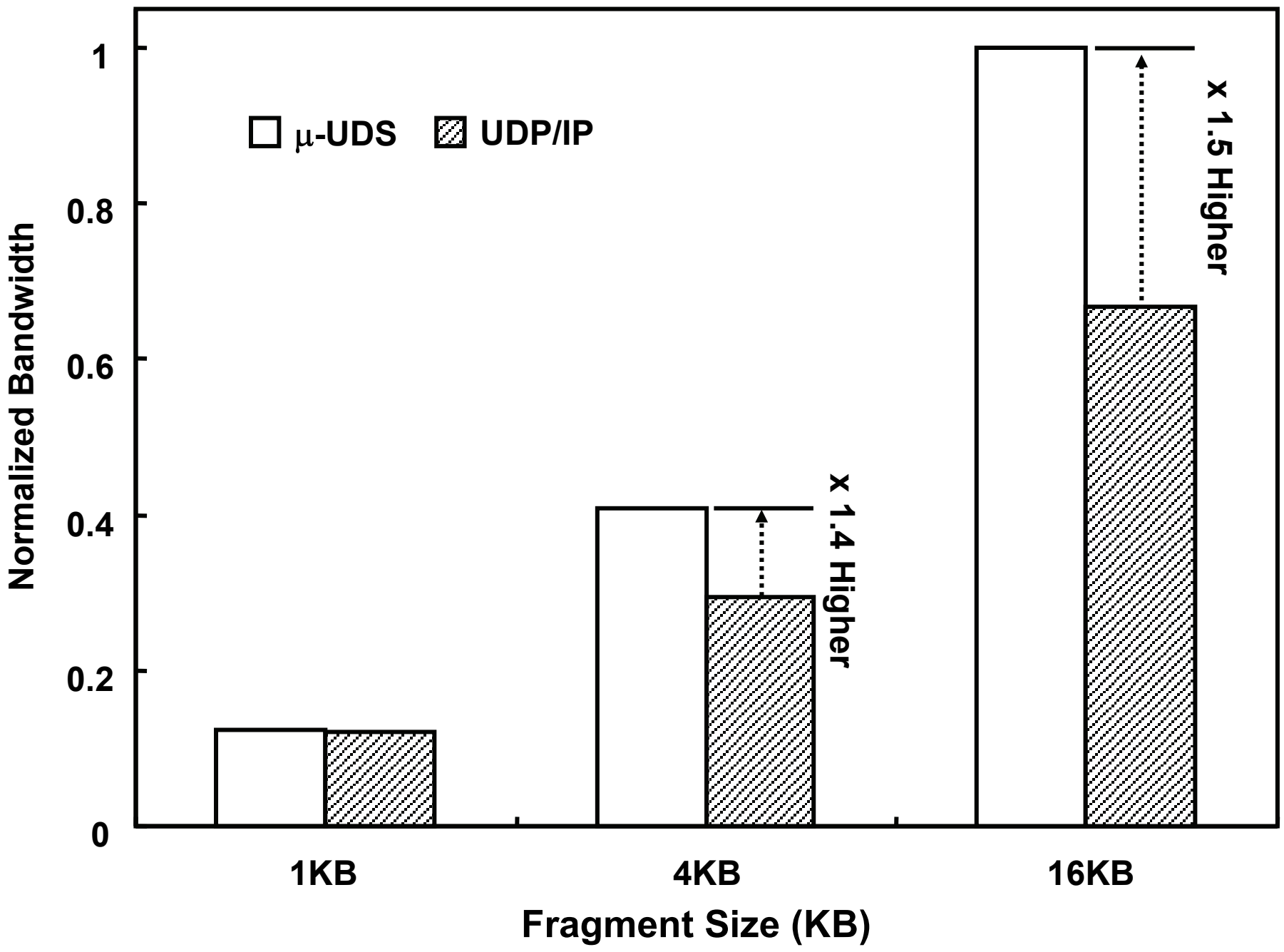

FIGURE 3.17: μ-UDS PERFORMANCE

In fragment sizes of 4KB and 16KB, our μ-UDS achieved roughly 1.5 times higher bandwidth than did kernel-based UDP/IP. This is because, instead of a complex protocol designed for the Internet, our μ-UDS internally uses a message queue object of System V IPC in order to transfer the data between two processors.

### 3.5.4 LINES OF CODE FOR SEAMLESS COMMUNICATION

In order to study the design effectiveness of both System V μ-IPC and μ-UDS, we counted their modified Lines of Code (LOC). First, we compared System V μ-IPC with two available references: a distributed OS, known as Distributed System V IPC in Locus [Fleisch 86], and a distributed middleware, known as DIPC [Sharif 99]. The modified LOC of System V μ-IPC was 3,853 LOC. Thus, System V μ-IPC is effectively implemented in the smallest size because the modified LOC of Locus was 5,559 LOC and the modified LOC of DIPC was 6,265 LOC. Next, we compared our μ-UDS with Internet domain sockets of Linux as a reference. It can be seen from the comparison that μ-UDS with 1,488 LOC is 13.7 times smaller than Internet domain sockets of Linux with 20,365 LOC.

## 3.6 SUMMARY

We have presented our seamless communication software, which enables legacy base applications designed for a single-core processor to be executed on AMP without any modifications of base applications and OSs. Its most important feature is the hooking and execution of OS system calls at the user level. In this way, the software maintains the API compatibility of IPC system calls on multiple processors. We have reported the detailed design of both System V μ-IPC and μ-UDSs as two most important IPCs. Moreover, our evaluations with respect to System V μ-IPC and μ-UDS have shown its effectiveness, demonstrating three fundamental features: a successful example with actual mobile terminal software, high performance and small code size.

# CHAPTER 4
# SECURE PROCESSOR PARTITIONING

This chapter presents a new hardware unit designed for secure processor partitioning, called a bus management unit. The unit enables OSs executed on separate processors to be mutually protected at the hardware level.

## 4.1 MOTIVATION

Current high-end embedded systems, such as mobile terminals, have already equipped with an execution environment designed for a closed class of open applications. For example, the approach of NTT DoCoMo enables applications written in Java to be executed on a Java VM [Gong 03]. The approach of KDDI enables pre-verified applications to be executed on a dedicated platform, called BREW [Qualcomm 04]. A similarity between both approaches is use of dedicated software platforms in order to form an execution environment.

The open applications that we target include general user-level programs and device drivers. This means that such new groups of open applications need to be directly executed on native OSs instead of traditional non-native software platforms. Sharing a native OS between base and open applications shows potential vulnerability since bugs and viruses include in open applications are likely to cause critical interference to base applications, such as unintentional access to prohibited areas, and loss of needed excess resource allocation. NTT DoCoMo, IBM, and Intel have, in fact, jointly announced specifications designed to encourage the development of mobile terminals having a security capability [NTT 04a] [NTT 04b].

Conventional secure techniques, which are implemented as software, help cover critical potential vulnerability. The secure techniques, however, cause additional potential vulnerability to systems in a vicious cycle since the software that implements the techniques itself might have new potential vulnerability. This means that one promising approach would seem to be the use of hardware.

Thus, we introduce secure processor partitioning, which enables native OSs executed on separate processors to be mutually protected by means of a new hardware unit, called a bus management unit. This unit prohibits a native OS compromised by malicious open applications from accessing other memory and I/O areas managed by other native OSs since processors are allowed to access only specified address ranges of memories and I/Os.

## 4.2 RELATED WORK

Our research differs in a number of respects from the current body of research on program security.

Major tool-level approaches include RATS [Fortify 09] and Splint [Evans 02]. While the approaches help find potential vulnerability by inspecting the source codes of open applications, the source codes of open applications are required before their execution.

Major library-level approaches include StackGuard [Cowan 98] and Libsafe

[Baratloo 00]. The libraries help detect malicious attacks, such as buffer overflow attacks, at run-time. Unlike tool-level approaches, these approaches require no source codes of open applications. There is, however, no guarantee that library-level approaches make it possible to detect all malicious attacks.

Major kernel-level domain approaches, such as SELinux [Loscocco 01] and Openwall [Openwall 01], allow both base and open applications to be run on a shared OS since a security module within the shared OS monitors system calls issued from all applications and imposes mandatory access control on all applications. It is, however, difficult to avoid security vulnerability in OSs and security modules.

Major virtualized domain approaches include LPAR [Armstrong 05], Xen [Barham 03], UML [Dike 00] and VMware [Sugerman 01]. Virtualized domains (i.e., virtualized native OSs) enhance system security since they allow base applications to be separated from open applications at the OS level. There is, however, a degree of security vulnerability in complex virtualization software [Hacker 07]. In accordance with the problem, recent work [Chen 08] [Seshadri 07] [Shinagawa 09] has proposed tiny virtualization software in order to reduce security vulnerability of virtualization software itself.

It should be noted that similar hardware approaches, such as SECA [Coburn 05] and distributed filters on Network-on-Chips (NoCs) [Fiolin 07], were proposed around the same time as our bus management unit in order to block illegal accesses at the hardware level. While our bus management unit helps enhance system-level program security, the similar approaches focus on only communication-level security.

## 4.3 BUS MANAGEMENT UNIT

FIGURE 4.1 outlines an example of AMP used for mobile terminals. In this figure, the AMP software structure equips with four domains: a base domain, an operator domain, a trusted domain, and an untrusted domain. Each domain has a native OS on a separate processor.

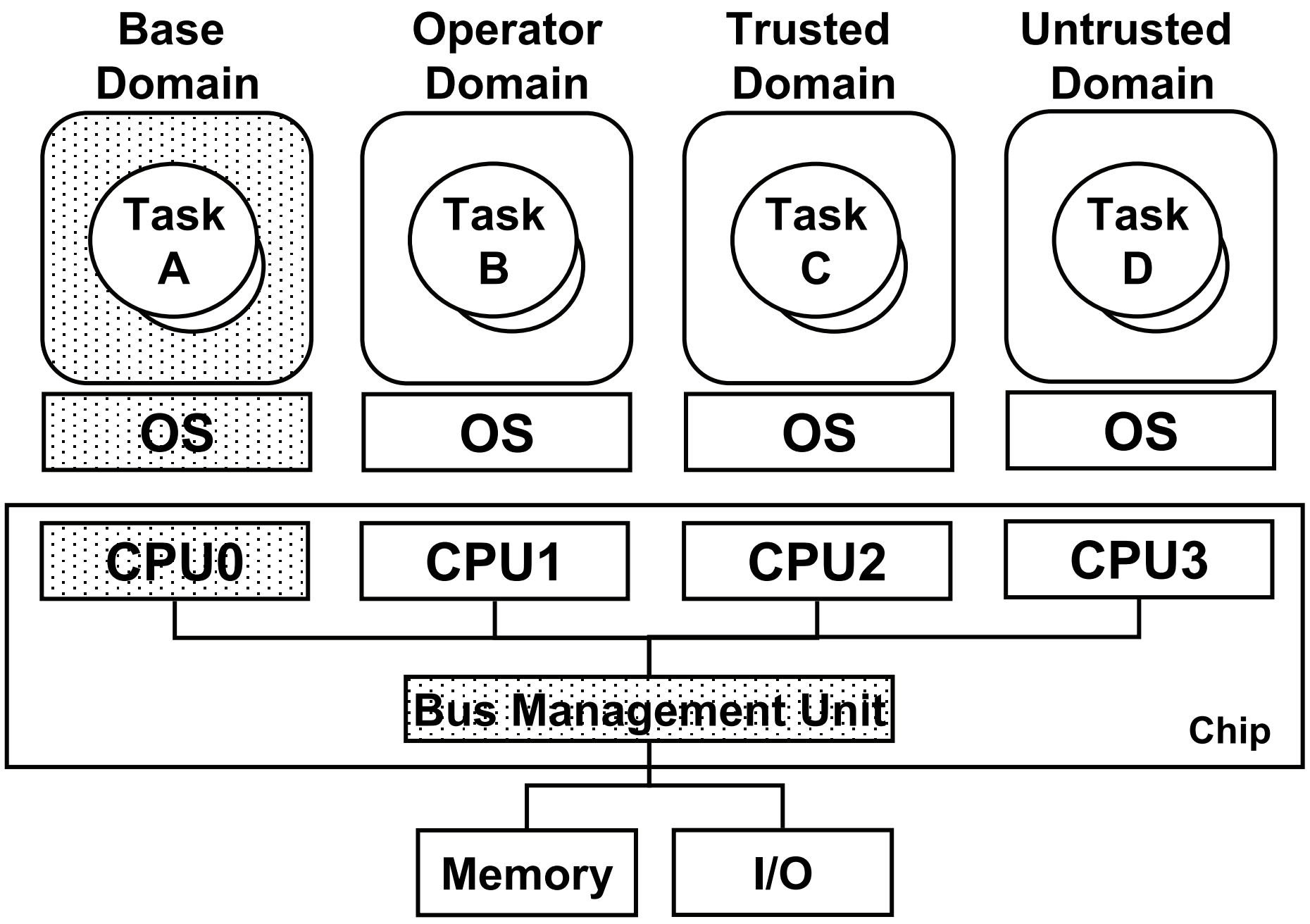

**FIGURE 4.1: AMP FOR MOBILE TERMINALS**

A base domain contains the base applications of a mobile terminal, such as a mailer and a browser. Open applications are never executed on a base domain. While an operator domain executes applications validated by an operator, a trusted domain executes applications validated by third parties. The open applications executed in both an operator domain and a trusted domain are assumed to be sufficiently trustworthy although they may include bugs. Finally, an untrusted domain executes all other open applications. The open applications might include viruses as well as bugs. In this way, users enable open applications to be installed to their corresponding classes of domains.

The requirements of our bus management unit are the achievement of two major functions under an important specification. Here, the two major functions of our bus management unit are 1) checking all accesses issued from processors that belong to open domains, and 2) deciding whether or not the accesses are illegal. In addition, the important specification is that only processors that belong to a base domain are allowed to change the control of the unit. This means that it is impossible for even incredibly malicious open applications to escape access checking since processors compromised by the open applications have no rights to change the control of our bus management unit. In this way, our platform ensures hardware-level hardened protection among domains. It should be noted that secure processor partitioning allows our bus management unit to be variously designed and implemented as far as the unit equips with the two major functions under the important specification.

FIGURE 4.2 describes an example of the design of our bus management unit. The unit contains two components: an access matrix and an access check component.

An access matrix stores the information on the accessibility of all possible combinations between processors and bus slaves (*e.g.*, memory or I/O). Then, only processors that belong to a base domain (*i.e.*, CPU0) are allowed to modify this access matrix.

An access check component checks the access information, such as processor ID, access type, and access address, of bus access issued from a processor to a bus slave. Then, the component determines whether the bus access should be granted, referring to an access matrix. In this figure, we assume that a base domain uses Slave A and Address Range 0, and both an operator domain and a trusted domain use Slave B and Address Range 1. While processors (*i.e.*, CPU1 and CPU2) executing in both an operator domain and a trusted domain are only allowed any access to resources used for both domains (*i.e.*, Slave B and Address Range 1), they are prohibited write access to resources used for a base domain (*i.e.*, Slave A and Address Range 0). Moreover, while processors (*i.e.*, CPU3) executing in an untrusted domain are only allowed read-only access to resources used for the operator domain and the trusted domain (*i.e.*, Slave B and Address Range 1), they are prohibited any access to resources used for the base domain (*i.e.*, Slave A and Address Range 0).

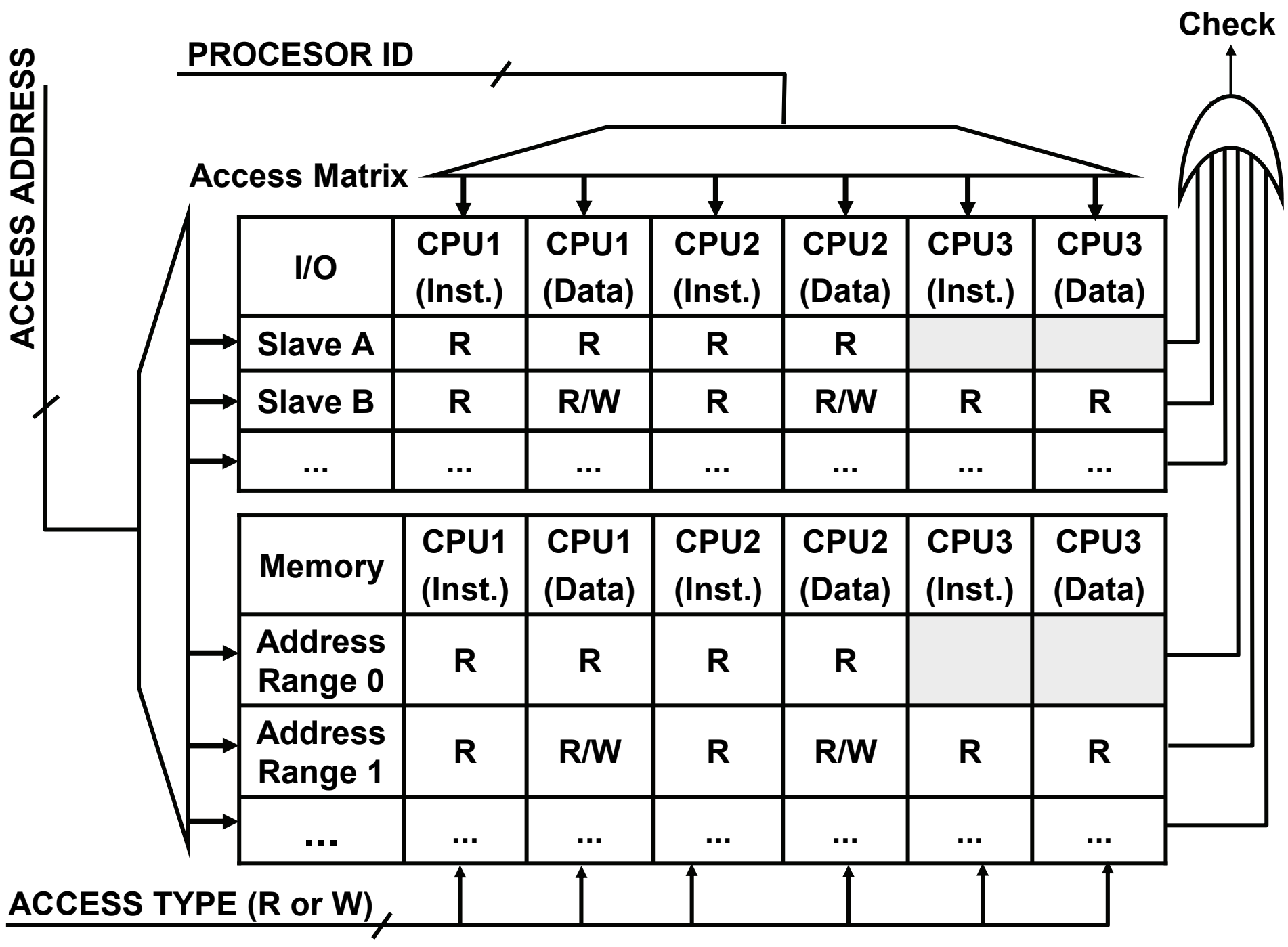

**FIGURE 4.2: BUS MANAGEMENT UNIT**

FIGURE 4.3 depicts an example of the internal design of a system bus with our bus management unit. Here, we assume that this system bus uses the AXI bus protocol [ARM 04].

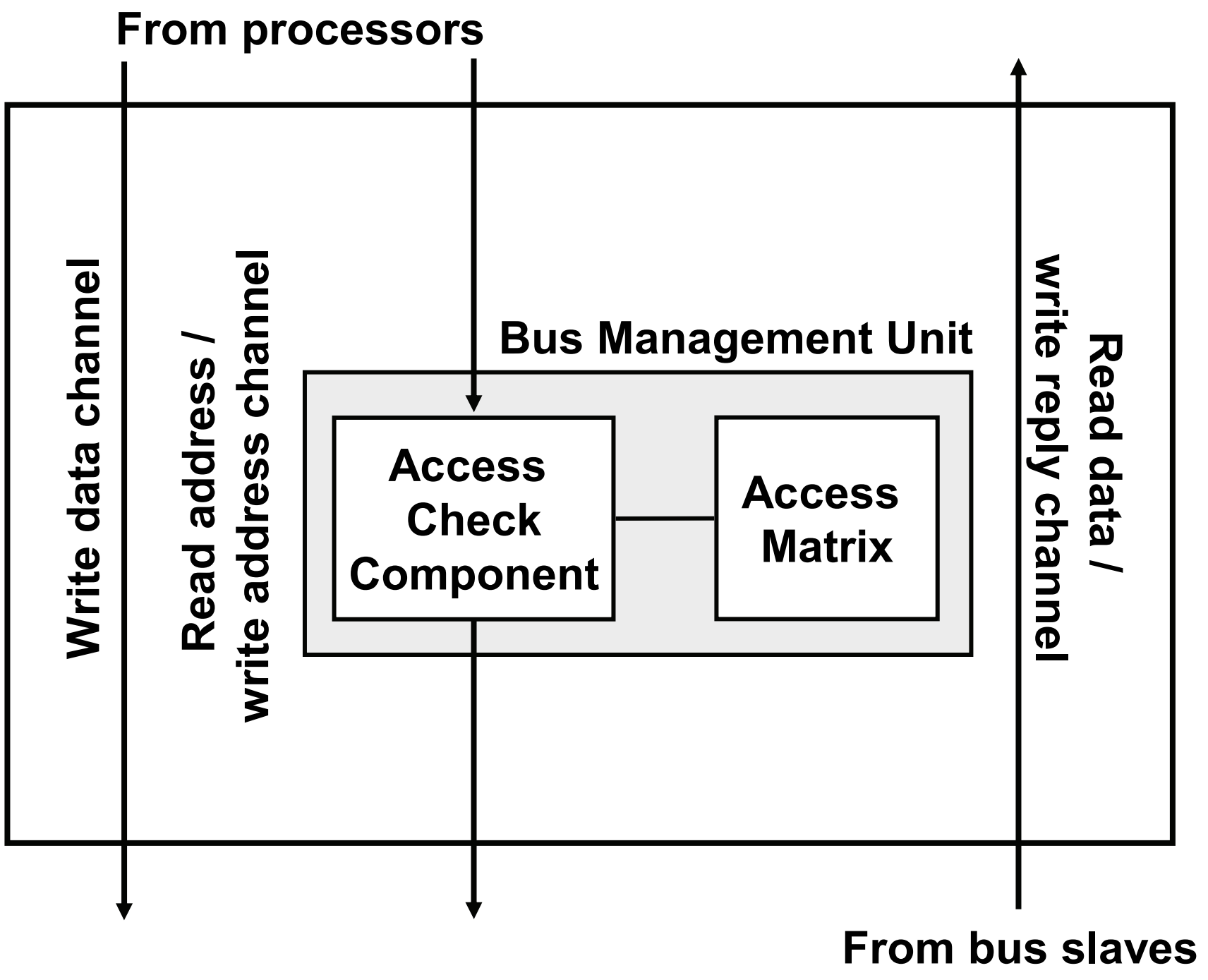

**FIGURE 4.3: INTERNAL DESIGN OF SYSTEM BUS**

As shown in this figure, only read address channels and write address channels are connected to an access check component while other channels (*i.e.*, write data channels, read data channels and write reply channels) are directly connected between processors and bus slaves. This is because an access check component needs two important access information: 32-bit address information (*i.e.*, ARADDR[31:0] and AWADDR[31:0]), and 4-bit ID information containing both processor ID and access type (*i.e.*, ARID[3:0] and AWID[3:0]). An access matrix contains 24 address range entries, allocating eight address range entries to each of three processors. If an access address is included in one of the address ranges, an access check component simply changes the access address to an invalid access address in order to cause a bus error on a system bus. Otherwise, the bus access directly passes the access component.

## 4.4 EVALUATION

Evaluations show that our bus management unit can be efficiently implemented (see Section 4.4.1), and the unit helps enhance system-level security (see Section 4.4.2).

### 4.4.1 HARDWARE SPECIFICATIONS

TABLE 4.1 summarizes hardware specifications of our bus management unit.

**TABLE 4.1: HARDWARE SPECIFICATIONS OF BUS MANAGEMENT UNIT**

| Item | Feature |
|---|---|
| Bus protocol | AXI |
| The number of input / output channels | 2 / 2 |
| The number of address ranges | 24 |
| Technology node | 130nm |
| Gate size | 53.4K gate |
| Delay between input and output channels | 1.28 ns |

FIGURE 4.4 shows its schematic block diagram. This unit is designed for ARM MPCore [ARM 06] with four processors. We synthesized this unit by Synopsys Design Compiler, minimizing the delay between input and output channels. These results indicate that the area and latency overhead of this unit is small enough to be incorporated into a system bus. In addition, FIGURE 4.5 describes the breakdown of the gate size of our bus management unit. An access matrix occupies 30% of the gate size since the access matrix is implemented as flip-flop arrays. Instead of flip-flops, use of an SRAM array would be a promising option in order to reduce the total gate size.

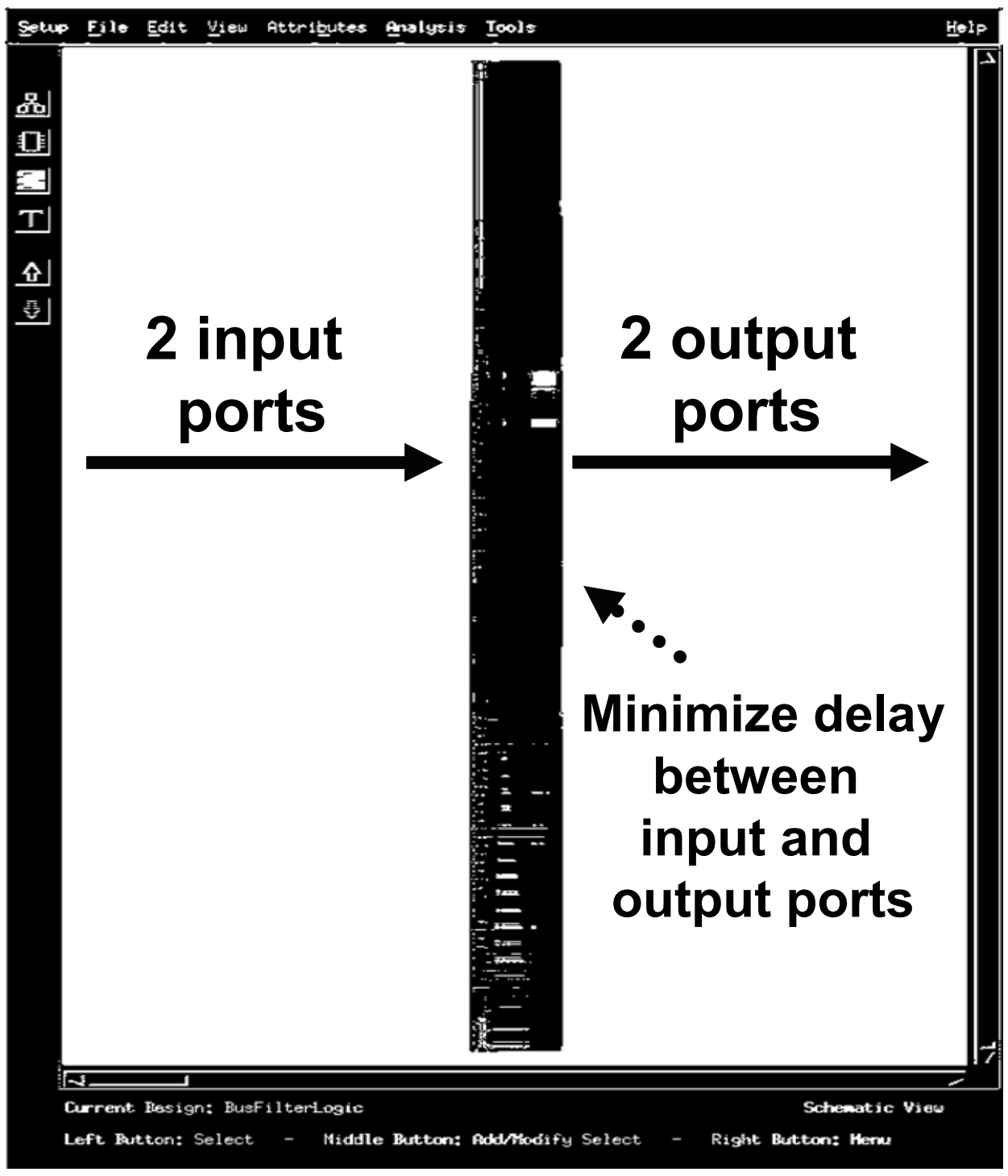

**FIGURE 4.4: SCHEMATIC BLOCK DIAGRAM OF BUS MANAGEMENT UNIT**

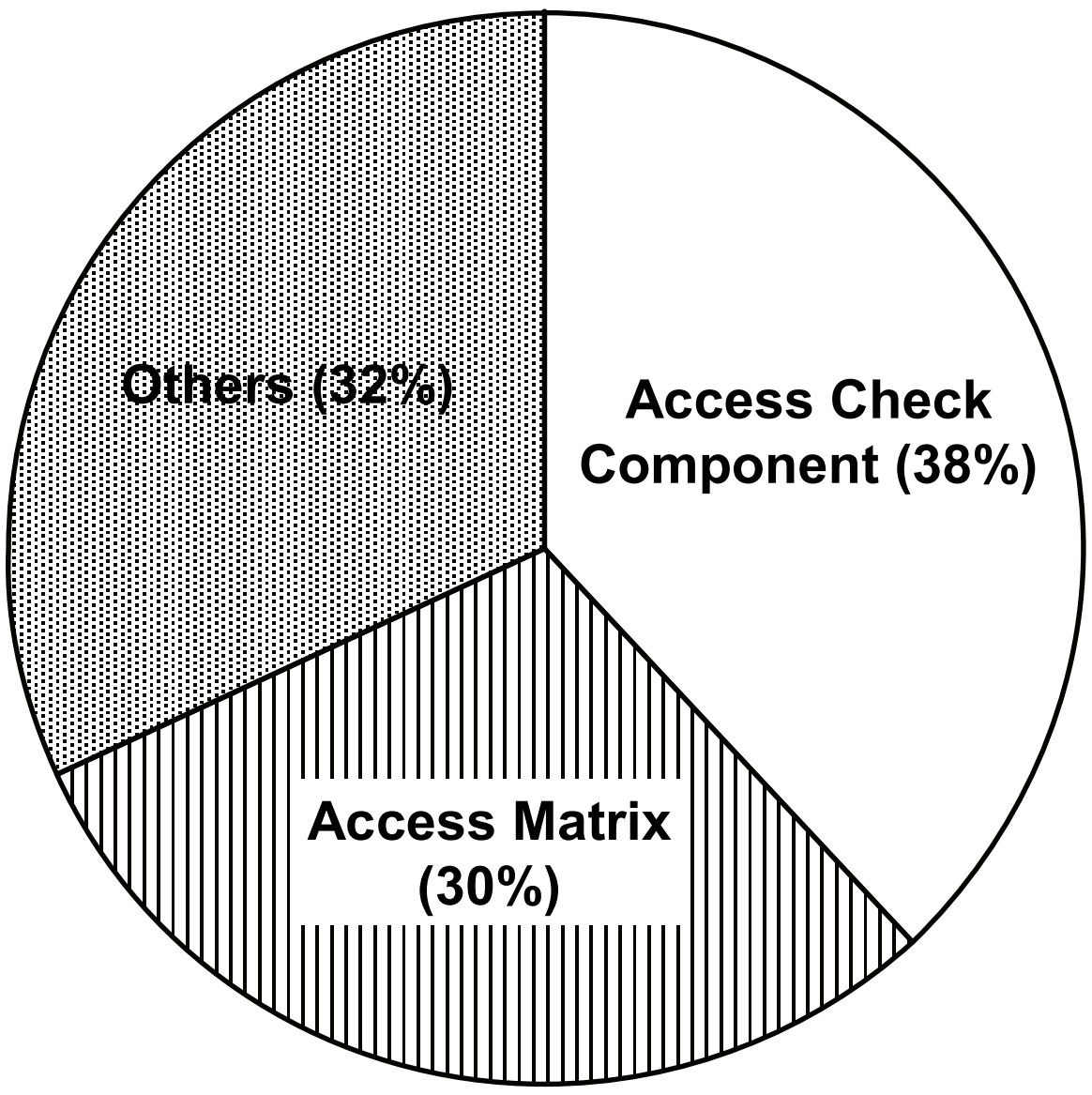

**FIGURE 4.5: BREAKDOWN OF GATE SIZE OF BUS MANAGEMENT UNIT**

### 4.4.2 SUCCESSFUL EXAMPLE FOR SECURE PARTITIONING

As a successful example of the application of our bus management unit, FIGURE 4.6 demonstrates the device coordination between a mobile terminal and an external projector.

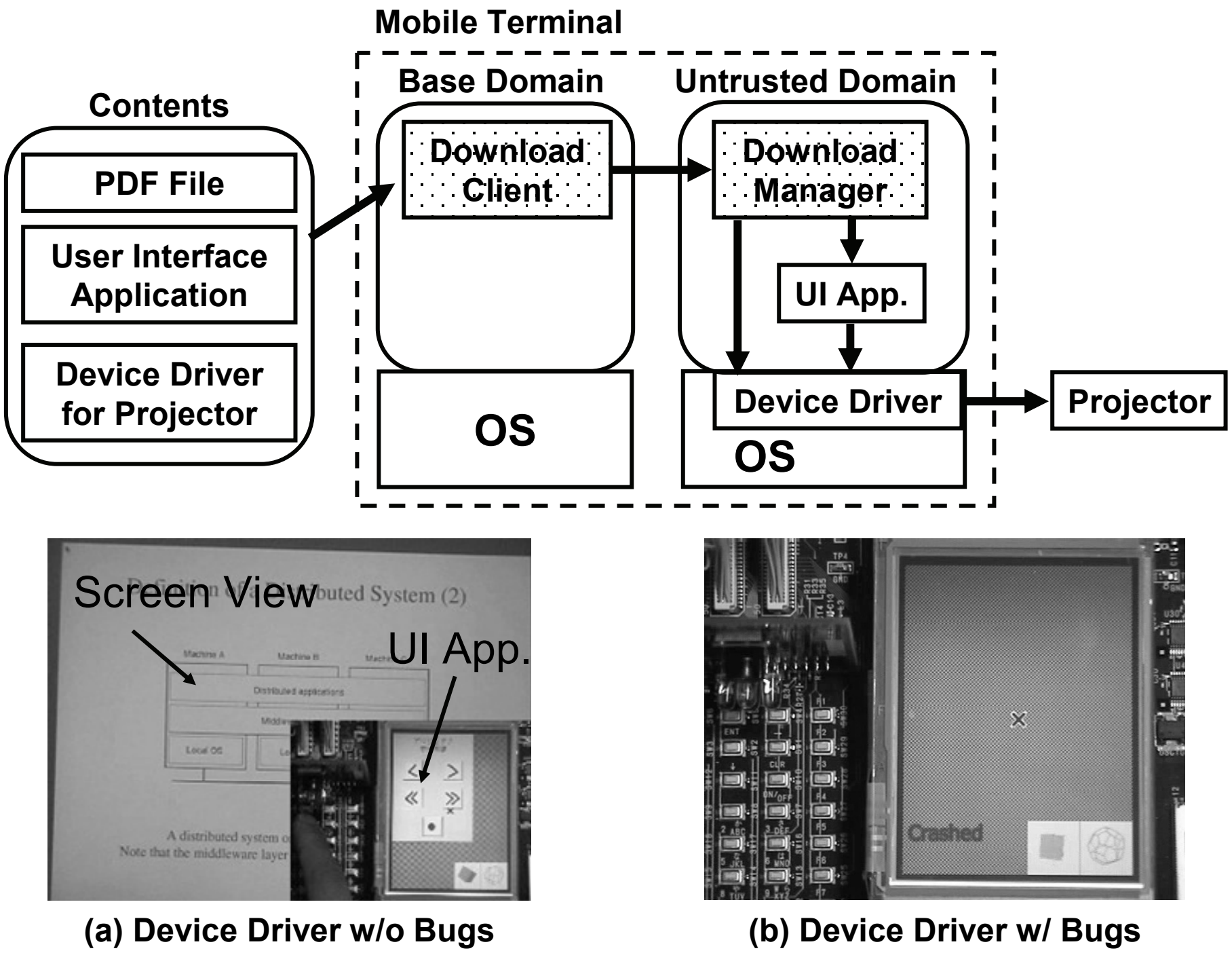

**FIGURE 4.6: DOWNLOADING PROJECTOR DEVICE DRIVER**

Here, a user displays a downloaded PDF file of his/her presentation through an external projector connected to a mobile terminal (see FIGURE 4.6 (a)). For the projector control, a projector device driver and a User Interface (UI) application are downloaded. A download client executed on a base domain manages the downloading of the projector device driver and the UI application. Then, a download manager executed on an untrusted domain installs the projector device driver into the OS of the untrusted domain, executing the UI application on the OS. The UI application and the projector device driver, however, might have critical bugs or viruses. For example, memory leakage bugs of the projector device driver result in crashing the OS of the trusted domain (see FIGURE 4.6 (b)). Our bus management unit maintains the protection of the base domain even during the OS crash, preventing illegal access issued by the crashed OS. In order to recover the untrusted domain, the base domain simply reboots a processor executing the crashed OS.

TABLE 4.2 summarizes the security level of secure platforms, reviewing the relationship between the execution of open applications and the protection of base domains. The underlined items indicate highly desirable characteristics. Our approach achieved the highest security level than did conventional approaches.

**TABLE 4.2: SECURITY LEVEL COMPARISON**

| **Software** | | **Kernel-level domain** | **Virtualized domain** | **Our approach** |
|---|---|---|---|---|
| Application | Bugs | No crash | No crash | No crash |
| | Viruses | Crash | No crash | No crash |
| Device driver | Bugs | Crash | No crash | No crash |
| | Viruses | Crash | Crash | No crash |

## 4.5 SUMMARY

We have presented secure processor partitioning supported by our bus management unit. The unit enables OSs executed on separate processors to be mutually protected at the hardware level. Its most important feature is the prevention of illegal access on a system bus. We have designed our bus management unit, referring to the AXI bus protocol. Evaluations have shown its effectiveness, demonstrating two fundamental features: excellent hardware specifications and a high security level.

# CHAPTER 5
# ASYMMETRIC VIRTUALIZATION

This chapter presents fast, secure virtualization, called asymmetric virtualization, by which many OSs over the number of processors are securely executed under secure processor partitioning.

## 5.1 MOTIVATION

Mobile Information Device Platform (MIDP) specifications [Sun 06] request General System for Mobile communications (GSM) / Universal Mobile Telecommunications System (UMTS) compliant devices to provide at least five protection domains in order to install downloaded applications: a base domain, an identified third party domain, an unidentified third party domain, an operator domain, and a manufacturer domain. The number of protection domains would seem to increase for the secure execution of various groups of open applications.

In order to cope with this issue, virtualization would seem to be one promising solution. Without virtualization, AMP needs to increase the number of processors in proportion to the number of required domains. Conventional virtualization technologies, however, have a degree of security vulnerability [Hacker 07]. In addition, the technologies unfit for embedded systems in terms of base features, such as performance overhead and memory footprint, since traditional virtualization technologies have been originally developed for computing systems.

We introduce a fast, secure virtualization technology, known as asymmetric virtualization, which utilizes secure processor partitioning (see Chapter 4). The most important feature of asymmetric virtualization is the achievement of both high performance and highly hardened security. In this way, open embedded systems enable any number of domains for the secure execution of many groups of open applications. It should be noted that Chapter 6 discusses the application of this asymmetric virtualization to SMP.

## 5.2 RELATED WORK

Our research differs in a number of respects from the current body of research on virtualization technologies. Virtualizing a processor needs to effectively trap sensitive instructions, which are defined as the instructions which would affect the allocation of system resources [Popek 74].

The major kernel-level Virtualization Machine Monitors (VMMs), known as type-I VMMs, include Xen [Barham 03], which employs para-virtualization [Whitaker 02] (*i.e.*, OS modifications). The approaches use a separate, additional processor mode (*e.g.*, a hypervisor mode) [Armstrong 05] [Neiger 06] in order to emulate sensitive instructions. They cause little performance degradation in many applications since the additional processor mode enables normal instructions to be directly executed. There remains, however, severe performance degradation in both the applications which use OS system calls and device drivers since sensitive instructions still need to be virtually emulated.

The major user-level VMMs, known as type-II VMMs, include User Mode Linux (UML) [Dike 00] (a port of Linux to run as a Linux process), and VMware [Sugerman 01]. The approaches emulate sensitive instructions by software-only mechanisms. While they need no additional processor mode, their software-only emulations result in severe performance degradation in many applications.

Further, the security level of both type-I and type-II VMMs depends on the implementation of VMM software since viruses make it possible to exploit security holes in the VMMs [Hacker 07]. FIGURE 5.1 describes the software structures of both type-I and type-II VMMs. Both VMMs have been widely accepted for use in computing systems with PCs and servers. This is because their high generality makes it possible to utilize conventional hardware.

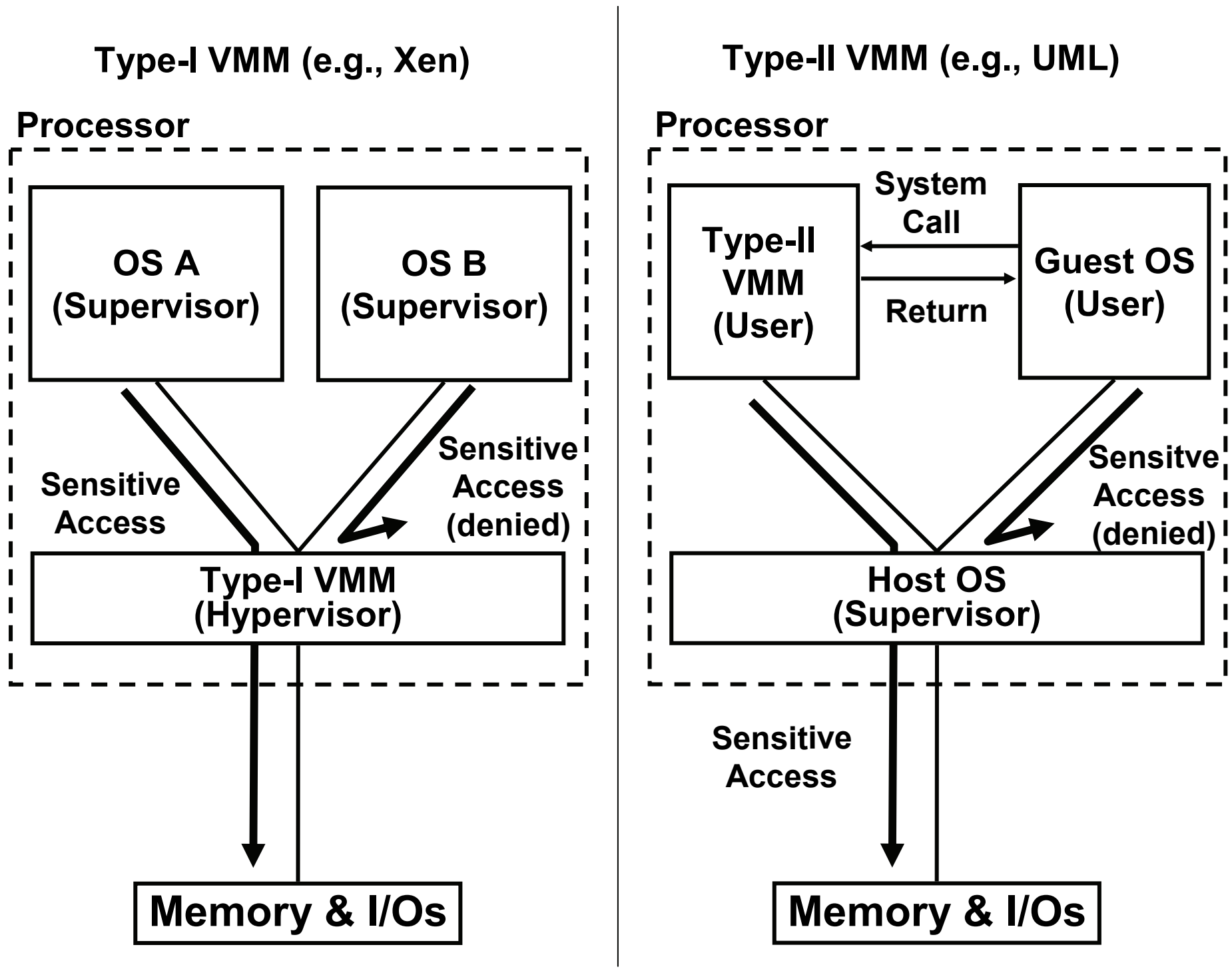

FIGURE 5.1: RELATED WORK: TYPE-I AND TYPE-II VMM

## 5.3 DESIGN PRINCIPLES

FIGURE 5.2 outlines an example of asymmetric virtualization on AMP used for mobile terminals. In this figure, the AMP software structure equips with five domains on three processors in accordance with MIDP specifications [Sun 06]: a base domain, an operator domain, a manufacturer domain, a trusted domain, and an untrusted domain. While a base domain executes base applications on a dedicated processor (*i.e.*, CPU0), the other domains execute open applications on any processor of the two remaining processors.

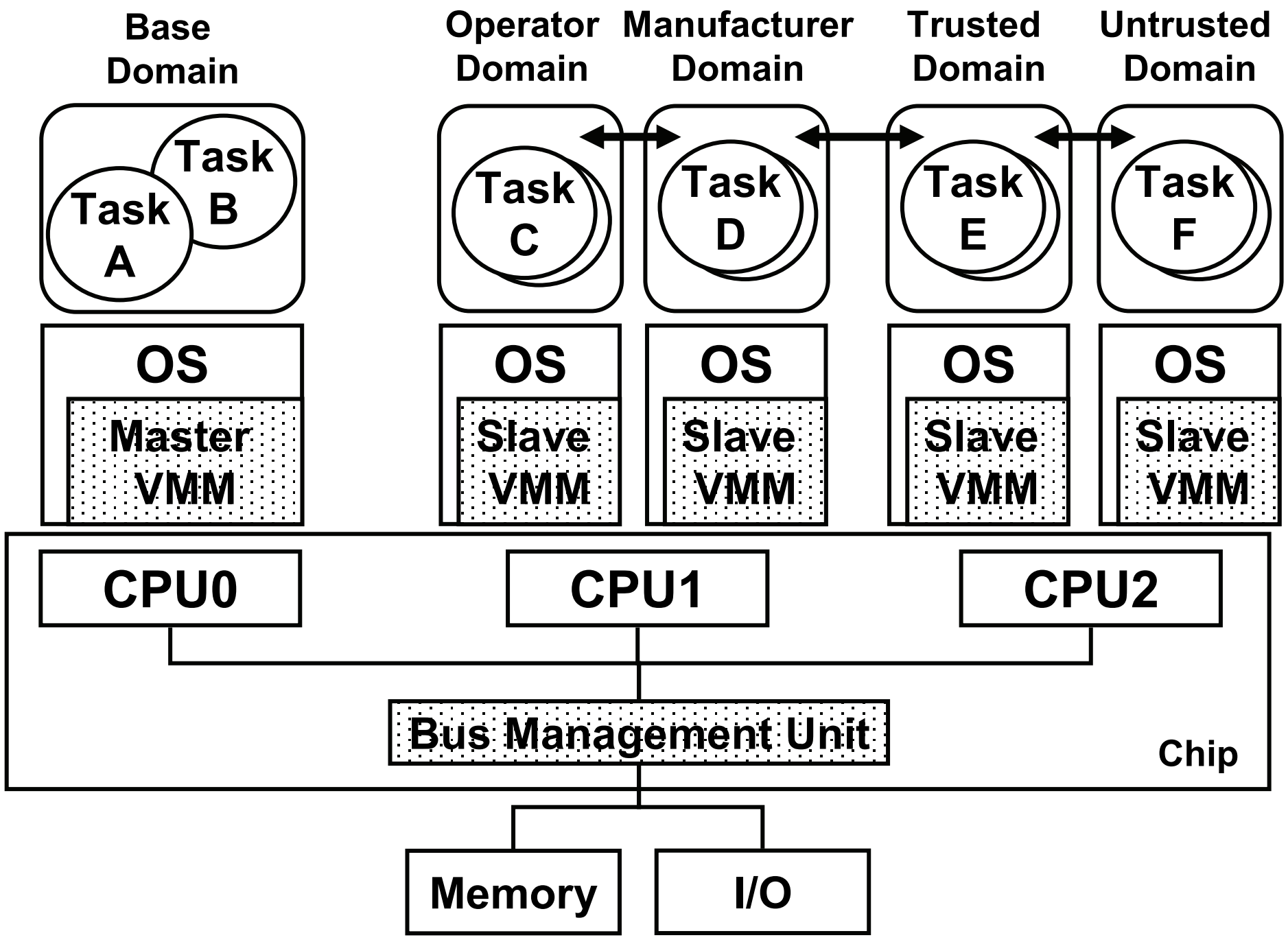

**FIGURE 5.2: VIRTUALIZED AMP FOR MOBILE TERMINALS**

Traditional VMMs deploy the architecture of parallel VMMs on a multi-core processor (*i.e.*, symmetric virtualization) [Armstrong 05]. Such symmetric virtualization, however, causes a deadlock problem which prevents non-compromised VMMs from accessing shared resources when a VMM compromised by open applications holds the shared resources [Coffman 71]. This results in separating the roles of virtualization software, called asymmetric virtualization.

FIGURE 5.3 illustrates the design principles of asymmetric virtualization. A traditional VMM supports three basic functions in it: domain scheduling, which decides the next executed domain; domain setting, which saves and restores processor contexts; and domain separation, which blocks illegal accesses issued from a domain. In our design principles, the three functions are allocated to three separate components: a Master VMM, a Slave VMM, and our bus management unit. Here, the function of domain scheduling is allocated to a Master VMM included in a base domain since a base domain needs to control domains in response to the flexible execution of open applications. The function of domain separation is allocated to our bus management unit since speed overhead of this function degrades the performance of executed domains. Further, the function of domain setting is allocated to a Slave VMM included in an open domain since a processor allocated for open applications requires the flexible execution of multiple domains.

Two most important factors of our design principles are both the allocation of a dedicated processor to a base domain (*i.e.*, use of a multi-core processor) and the domain protection of our bus management unit. The allocation of a dedicated processor to a base domain guarantees the protection of processor resources, such as mode registers and system registers, since the processor resources are never shared with open domains. In addition, our bus management unit enables domain resources, such as memories and I/Os, to be protected among domains at the hardware level. Thus, our asymmetric virtualization needs no emulation of sensitive instructions because the two factors avoid resource interference among virtualized domains, maintaining highly hardened security.

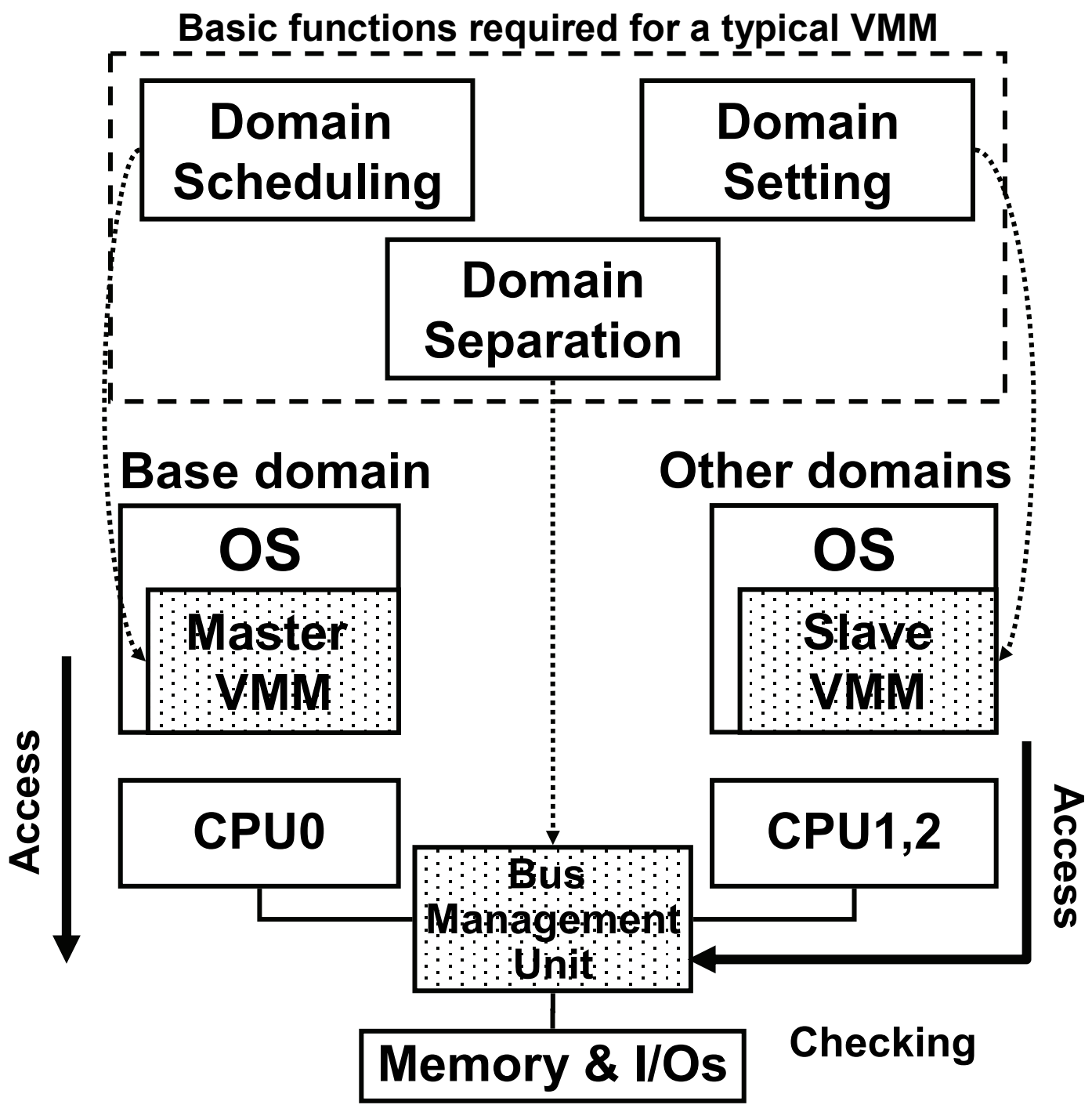

FIGURE 5.3: DESIGN PRINCIPLES OF ASYMMETRIC VIRTUALIZATION

TABLE 5.1 compares our asymmetric virtualization with other approaches. In the table, the best characteristic is underlined for each item. Type-II VMMs have the advantage of high generality. Type-I VMMs have the advantage of higher performance and security than do type-II VMMs at the expense of a new processor mode. Our asymmetric virtualization has the advantages of highest performance and security at the sacrifice of generality (*i.e.*, the fixed execution of a base domain on a dedicated processor). In other words, at the sacrifice of generality, the allocation of a dedicated processor to a base domain helps achieve the secure control of our bus management unit.

Then, our bus management unit provides highly hardened security with fast access checking at the hardware level.

**TABLE 5.1: COMPARISON WITH OTHER VIRTUALIZATION APPROACHES**

| **Item** | **Type-I VMM** | **Type-II VMM** | **Our approach** |
|---|---|---|---|
| Security level | Moderate | Low | High |
| Performance | Moderate | Low | High |
| Scope of applications | General | General | Specific |

It should be noted that virtualization-assist hardware, such as Intel VT [Neiger 06], has been proposed in order to efficiently support processor virtualization. Adams and Agesen [Adams 06], however, have shown that the hardware fails to provide an unambiguous performance advantage for two primary reasons: 1) no support for MMU virtualization, and 2) failure of co-existing with existing software techniques for MMU virtualization. By way of contrast, our bus management unit results in faster virtualization, eliminating MMU virtualization.

### 5.3.1 MASTER VMM

A Master VMM schedules domains on a base domain in response to two APIs: a context-setting API, which sets a domain context, and a context-switching API, which notifies a Master VMM of the execution of a specified domain. Here, a domain context is defined as a set of register values used to restore a processor. For example, a domain context of an ARM processor contains one register bank with eight registers, two register banks with five registers, six register banks with two registers, one current processor status register, one saved processor status register, and CP15 registers [ARM

05].

FIGURE 5.4 describes the design of a Master VMM. It consists of three components: 1) a domain context manager, 2) an inter-VMM communication handler, and 3) a domain scheduler.

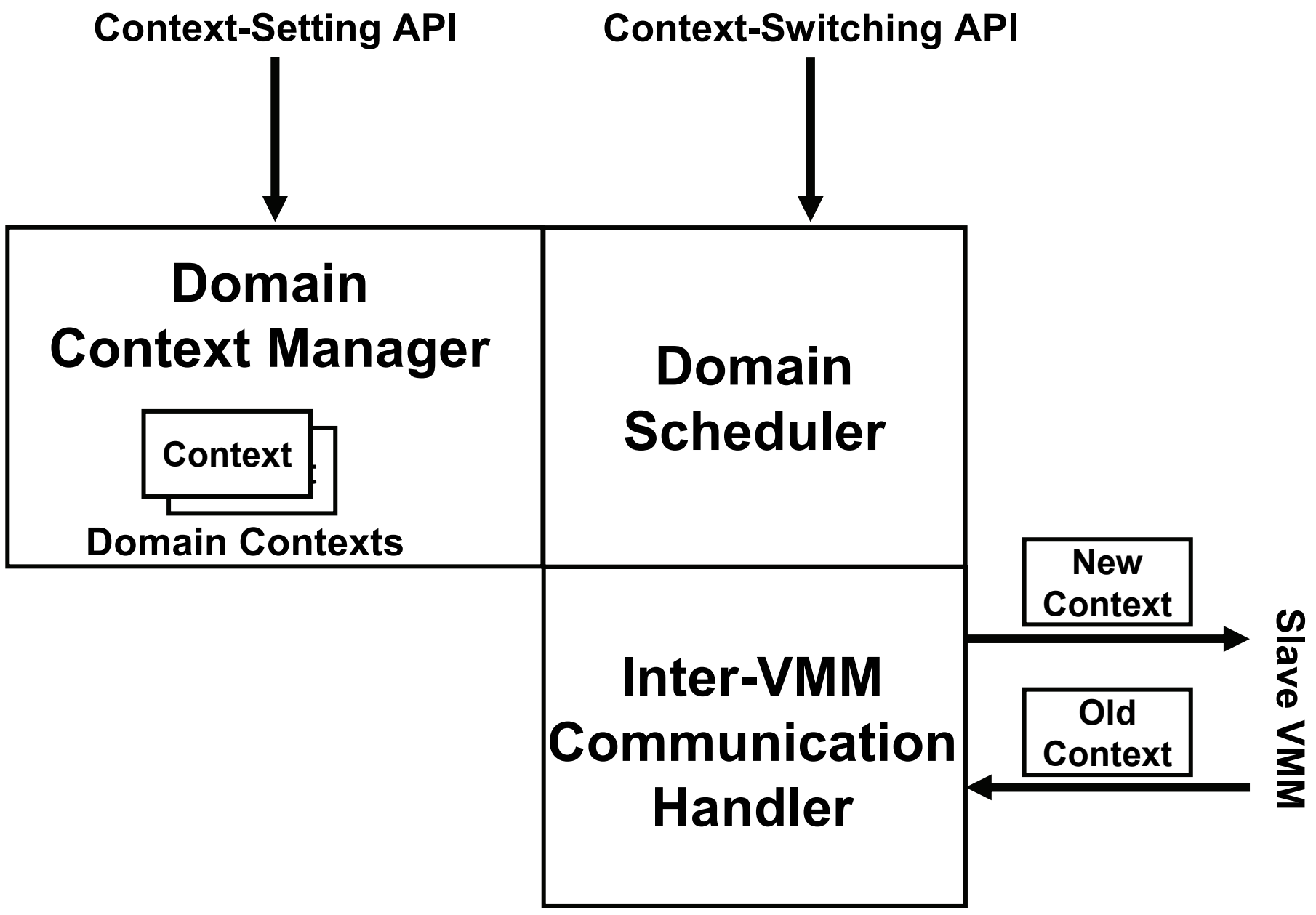

**FIGURE 5.4: DESIGN OF MASTER VMM**

A domain context manager governs domain contexts, which are connected to a double-linked list with a hash table. This domain context manager adds a new domain context given from a context-setting API to the double-linked list. In addition, the manager sends a domain context to a domain scheduler, and receives a domain context from a domain scheduler.

An inter-VMM communication handler sends a domain context to a Slave VMM and receives a domain context from a Slave VMM via shared memory and an inter-processor interrupt. The detailed design of this handler is described in FIGURE 5.7.

A domain scheduler controls all domains in coordination with a domain manager and an inter-VMM communication handler. When a domain scheduler is invoked by a context-switching API, the domain scheduler decides to invoke a domain specified by the API. Then, the domain scheduler receives the context of the specified domain from a domain context manager, and sends the domain context to a Slave VMM via an inter-VMM communication handler. In addition, the domain scheduler changes an access matrix of our bus management unit, allowing a processor executing the specified domain to access its domain resources. Finally, the domain scheduler receives from the inter-VMM communication handler a domain context which the Slave VMM previously executed, giving the domain context to the domain context manager.

### 5.3.2 SLAVE VMM

A Slave VMM switches multiple domains on a processor in response to a request from a Master VMM. This Slave VMM emulates no sensitive instructions, unlike other VMM software.

FIGURE 5.5 describes the design of a Slave VMM. It consists of two components: 1) an inter-VMM communication handler, and 2) a domain switcher. An inter-VMM communication handler receives a domain context from a Master VMM, and sends a domain context to a Master VMM via shared memory and an inter-processor interrupt. The detailed design of this handler is described in FIGURE 5.7. A domain switcher controls both the domain context that was previously executed on a Slave VMM and the domain context that will be executed next on the Slave VMM. When a domain scheduler receives a new domain context from a Master VMM, it sends back an old domain context executed on the domain scheduler via an inter-VMM communication handler. After that, the domain scheduler sets a new domain context to a processor on it.

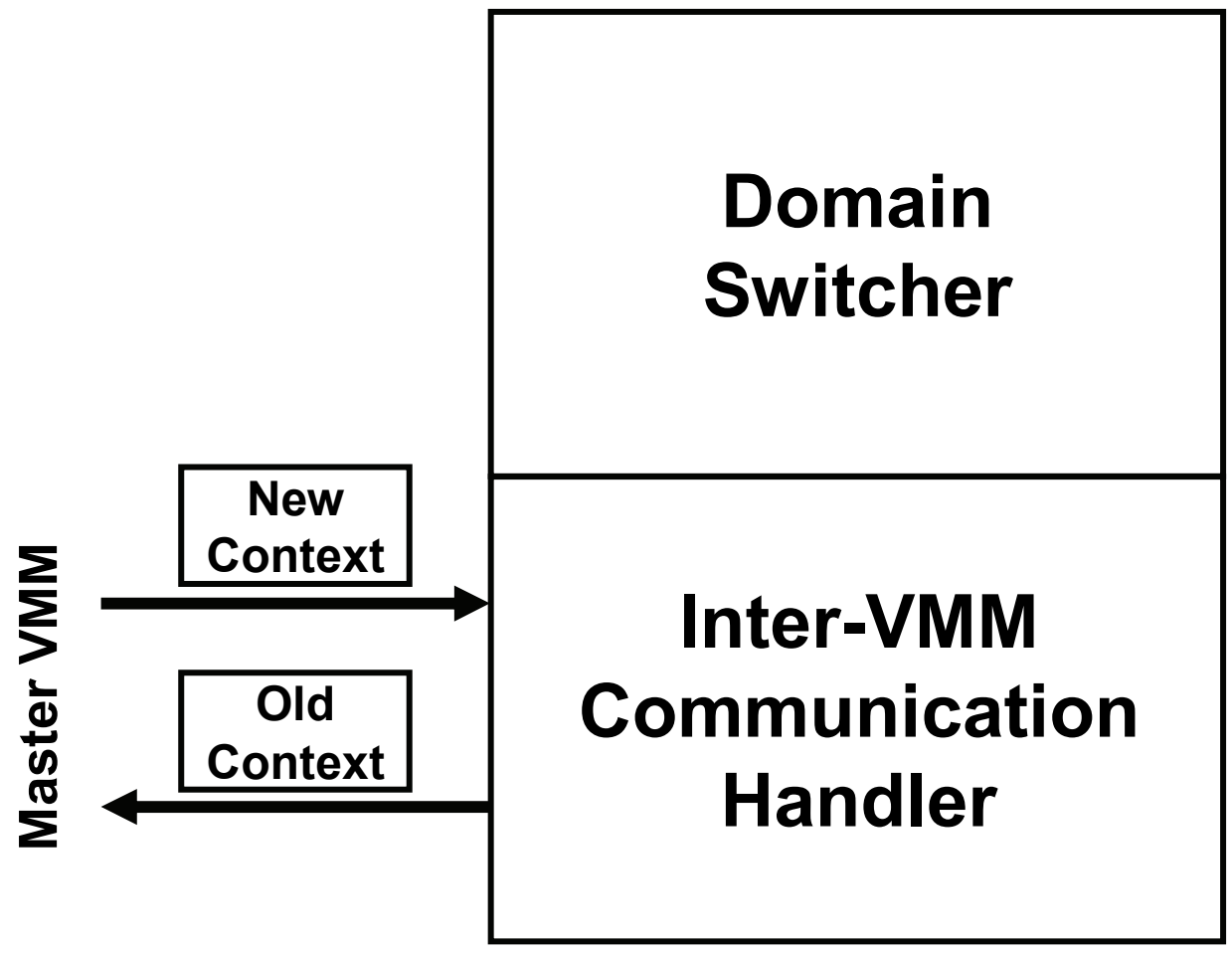

**FIGURE 5.5: DESIGN OF SLAVE VMM**

It should be noted that the execution of a Slave VMM is guaranteed. Typical RISC processors employ Harvard architecture [Hennessy 07], whose processors have separated caches for instructions and data. In order to modify any instructions of a Slave VMM, a processor must issue a data access to the text section of the Slave VMM. This means that the processor results in causing a write miss on the data cache even though instructions of the slave VMM have been already stored to the instruction cache. Since our bus management unit allows the processor only to fetch instructions through the instruction bus, it can block a cache-refill data access on the data bus. In this way, our bus management unit prevents instructions of Slave VMMs from being modified by compromised OSs.

### 5.3.3 INTER-VMM COMMUNICATION

FIGURE 5.6 illustrates the design of inter-VMM communication between a Master VMM executed on processor 0 and a Slave VMM executed on processor K. Two buffers, called “previous”, and “next”, characterize this inter-VMM communication.

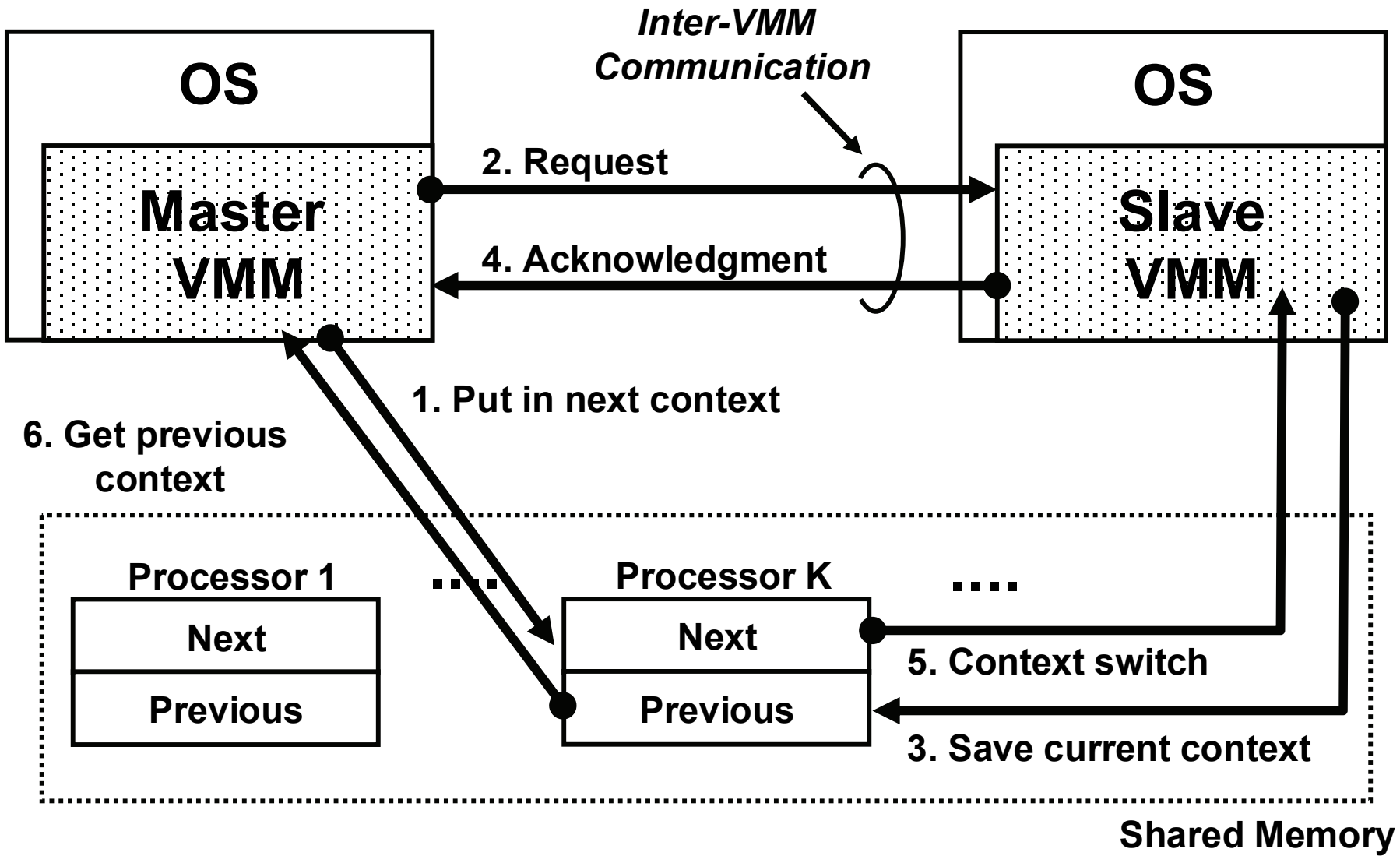

**FIGURE 5.6: INTER-VMM COMMUNICATION**

First, a Master VMM puts a domain context into the “next” buffer allocated for processor K on shared memory. Second, the Master VMM sends a context switch request with an inter-processor interrupt to a Slave VMM executed on processor K. After that, the Slave VMM saves the old domain context executed on processor K to the “previous” buffer. Moreover, the Slave VMM replies an acknowledgment to the Master

VMM, switching to the requested domain context. Here, the Slave VMM disables any interrupts in order to maintain the atomicity between the acknowledgment and the context switch. Finally, the Master VMM gets the old domain context of processor K from the "previous" buffer.

## 5.4 INTER-DOMAIN COMMUNICATION

FIGURE 5.7 shows the design of Inter-Domain Communication (IDC), which helps achieve communication between an application on a running domain and an application on a dormant domain (*i.e.,* a non-running domain), since the dormant domain must be activated on demand. Here, we assume that an application on domain 1 communicates with an application on domain 2. In the case that domain 2 has already run on a processor, a device driver included in domain 1 simply sends data to the corresponding device driver included in domain 2. Otherwise, a device driver included in domain 1 sends a context switch request and data to the master device driver included in a base domain. A kernel thread of the master device driver invokes domain 2 through a Master VMM, retransferring the data to the device driver included in domain 2.

It should be noted that these device drivers use inter-processor interrupts for IDC. A large number of interrupts from compromised OSs have the potential risk to cause Denial-of-Service (DoS) attacks to a base domain. Interrupt masking or an interrupt controller with a Quality-of-Service (QoS) mechanism, however, helps protect such DoS attacks.

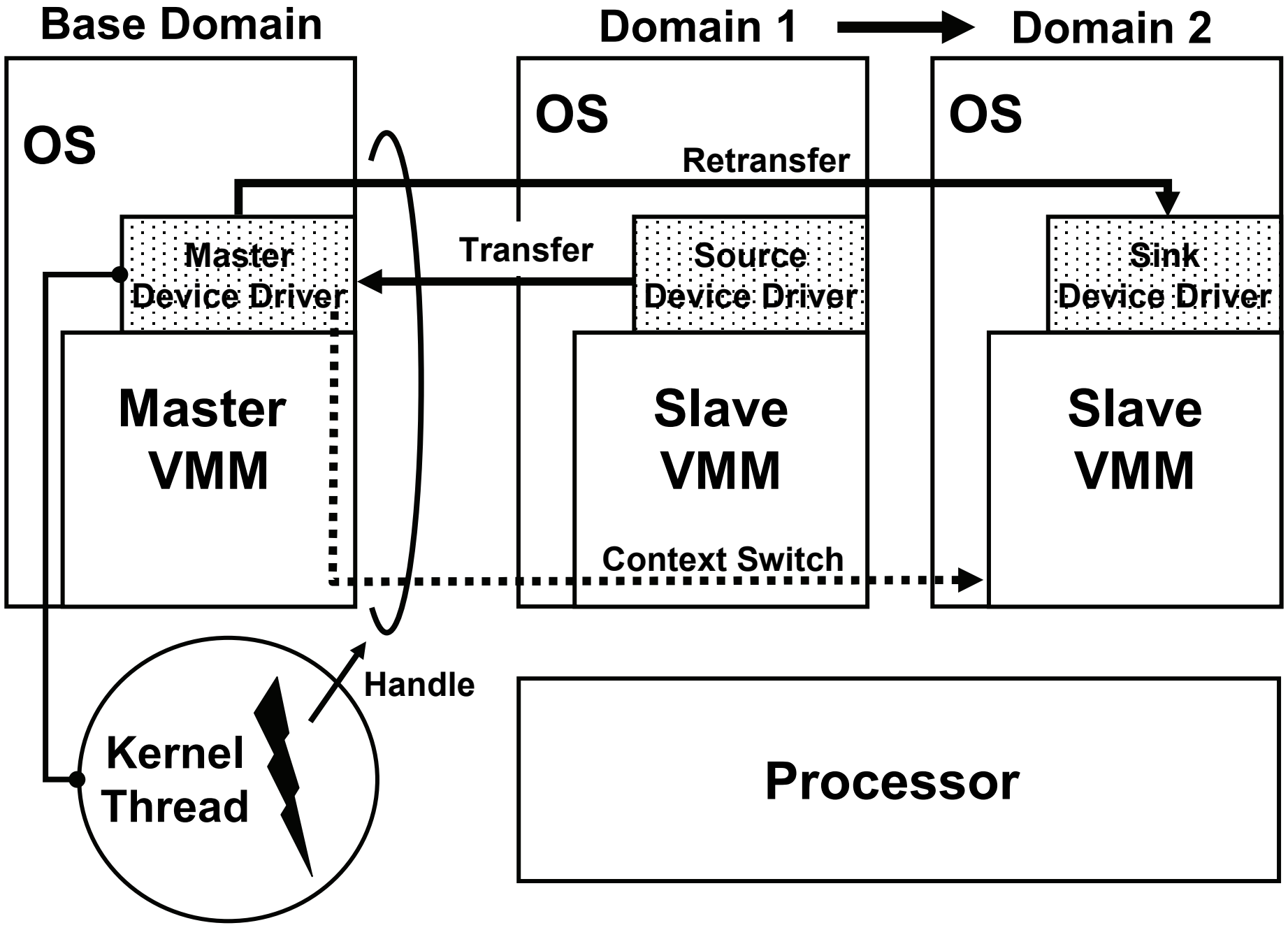

FIGURE 5.7: INTER-DOMAIN COMMUNICATION

## 5.5 EVALUATION

We used the same MP211 evaluation environment as Section 3.5. In this section, a Master VMM and a Slave VMM are implemented as a kernel module and an interrupt handler, respectively. Evaluations show that our asymmetric virtualization actually worked on AMP (see Section 5.5.1). In addition, the virtualization achieved higher performance (see Section 5.5.2, and 5.5.3) and lower code size (see Section 5.5.4) than did other approaches.

### 5.5.1 SUCCESSFUL EXAMPLE FOR VIRTUALIZATION

As a successful example of the application of our asymmetric virtualization, FIGURE 5.8 demonstrates that five Linux OSs run on three ARM processors. Processor 1 has two domains 1 and 3, and processor 2 also has two domains 2 and 4. In the lower half of the figure, 3D graphic processes show the operating states of each domain. In the top half of the figure, a control board process on domain 0 manages running domains by IDC. In this figure, domains 3 and 4 are running, and domains 1 and 2 are dormant.

The time required for an OS switch was less than 0.5ms. This OS switch overhead is not a problem with Linux-based mobile terminals since the time quantum of a process on embedded Linux (e.g., 150ms) is much longer than the OS switch overhead.

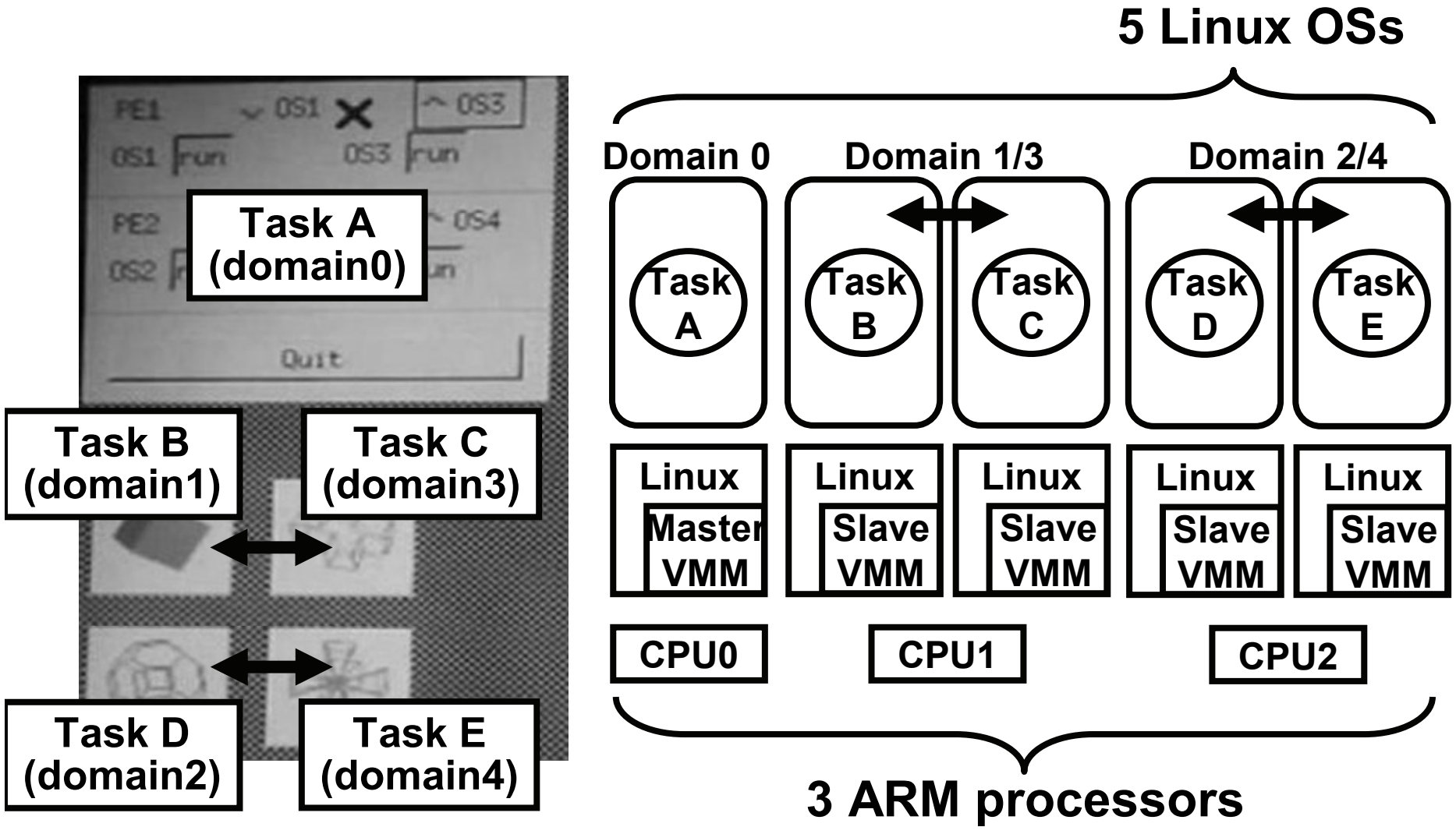

**FIGURE 5.8: VIRTUALIZING FIVE DOMAINS ON THREE PROCESSORS**

## 5.5.2 VIRTUALIZATION PERFORMANCE ON OPEN DOMAIN

In order to prove that our approach achieved faster virtualization than did other approaches, we executed two micro-benchmarks of LMbench [Mcvoy 96] on an open domain: processes and context switching micro-benchmarks. LMbench is a typical benchmark in order to measure single OS performance on PCs. While the processes micro-benchmarks are used to measure the basic process primitives, such as creating a new process and running a different program, the context switching micro-benchmarks are used to measure the time needed to save the state of one process and restore the state of another process. Here, we were unable to evaluate application-level benchmarks because of two primary reasons: 1) the small memory capacity of our evaluation board and 2) no standard application-level benchmarks for the measurement of virtualization overhead.

FIGURE 5.9 and FIGURE 5.10 describes the evaluation results of two micro-benchmarks. We selected 13 items, which are especially important to measure the efficiency of processor virtualization itself (*i.e.*, relatively less independent from I/Os and file systems), from 24 items shown in [Barham 03]. Further, in reference to [Barham 03], the results of other VMMs, Xen (a type-I VMM) and UML (a type-II VMM), are also shown in the figures. Here, respectively in two different evaluation environments, each performance of the micro-benchmarks executed on non-virtualized Linux is normalized to 1 for the sake of relative comparison.

In process micro-benchmarks measuring system call performance, the average virtualization overhead of our approach was only 3.2%. For reference, our approach achieved 1.2 times higher average performance than did Xen and 58 times higher one than did UML. In particular, our approach improved "fork" and "exec" performance because an OS on a Slave VMM is able to update large numbers of a page table without any emulation. While Xen software needs to check whether the page table updates are illegal, a Slave VMM does not need to check the updates because of our bus management unit. Here, one small anomaly seen here, “sh proc” in Xen, presumably occurred due to a fortuitous cache alignment [Barham 03] since “slct TCP” in Xen produces better performance than does non-virtualized Linux.

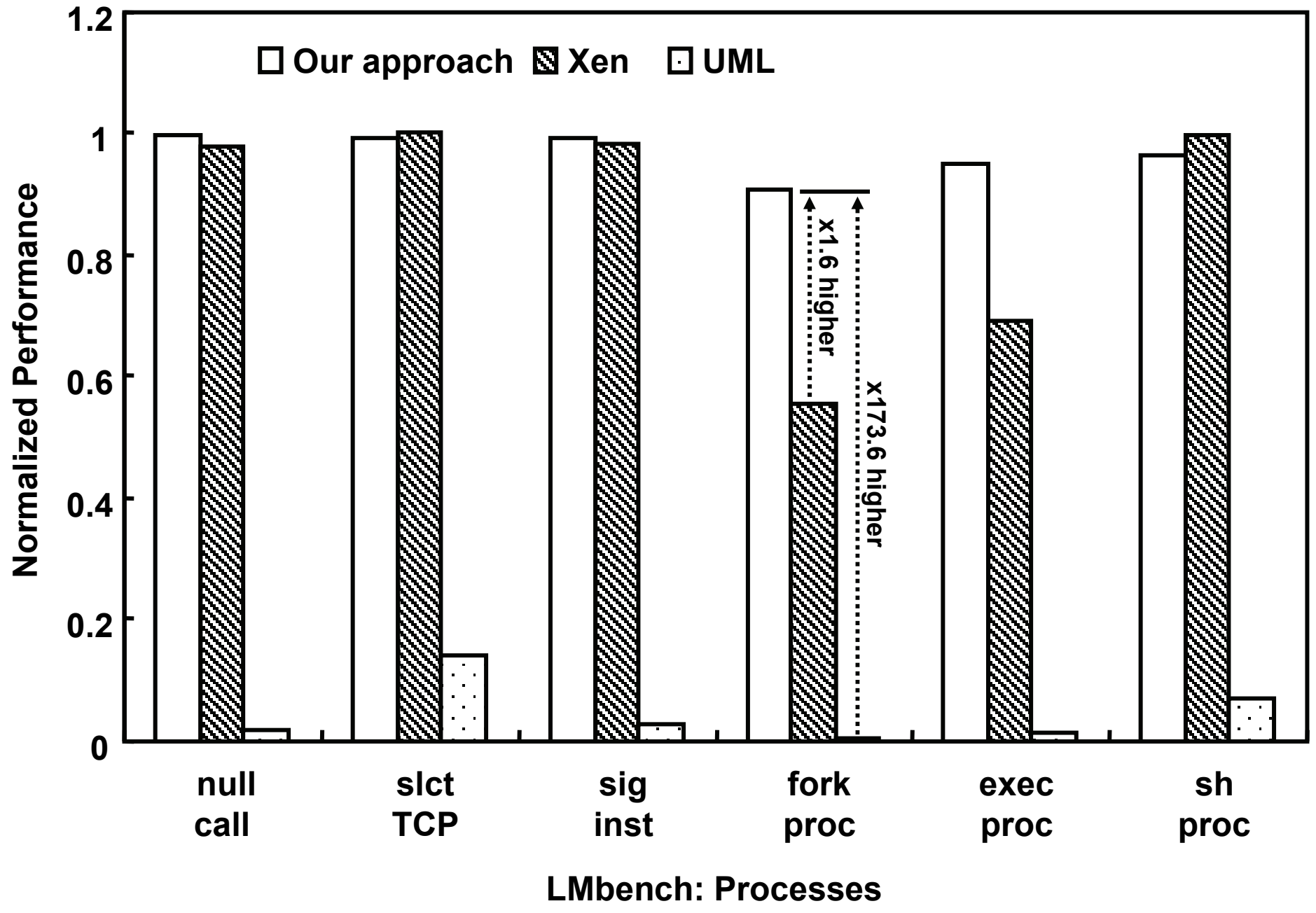

**FIGURE 5.9: PROCESS MICRO-BENCHMARKS ON OPEN DOMAIN**

In context switching micro-benchmarks measuring context switching time between different numbers of processes with different working set sizes, the average virtualization overhead of our approach was only 2.1%. For reference, our approach achieved 2.1 times higher average performance than did Xen and 20 times higher one than did UML. In particular, our approach significantly improved the context switch time for smaller working set sizes. This is because an OS on a Slave VMM is able to change a page table base without any emulation. While Xen software needs to check whether the changed page table base is illegal, a Slave VMM does not need to perform the checking because of our bus management unit.

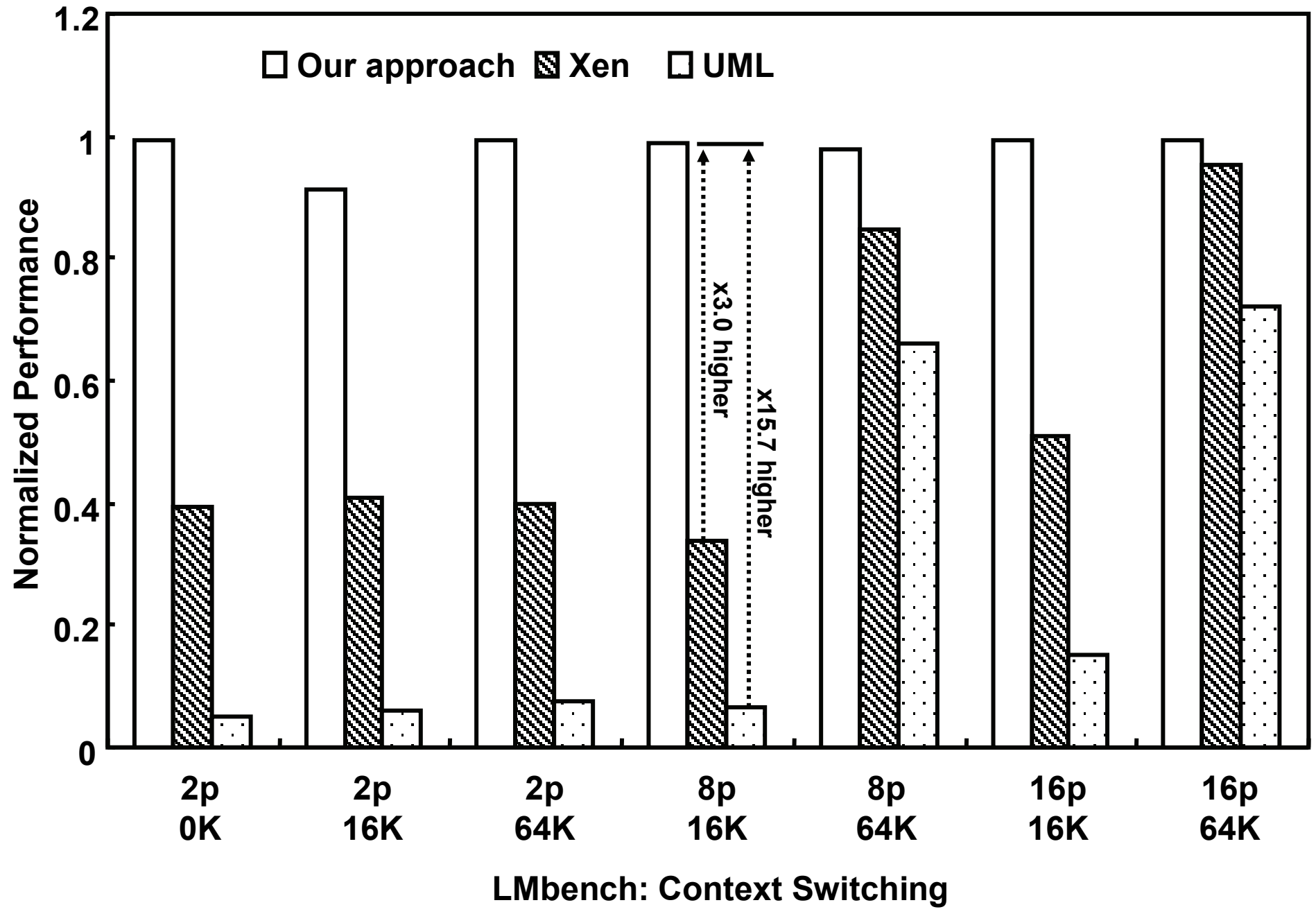

**FIGURE 5.10: CONTEXT SWITCHING MICRO-BENCHMARKS ON OPEN DOMAIN**

The above measurements in terms of OS-level virtualization overhead imply application-level virtualization overhead. In terms of benchmarks which use few system calls, such as SPEC INT [Henning 06], the virtualization overhead of our approach would be almost the same as that of Xen or UML. In open embedded systems, our target applications, however, would seem to use a lot of system calls in order to achieve new services, such as device coordination among appliances. This means that our approach would greatly improve the performance of such applications in open embedded systems.

### 5.5.3 INTER-DOMAIN COMMUNICATION BANDWIDTH

In order to examine IDC bandwidth degradation, we evaluated TCP bandwidth between two domains by using the ttcp benchmark [Muuss 85]. Both sender (TX) and receiver (RX) applications were configured using a socket buffer size of 128KB with a Maximum Transfer Unit (MTU) of 1500 bytes to transfer 8 MB of data. In our approach, the virtualization overhead of network device drivers was less than 5%. Thus, the network device driver achieved almost the same bandwidth as the network device driver for non-virtualized Linux. It should be noted that we were not able to compare our approach with Xen and UML because the transferred data size configured in [Barham 03] of 400MB was immeasurable in our evaluation environment.

### 5.5.4 LINES OF CODE FOR VIRTUALIZATION

In order to show the efficient implementation of our asymmetric virtualization, FIGURE 5.11 illustrates the Lines Of Code (LOC) of our approach, compared with those of other VMMs, Xen and UML. The modified LOC of virtualization software in our approach are 1.7 times smaller than that of Xen and 7.4 times smaller than that of UML. In addition, the modified LOC of network drivers in our approach are 2.0 times smaller than that of Xen and 3.7 times smaller than that of UML. This is because our bus management unit helps simplify virtualization software while other VMMs need extra functions, such as page table update, to handle multiple domains.

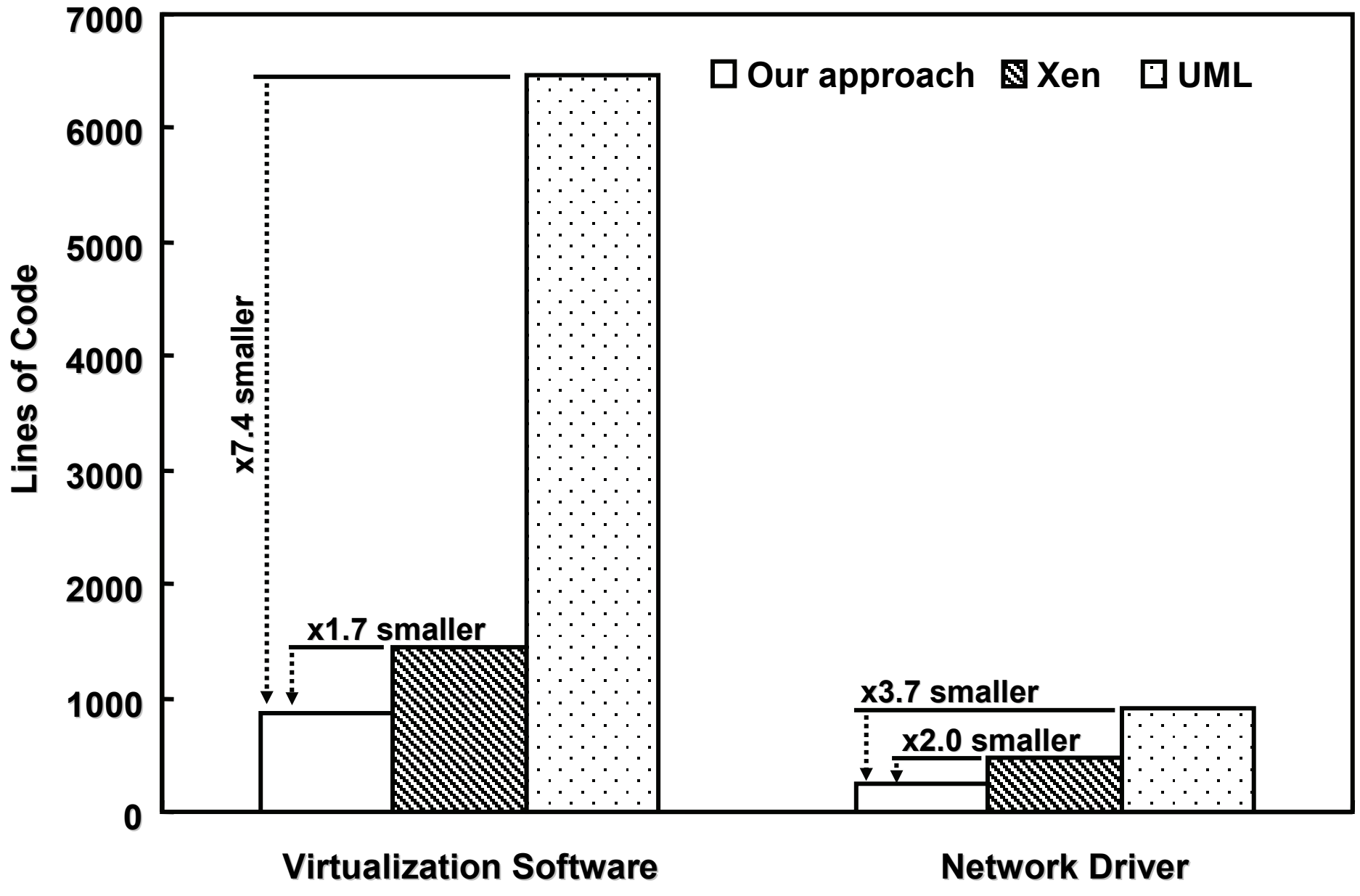

FIGURE 5.11: LINES OF CODE FOR VIRTUALIZATION

## 5.6 SUMMARY

We have presented our asymmetric virtualization, by which many OSs over the number of processors are securely executed under secure processor partitioning. Its most important feature is the achievement of both the allocation of a dedicated processor to a base domain and the domain protection of our bus management unit. In this way, asymmetric virtualization achieves fast, secure virtualization. We have designed asymmetric virtualization based on both ARM processors and Linux OSs. Moreover, our evaluations have shown its effectiveness, demonstrating three fundamental features: a successful example on three ARM processors, high performance and small code size.

# CHAPTER 6
# SECURE DYNAMIC PARTITIONING

This chapter presents secure dynamic partitioning, by which the number of processors allocated to individual OSs makes it possible to be dynamically changed on SMP under secure processor partitioning.

## 6.1 MOTIVATION

Traditional SMP fails to execute multiple domains since they support only a single domain. Our secure dynamic partitioning helps support multiple domains required for open embedded systems on SMP. As mentioned in Chapter 5, conventional virtualization technologies, however, have a degree of security vulnerability [Hacker 07]. In addition, the technologies unfit for embedded systems in terms of base features, such as performance overhead and memory footprint, since traditional virtualization technologies have been originally developed for computing systems.

We introduce secure dynamic partitioning for SMP, which utilizes both secure processor partitioning (see Chapter 4) and asymmetric virtualization (see Chapter 5). The most important feature of secure dynamic partitioning is dynamic processor allocation, by which the number of processors allocated to individual OSs is dynamically changed under secure processor partitioning. In this way, open embedded systems enable any number of domains to be securely executed even on SMP. It should be noted that asymmetric virtualization enables the number of processors to be dynamically changed by switching domains on individual processors of AMP.

## 6.2 RELATED WORK

Our research differs in a number of respects from the current body of research on dynamic partitioning.

Major OS-level approaches include eSOL eT-Kernel [Gondo 07] and QNX BMP [Johnson 07]. Since these approaches use processor affinity settings to allow applications executed on an SMP OS to be run only on specified processors, they make it possible to change the number of processors assigned to an application. Thus, these approaches impose no performance overhead on applications. The execution of base and open applications on the same SMP OS, however, will result in critical security vulnerability.

The major VMM-level approaches, as described in Section 4.2, include Xen [Barham 03] and LPAR [Armstrong 05]. These approaches provide any number of processors to applications through processor virtualization features. In addition, the approaches help enhance system security since they allow base applications to be separated from open applications at the OS level. Virtualization, however, results in unignorable performance degradation [Adams 06] [Barham 03], and there is a degree of security vulnerability in complex virtualization software [Hacker 07]. In accordance with the problem, recent work [Chen 08] [Seshadri 07] [Shinagawa 09] has proposed tiny virtualization software in order to reduce security vulnerability of virtualization software itself.

## 6.3 DESIGN PRINCIPLES

FIGURE 6.1 shows an example of SMP used for mobile terminals. This example has three domains, a base domain for the execution of base applications and two open domains (A/B) for the execution of open applications, such as an operator domain and a manufacturer domain [Sun 06]. The base domain maintains at least one processor for its executions.

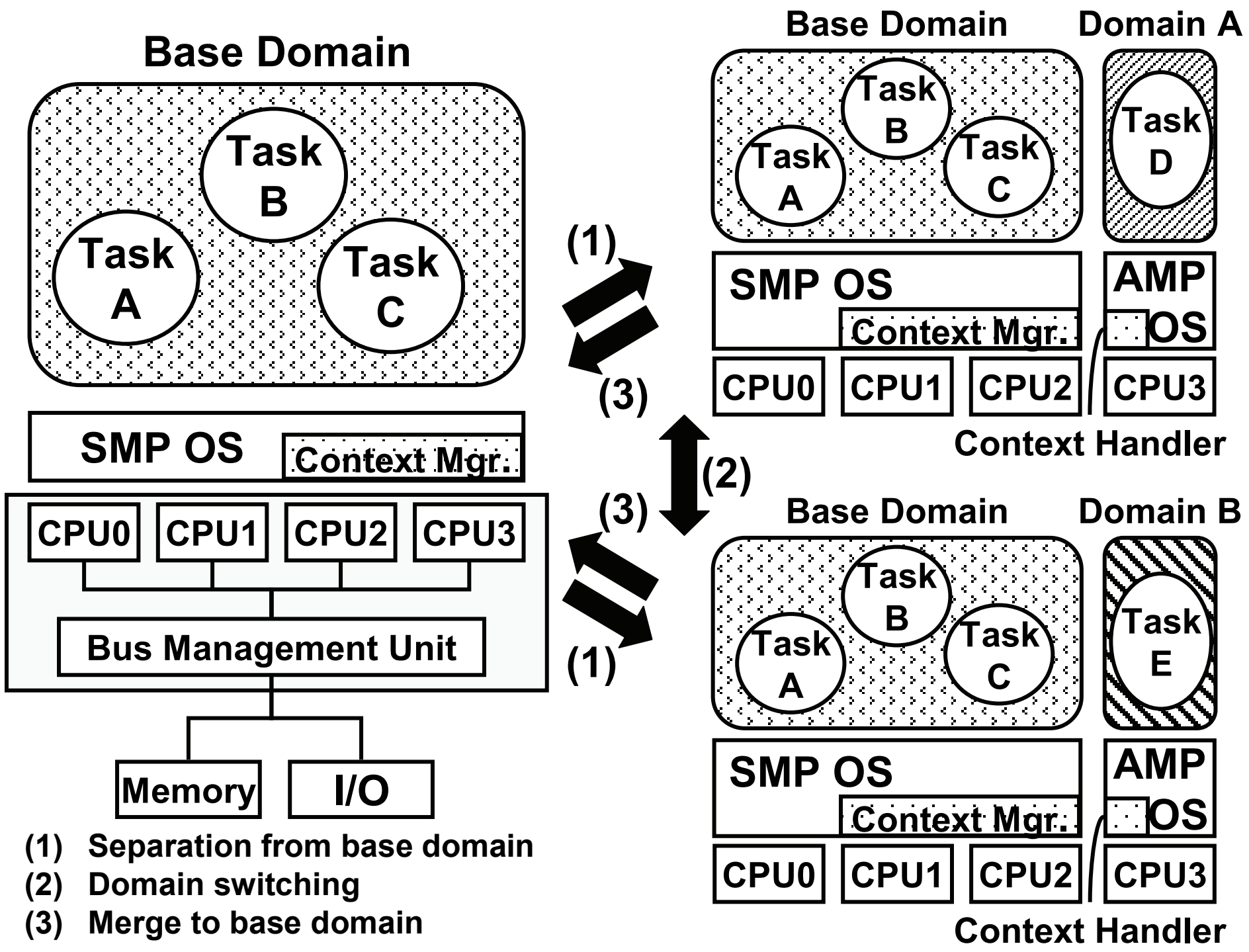

**FIGURE 6.1: SMP FOR MOBILE TERMINALS**

For the scalable extension of base applications, all four processors are allocated to the base domain on the SMP OS. Further, where coordination is required between base and open applications, a processor allocated to the base domain (*e.g.*, CPU3) will be yielded to an open domain. As shown on the right-hand side of the figure, base applications can, for example, be executed separately on an SMP OS with three processors, while open applications are executed with the remaining processor.

The SMP software structure includes two components: a context manager and a context handler. Here, a context refers to the register values required to restore a processor state. It should be noted that a context manager and a context handler are respectively extensions of a Master VMM and a Slave VMM (see Section 5.3.1 and 5.3.2) in order to control SMP. In addition, inter-VMM communication (see Section 5.3.3) is used between a context manager and context handlers.

A context manager is run only on the base domain, and it manages base domain contexts, which are required to restore to the base domain any processors previously allocated to domains. It also manages all open domain contexts. Further, it controls self-transition management (see Section 6.3.1). It also sends to a context handler the context of a domain in which an execution is to be performed, ordering that the execution be made.

A context handler is run on each open domain, and it conducts domain switching, from a current domain to a domain specified by the context manager.

In transitions from, for example, a state with only a base domain to one with both a base domain and an open domain (*e.g.*, Domain A in state transition (1) in FIGURE 6.1), the context manager saves the SMP OS context of the processor (here, CPU3) allocated for the execution of Domain A. Moreover, it restores the context of Domain A to the processor. As a result, the number of processors allocated to the base domain is dynamically reduced from four to three. FIGURE 6.2 summarizes this separation transition with required OS contexts.

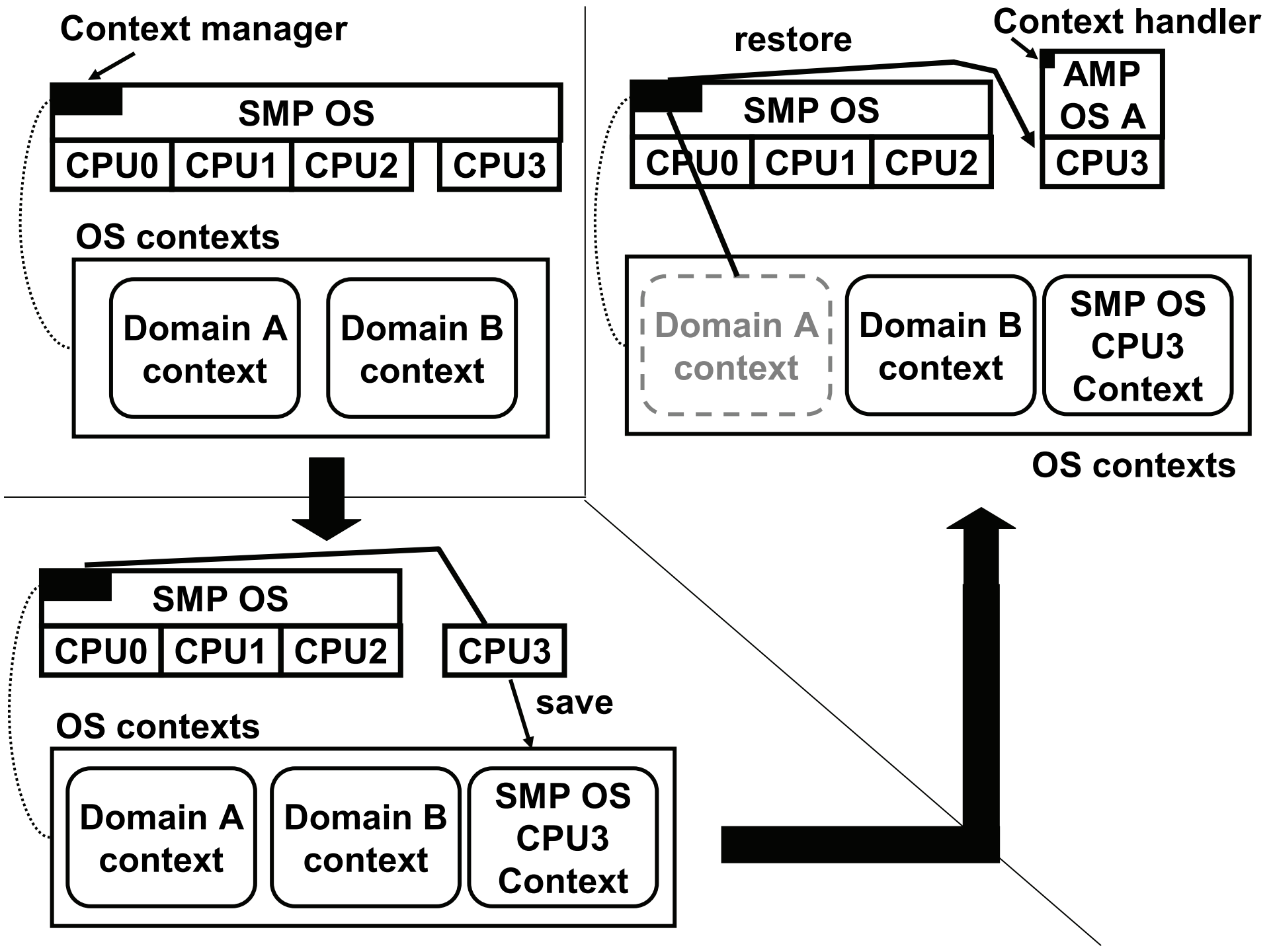

**FIGURE 6.2: SEPARATION FROM BASE DOMAIN**

In state transition (2) in FIGURE 6.1, *e.g.*, in switching from Domain A to Domain B, the context manager sends the Domain B context to the context handler of Domain A, using inter-VMM communication. Moreover, it receives from the context handler the Domain A context, which had previously been saved. FIGURE 6.3 summarizes this switching transition with required OS contexts. It should be noted that this OS switching is equivalent to the function of asymmetric virtualization designed for AMP, as described in Chapter 5.

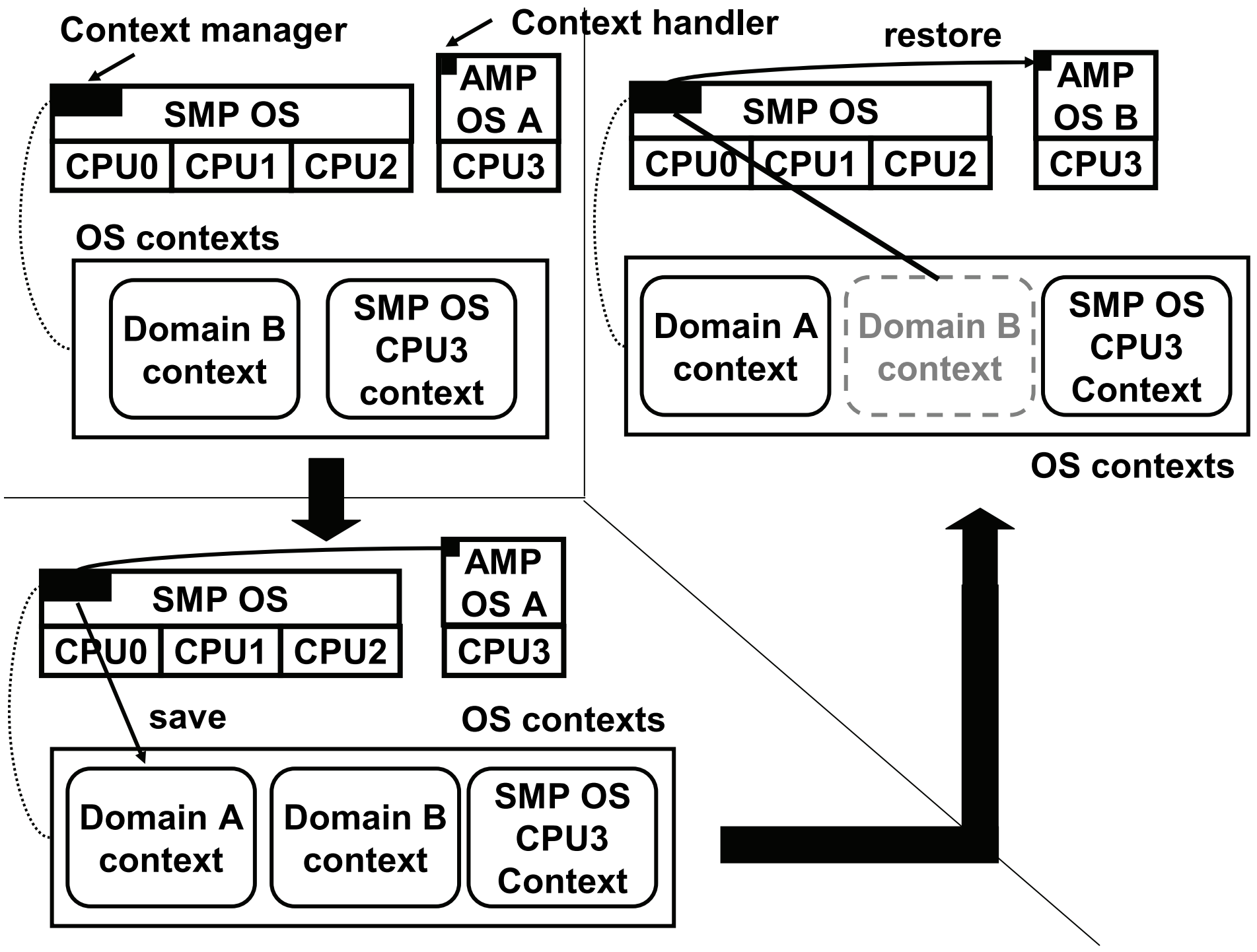

**FIGURE 6.3: DOMAIN SWITCHING**

Finally, in state transition (3) in FIGURE 6.1, *e.g.*, in switching from a state with both a base domain and an open domain to one with only a base domain, the context manager sends the SMP OS context of the processor performing executions in Domain B (*i.e.*, CPU3) to the context handler of Domain B. Moreover, it receives from the context handler the Domain B context, which had previously been saved. As a result, the processor allocated for executions in a domain (*i.e.*, CPU3) is restored to the base domain, and the number of processors allocated to the base domain is dynamically increased from three to four. FIGURE 6.4 summarizes this merge transition with required OS contexts.

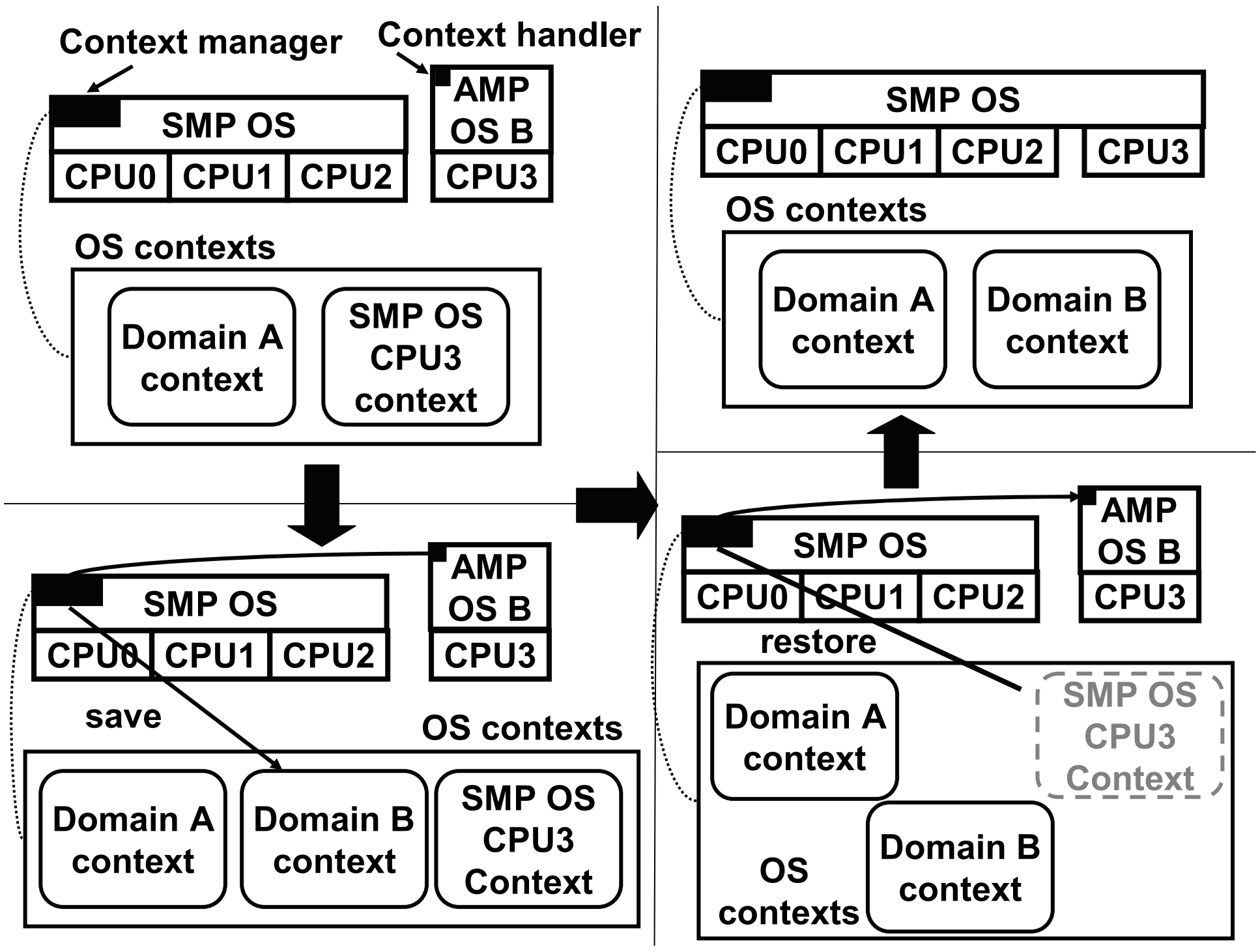

**FIGURE 6.4: MERGE TO BASE DOMAIN**

### 6.3.1 SELF-TRANSITION MANAGEMENT

For separating a processor from a base domain (state transition (1) in FIGURE 6.1) and merging a processor back to the base domain (state transition (3) in FIGURE 6.1), a new operational control of processors, called self-transition management, is newly required. Self-transition management utilizes CPU Hotplug technology [Mwaikambo 04] in order to control a base domain. CPU Hotplug technology, originally developed by Russell *et al.*, is used to remove faulty processors from a system and add new processor substitutes to that system without stopping on-line operations. In ARM MPCore Linux, this technology allows unused processors to be put into a low power mode in order to reduce power consumption, as outlined in FIGURE 6.5. In other words, it simply suspends use of the processors with clock gating.

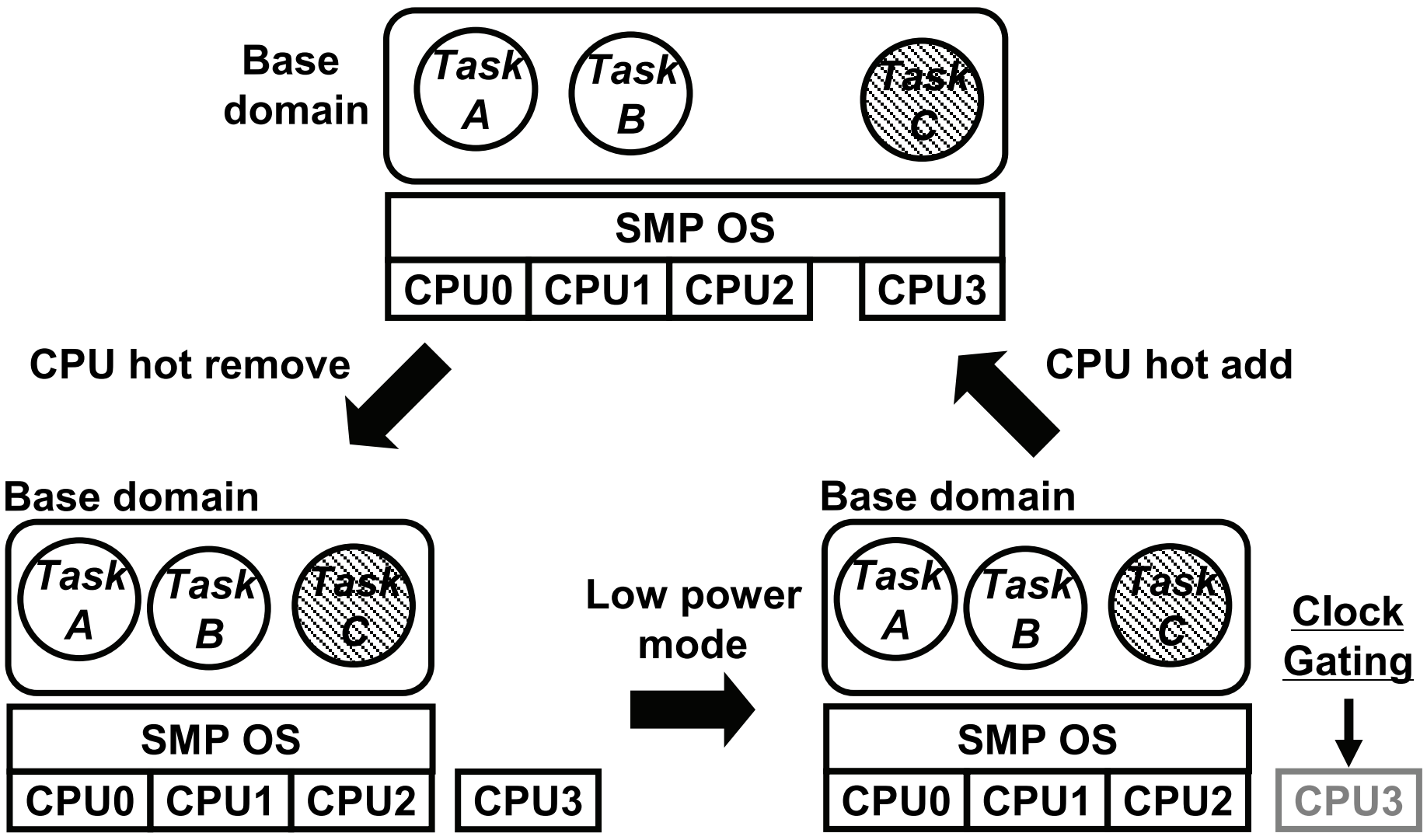

**FIGURE 6.5: CPU HOTPLUG TECHNOLOGY**

FIGURE 6.6 shows the relationship between our self-transition management and CPU Hotplug technology. While the white boxes indicate the conventional operations of CPU Hotplug technology, the gray boxes indicate newly-added operations for self-transition management. When a processor in an idle state is put into a low power mode, the CPU Hotplug technology (*i.e.*, the white boxes) implemented on ARM MPCore Linux requests the execution of a "CPU Hot Remove" processing. The "CPU Hot Remove" processing might involve, for example, 1) the migration of processes previously executed on that processor to other live processors, 2) a change in interrupt distribution to the processor, 3) deactivation of cache coherence (setting to the AMP mode), or 4) a processor's waiting for an interrupt while clock gating is being conducted. After the processor receives a wake-up interrupt, the technology requests the execution of a "CPU Hot Add" processing, such as the re-activation of cache coherence (setting to the SMP mode) or the return of a processor to an idle state.

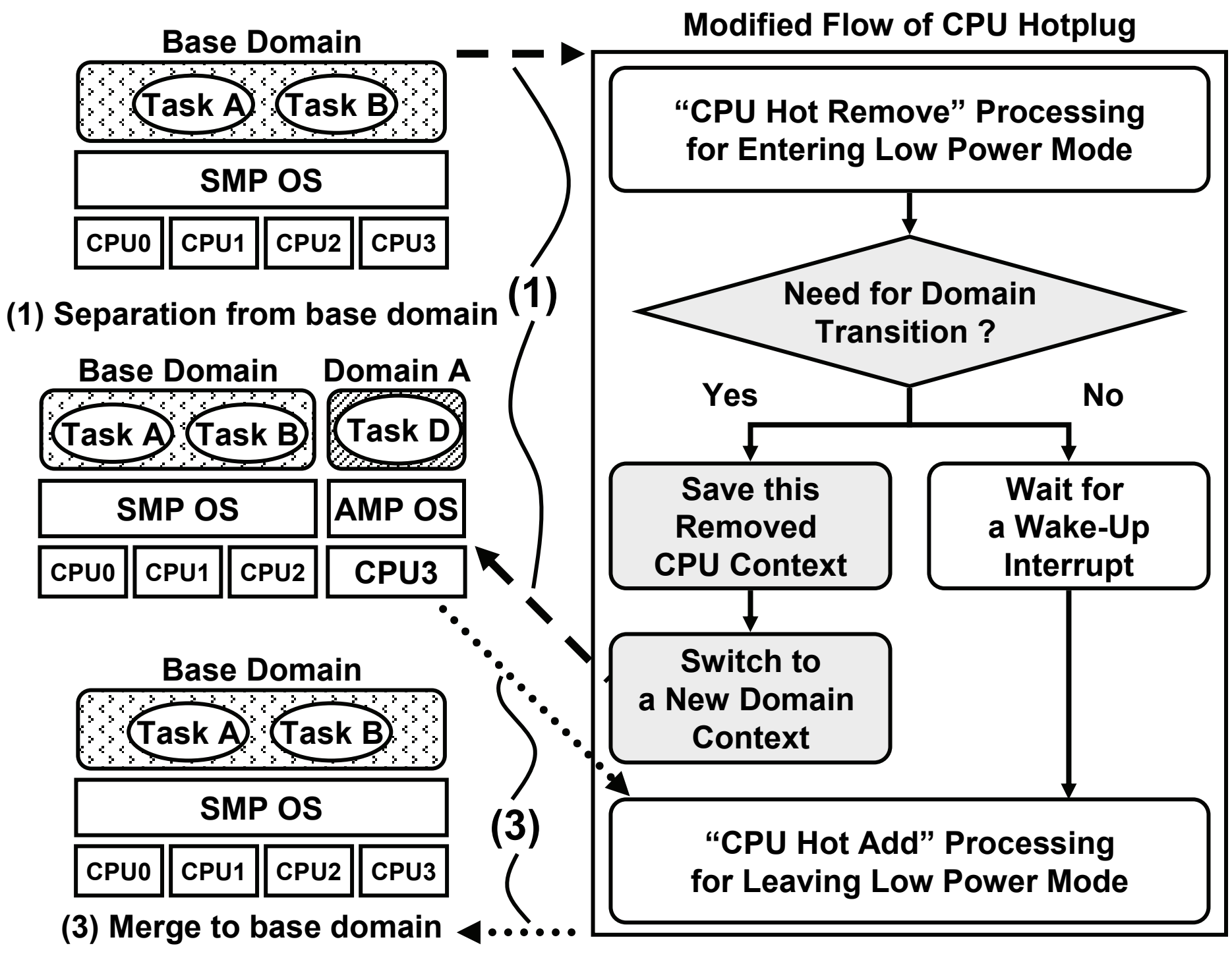

**FIGURE 6.6: SELF-TRANSITION MANAGEMENT**

Our self-transition management modifies the operational flow of CPU Hotplug technology instead of suspending processors. A context manager equips with three base domain context buffers in order to save the processor contexts corresponding to the three CPUs which make it possible to be separated from the base domain. In the case of separating a processor from the base domain and allocating it to an open domain (*i.e.*, state transition (1) in FIGURE 6.6), it calls the *cpu_down* function (the API for a "CPU Hot Remove" processing) in order to request the processor to execute a "CPU Hotplug Remove" processing. Instead of making the processor wait for an interrupt, it saves the base domain context to the base domain context buffer corresponding to the processor.

In addition, it restores an open domain context to the processor in the *platform_cpu_die* function (the function that enters into a low power mode) called from the *cpu_down* function to the processor. In this way, our self-transition management enables processors which previously executed functions in the base domain to start to execute in domains. *The key feature in our self-transition management is changing the value of the program counter saved in a base domain context to the program address which corresponds to the point at which waiting for an interrupt has been completed, i.e., the address that corresponds to the point just before "CPU Hot Add" processing commences.*

In the case of merging a processor from a domain back into the base domain (*i.e.*, state transition (3) in FIGURE 6.6), the self-transition management requests the context manager to perform a domain context switch. The context manager orders the processor's context handler to perform a domain context switch, providing the base domain context buffer to the context handler. The context handler then conducts a domain switch from the current domain to the base domain. Here, as mentioned earlier, since the value of the program counter is changed to the address directly preceding a "CPU Hot Add" processing, the processor executes the "CPU Hot Add" processing in the *platform_cpu_die* function. After that, it returns to an idle state in the base domain *as if it had received a wake-up interrupt.* In this way, the self-transition management enables processors which previously executed functions in domains to resume making executions in the base domain.

### 6.3.2 UNIFIED VIRTUAL ADDRESS MAPPING

For a state transition between domains, all registers in a processor have to be set with the register values of a new domain context. Traditional embedded processors, including ARM MPCores [ARM 06], generally do not allow mode registers or control registers, such as a pointer register for use with a page table, to be simultaneously restored. Here, the domain switching code achieves state transitions between domains by setting all processor registers required for an OS (*i.e.*, the OS context) to a processor. Thus, the code requires the execution of a consistent program flow even if the domain switching code executed in an OS *before* a state transition uses virtual addresses different from those in an OS *after* the state transition. In other words, unlike AMP software structures, it is difficult for the domain switching code to share with the same virtual addresses between OSs since a base domain and an open domain use different OSs, such as an SMP OS and an AMP OS.

FIGURE 6.7 explains the issue with the domain switching code to achieve state transitions between domains. For example, in switching from Domain A to the base domain, the domain switching code changes the memory map of Domain A to that of the base domain by using a system instruction (e.g. the instruction at 0xC0001008). Because the virtual addresses to which the domain switching code of Domain A refers are not always pointing to those used for the domain switching code of the base domain, the domain switching code starts to execute an unexpected program flow after the code has changed the memory map. This, thus, results in unstable state transitions caused by losing control of the processors.

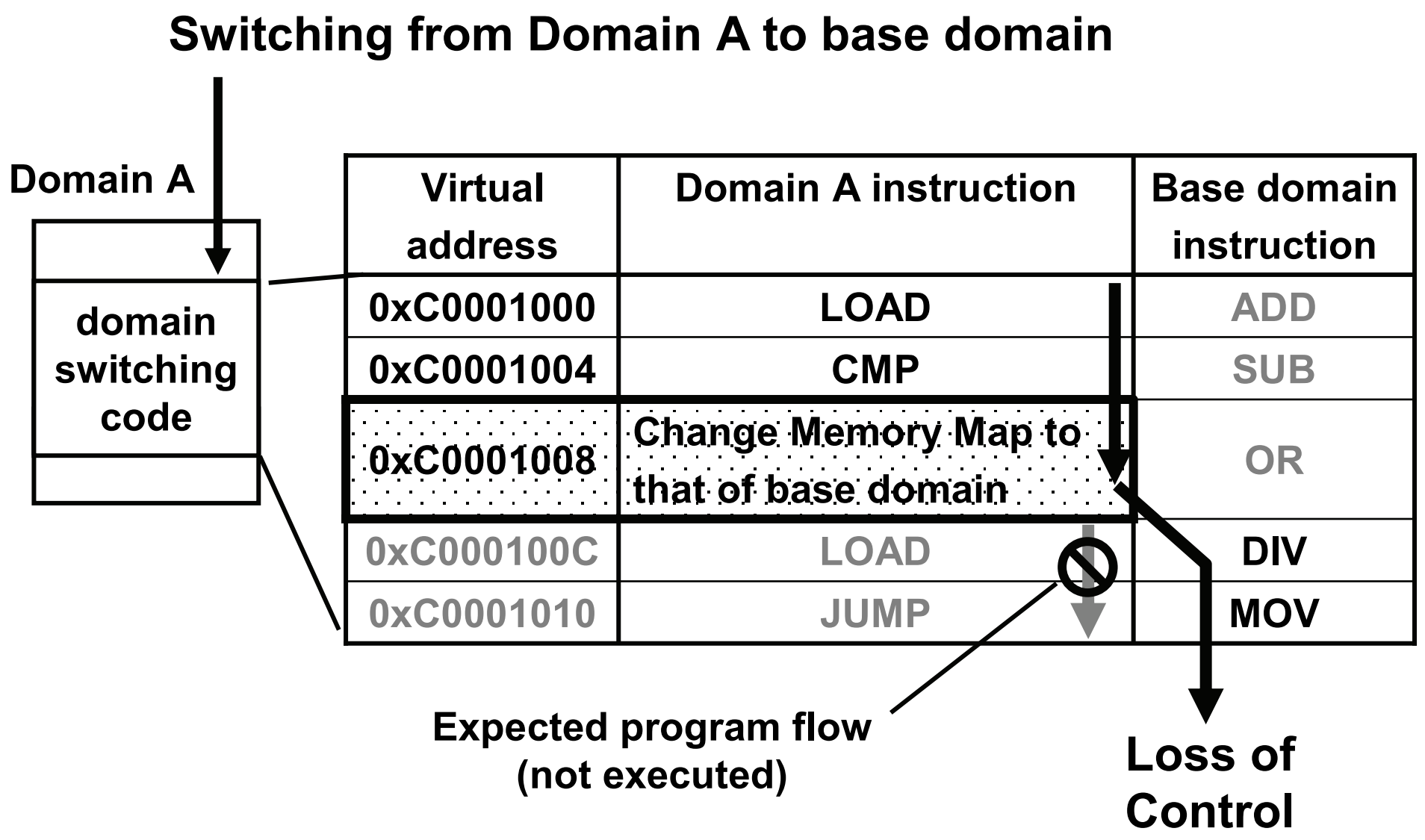

**FIGURE 6.7: ISSUE ON DOMAIN TRANSITION**

In order to avoid this situation and achieve stable state transitions, we employed unified virtual address mapping, a technology for matching virtual addresses in the domain switching code shared between an OS used before a state transition and an OS used after that state transition. It simply enforces each processor to set the memory map of the domain switching code to a specific address. FIGURE 6.8 shows the mapping between physical addresses and virtual addresses in terms of both the SMP OS of the base domain and AMP OSs of open domains (*i.e.*, Domains A and B). Unified virtual address mapping arranges common instructions and data used for the domain switching code in an area of the physical memory (*e.g.*, 0x0e001000 for the common instructions in FIGURE 6.8 and 0x0f000000 for the common data) that is separate from areas of the physical memory used by the SMP OS and AMP OSs. Further, it assigns the common instructions and data to virtual addresses that are the same in both the SMP OS and the AMP OSs (*e.g.*, 0x0ffb0000 in FIGURE 6.8). In this way, unified virtual address mapping achieves stable operations. For example, the mapping enables a processor executing the domain switching code to fetch correct instructions or read correct data even after the setting of a pointer register to a page table. This is because the instructions and data used for the domain switching code are assigned to the same virtual addresses as those in both the OS used before a state transition and the OS used after that state transition.

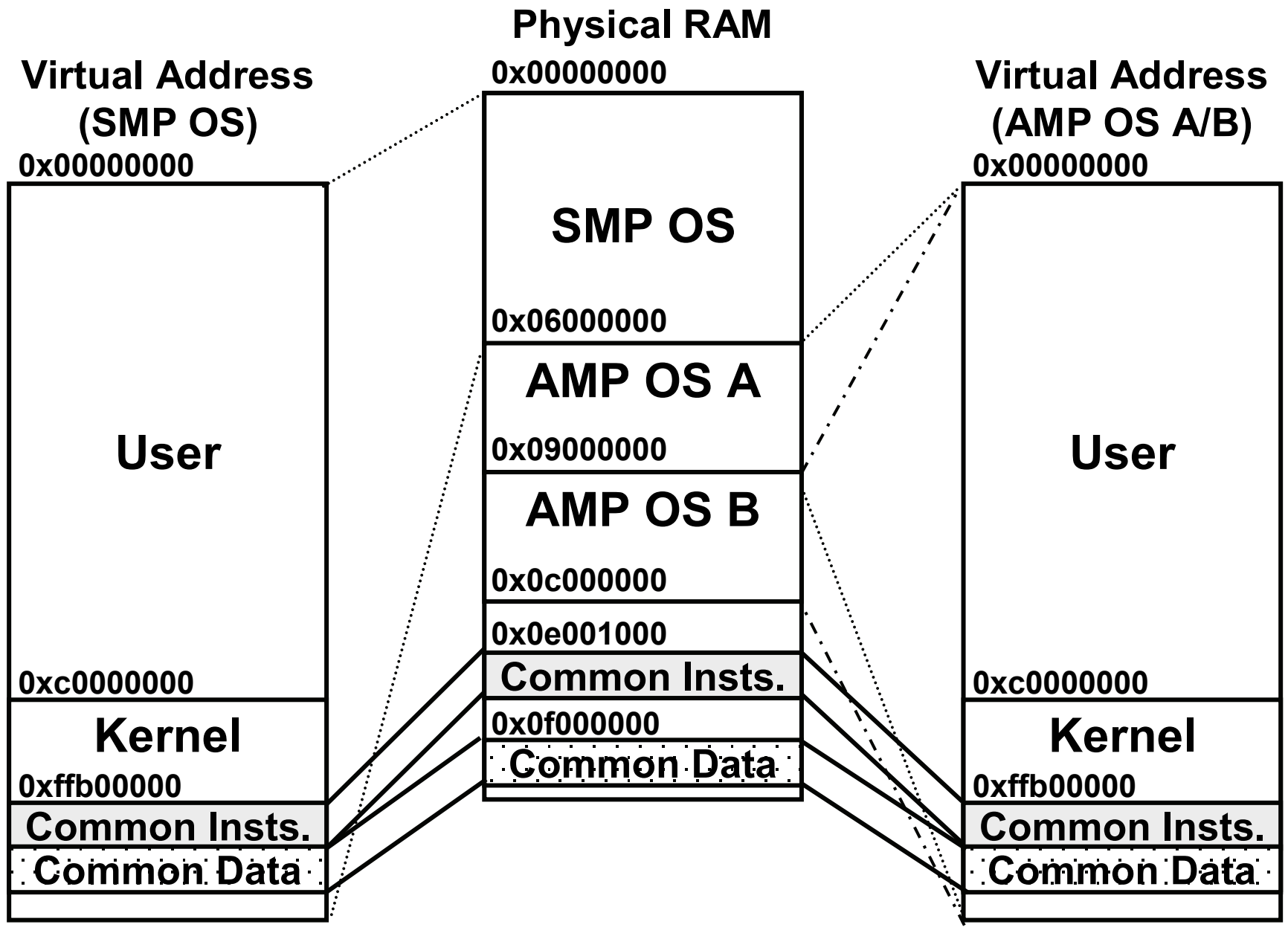

**FIGURE 6.8: ADDRESS MAP FOR UNIFIED VIRTUAL ADDRESS MAPPING**

In addition, unified virtual address mapping is designed to prevent extra virtual addresses from being newly allocated to OSs. *The key feature in unified virtual address mapping is the utilization of unused virtual addresses within the virtual address ranges allocated to I/O devices, which is possible because the size of common instructions and* *data is a small number of kilobytes*. Thus, no extra virtual addresses are required for mapping the domain switching code.

It should be noted that the only security risk of our secure dynamic partitioning is the transition to merge from a domain to the base domain. If a malicious domain should compromise the domain switching code, the malicious domain would be merged to the trusted base domain. However, as illustrated in Section 4.3 and Section 5.3.2, our bus management unit enables malicious domains to be prevented from modifying the code through data cache and from changing restored OS contexts.

## 6.4 EVALUATION

TABLE 6.1 summarizes our SMP evaluation environment, called MPCore [ARM 06]. MPCore is an embedded symmetric multi-core processor which equips with four ARM processors. FIGURE 6.9 shows our evaluation board with an MPCore processor.

**TABLE 6.1: SMP EVALUATION ENVIRONMENT**

| Item | Feature |
|---|---|
| SoC | MPCore (MP11 CPU x 4) @ 130nm |
| Cache | I$: 32KB, D$: 32KB per MP11 CPU |
| Clock frequency | ARM: 240MHz, Bus: 35MHz |
| OS | Linux 2.6.7 / SMP OS x 1, AMP OS x 2 |

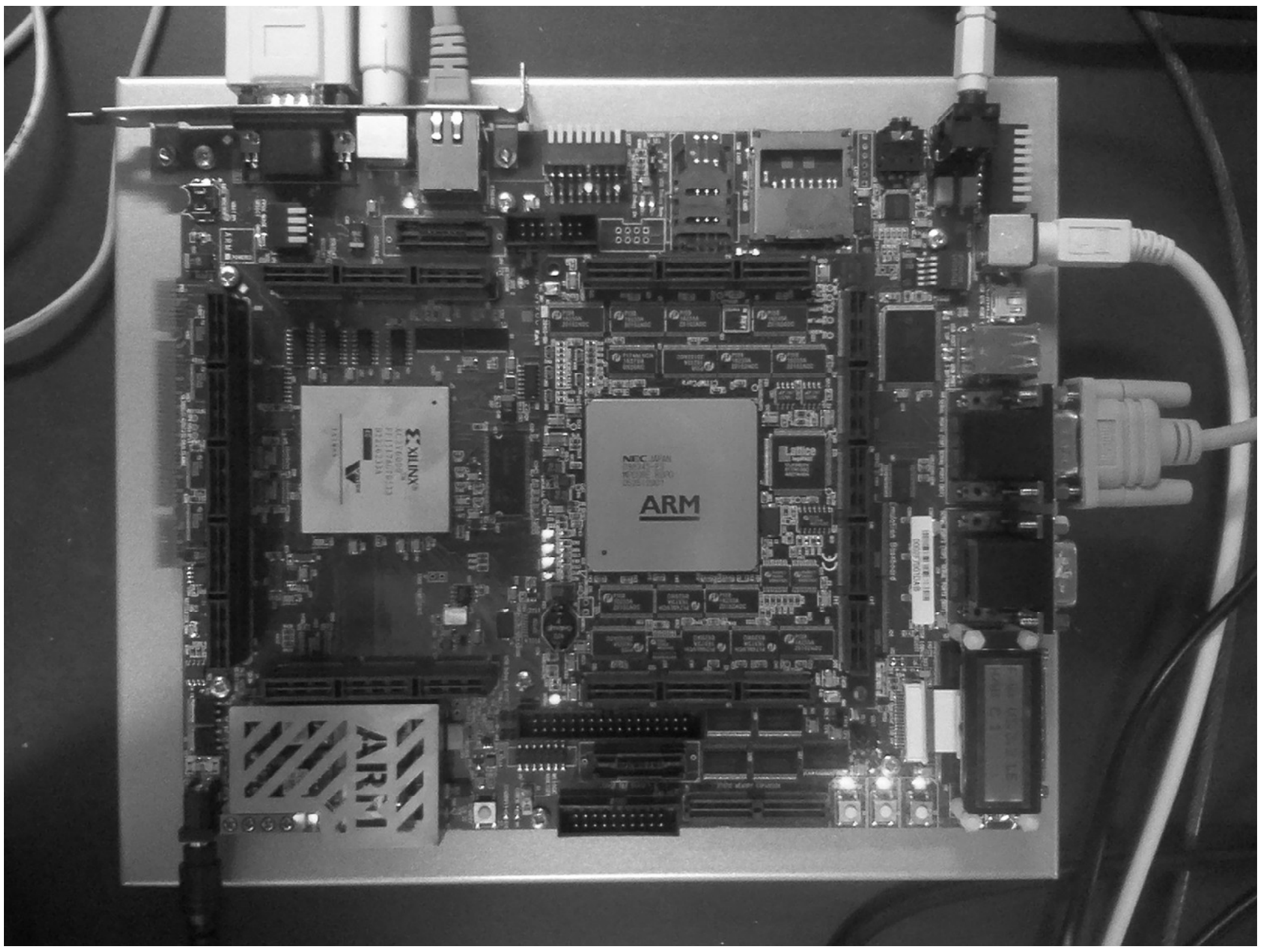

**FIGURE 6.9: MPCORE AS SMP EVALUATION ENVIRONMENT**

FIGURE 6.10 outlines the evaluation implementation of our secure dynamic partitioning, which supports one SMP Linux OS (*i.e.*, the base domain for base applications) and two AMP Linux OSs (*i.e.*, two open domains) on an MPCore. Thus, it enables two processors to be flexibly allocated from the SMP OS to AMP OSs or to be de-allocated from AMP OSs to the SMP OS. Evaluations show that our secure dynamic partitioning actually worked on SMP (see Section 6.4.1). In addition, the dynamic partitioning achieved higher performance (see Section 6.4.2, and 6.4.3), faster state transition time (Section 6.4.4), and lower code size (see Section 6.4.5) than did other approaches.

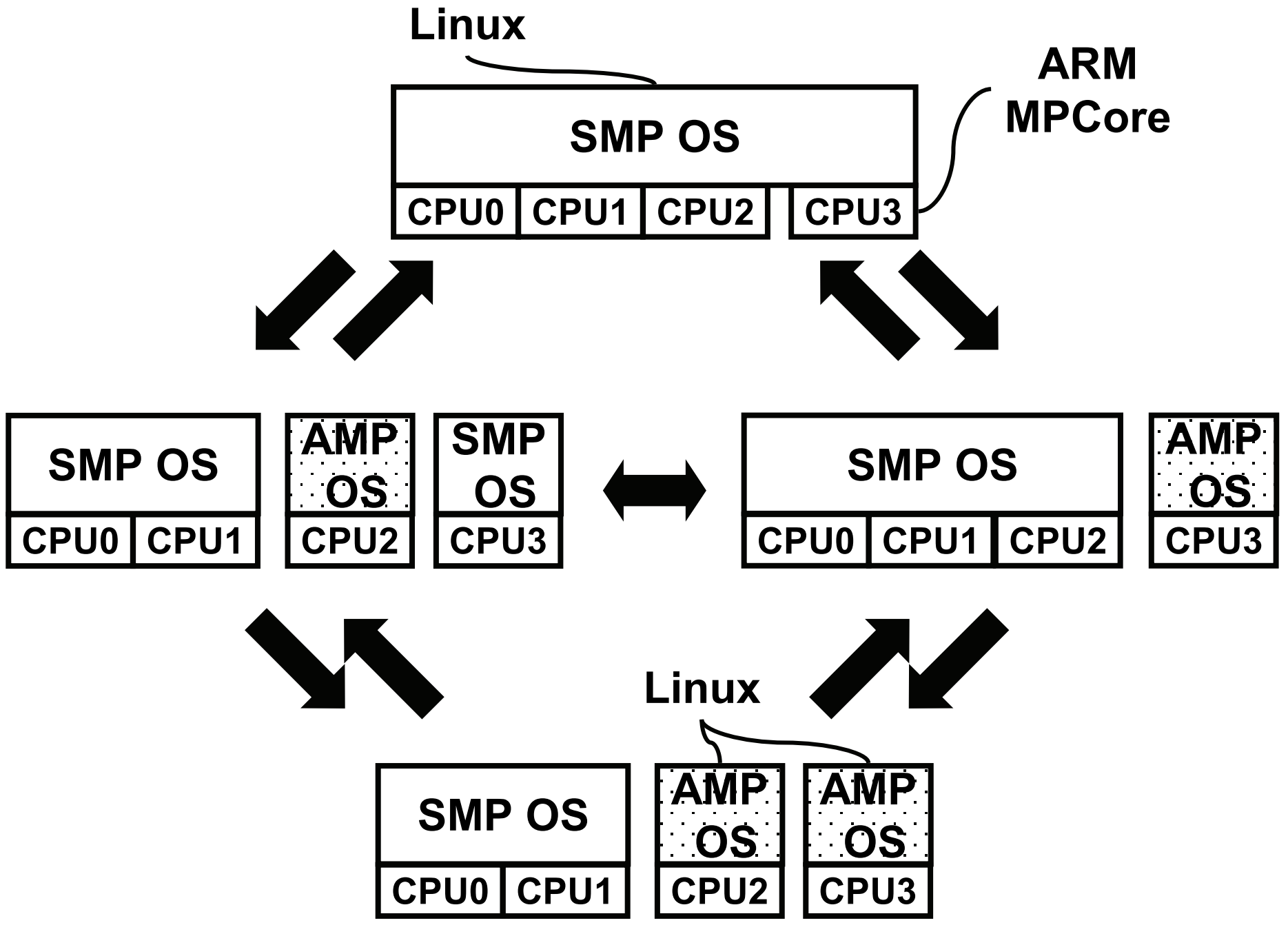

FIGURE 6.10: EVALUATION IMPLEMENTATION OF DYNAMIC PARTITIONING

### 6.4.1 SUCCESSFUL EXAMPLE FOR DYNAMIC PARTITIONING

FIGURE 6.11 demonstrates a successful example of the application of our secure dynamic partitioning. In normal operation, a large-screen video application (*i.e.*, a base application) which requires the full performance of four CPUs runs on the base domain. In response to a user request for the execution of an open application, a control application removes a CPU from the base domain, allocating the CPU to an open domain. Then, the video application on the base domain is scaled to a medium-screen one which requires only the performance of three CPUs. On the other hand, the open domain to which the CPU is allocated executes the requested open application.

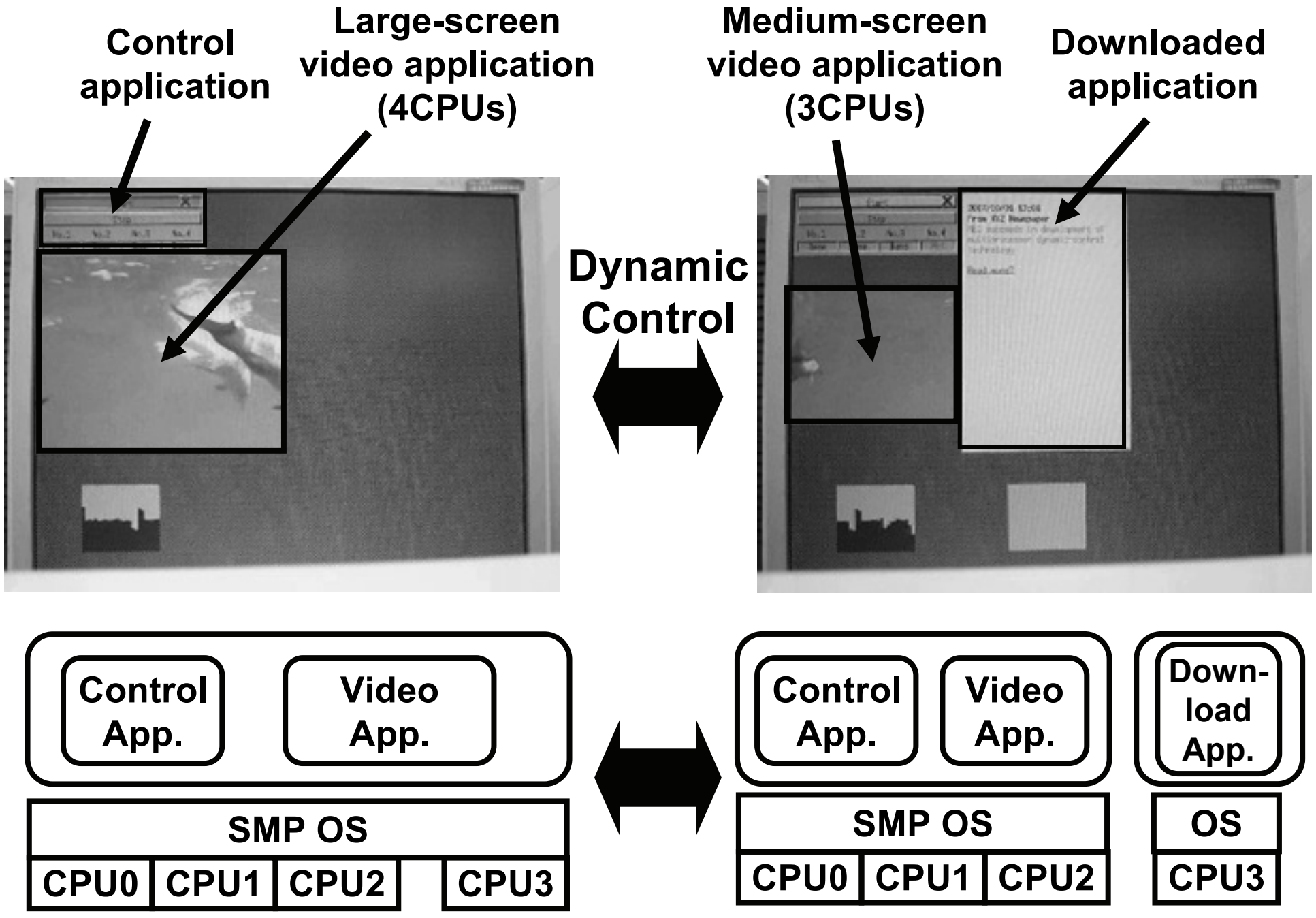

**FIGURE 6.11: SECURE DYNAMIC PARTITIONING ON MPCORE**

### 6.4.2 SCALABLE EXTENSION OF BASE APPLICATIONS

In order to study that a base domain has the ability on dynamically changing its performance for the scalable extension of base applications, FIGURE 6.12 shows allocation of MultiProcessor Dhrystone MIPS (MP DMIPS) to the base domain and an open domain. We used MP DMIPS in order to check whether our approach correctly allocates a CPU to an open domain or de-allocates a CPU from the domain because the MP DMIPS is simply proportional to the number of CPUs contained in an OS. As a result, we have confirmed that state transition (1) shown in FIGURE 6.1 reduces total performance in the base domain by an amount corresponding to that of a single processor while the amount of reduced performance is gained in the open domain. Further, with state transition (3) in FIGURE 6.1, performance in the base domain is increased back to the previous level.

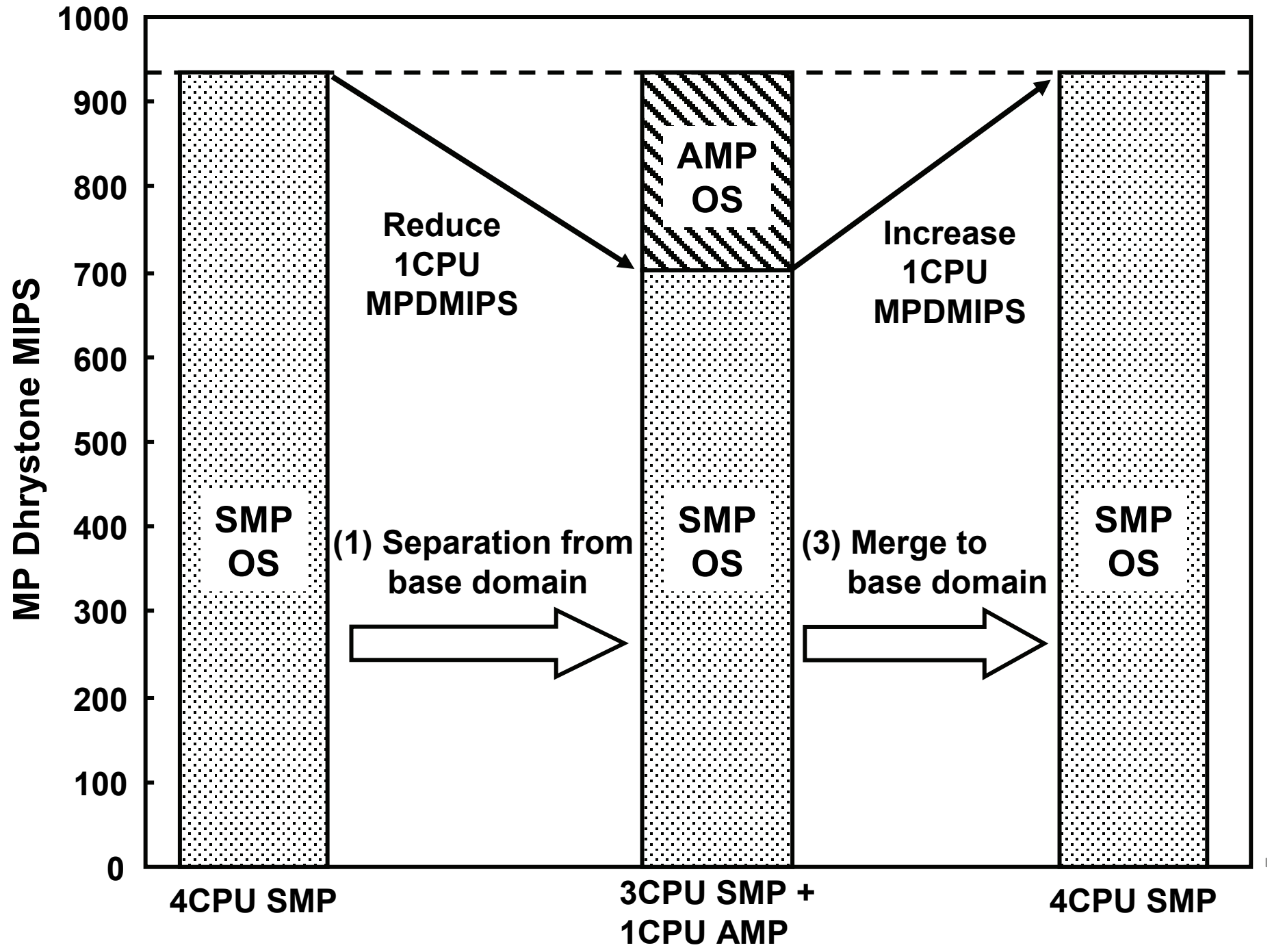

FIGURE 6.12: SCALABLE EXTENSION OF BASE APPLICATIONS

### 6.4.3 VIRTUALIZATION PERFORMANCE ON BASE DOMAIN

In order to examine that secure dynamic partitioning affects the performance of a base domain, FIGURE 6.13 and FIGURE 6.14 show the evaluation results for LMbench [Mcvoy 96] processes and context switching micro-benchmarks executed in a base domain. The evaluation conditions are the same as Section 5.5.2. As shown in the figures, the base domain achieves nearly the same performance as does the base-reference SMP Linux. This cannot be said for conventional virtualization software.

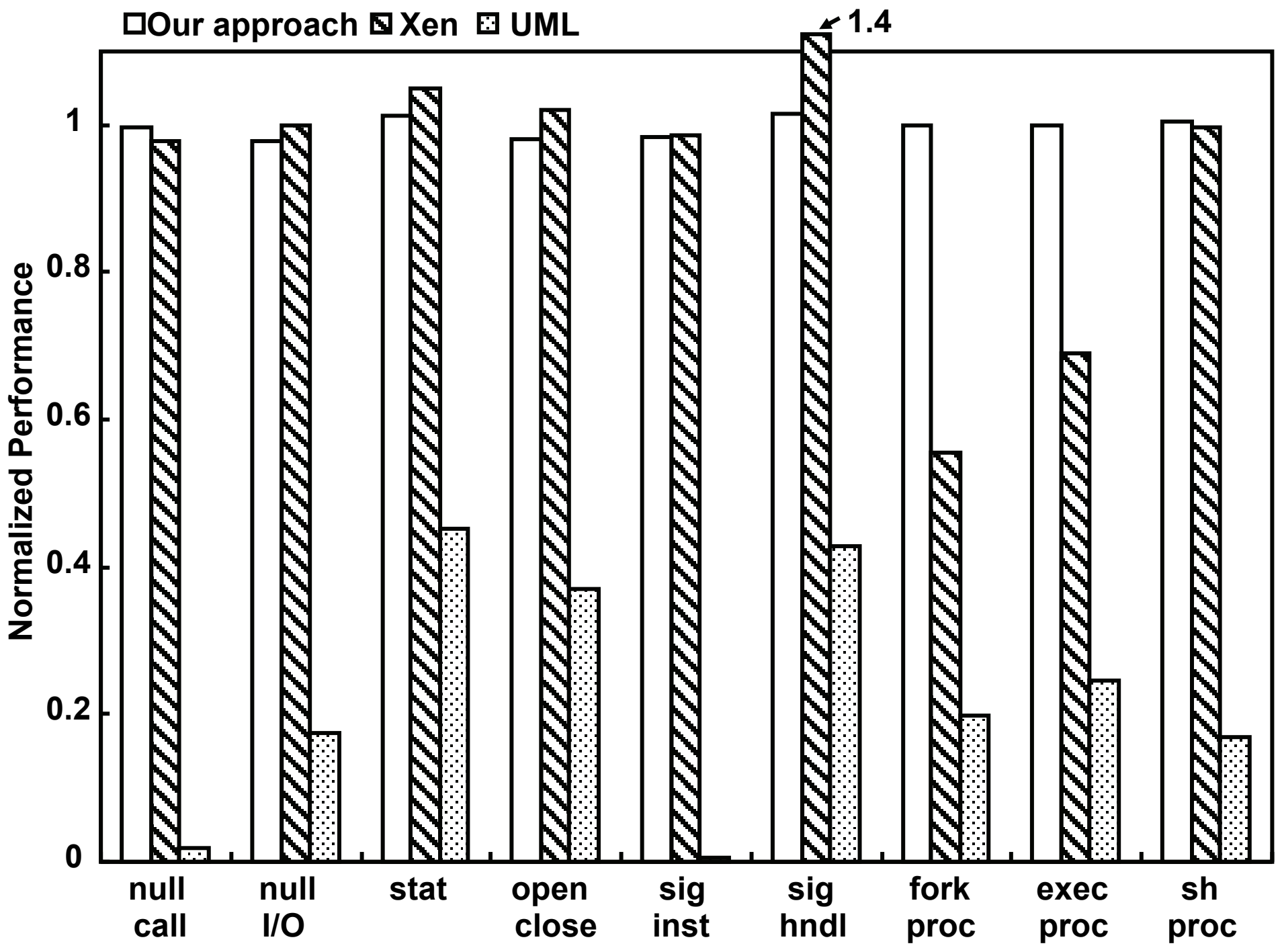

FIGURE 6.13: PROCESS MICRO-BENCHMARKS ON BASE DOMAIN

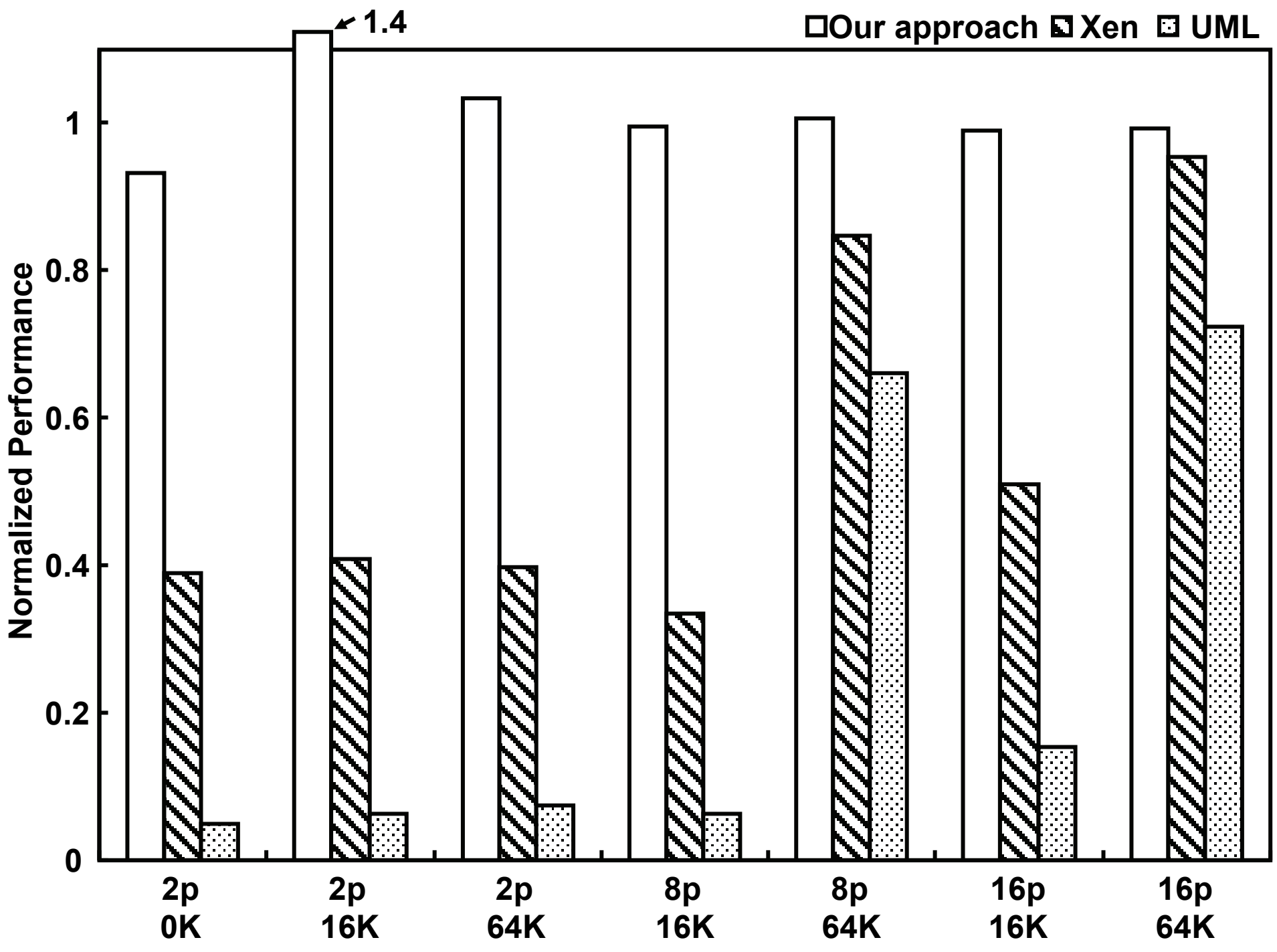

FIGURE 6.14: CONTEXT SWITCHING MICRO-BENCHMARKS ON BASE DOMAIN

Two small anomalies seen here, signal handling in Xen and two processes of 16KB array size each with our approach, presumably occurred due to a fortuitous cache alignment [Barham 03]. It should be noted that we have confirmed that the performance of an open domain on this SMP achieved the same performance as did asymmetric virtualization on AMP.

### 6.4.4 DOMAIN TRANSITION TIME

TABLE 6.2 shows times required for the state transitions shown in FIGURE 6.1, compared with that of CPU Hotplug technology.

**TABLE 6.2: STATE TRANSITION TIME OF DYNAMIC PARTITIONING**

| Item | Time | |
|---|---|---|
| | Proposed | CPU Hotplug |
| Separation from the base domain | 2.5ms | 1.5ms |
| Switching to an open domain | 0.5ms | N/A |
| Merge to the base domain | 4.5ms | 2.5ms |

Here, “separation from the base domain,” “switching to an open domain,” and “merge to the base domain” correspond, respectively to state transitions (1), (2) and (3) in FIGURE 6.1. The time required for state transitions is quite low (of a single-millisecond order). Further, the greatest time difference with CPU Hotplug technology is only 2.0ms. This small transition time of the order of a millisecond helps execute soft real-time applications, such as video applications, continuously, as shown in FIGURE 6.11.

### 6.4.5 LINES OF CODE FOR DYNAMIC PARTITIONING

TABLE 6.3 demonstrates that our secure dynamic partitioning is implemented with a small binary code size (*i.e.*, less than 40KB). The binary code size for common text (instructions) and data is also small, being implemented in only 9.2KB. The increases in binary code size of SMP Linux and AMP Linux are only 1.5% and 1.3%, respectively, over that for un-modified OSs. In terms of Lines of Code (LOC), the modified LOC values of SMP Linux and AMP Linux are 1549 LOC and 1145 LOC, respectively. This means that their modified LOC are almost the same as or less than the modified LOC of virtualization software without any functions for changing the number of processors within a domain (*e.g.,* it is almost the same as the LOC of Xen and 4 times smaller than that of UML).

**TABLE 6.3: LINES OF CODE FOR DYNAMIC PARTITIONING**

| Linux | Text | Data | BSS | Common text | Common data | Total |
|---|---|---|---|---|---|---|
| SMP | +11.2 | +1.6 | +16.2 | +0.3 | +8.9 | +38.2 |
| AMP | +6.6 | +1.1 | +16.1 | | | +32.9 |

## 6.5 SUMMARY

We have presented our secure dynamic partitioning, by which the number of processors allocated to individual OSs makes it possible to be dynamically changed on SMP. Its most important feature is dynamic processor allocation utilizing secure processor partitioning. In this way, secure dynamic partitioning achieves fast, flexible secure partitioning even on SMP. We have designed secure dynamic partitioning based on both ARM symmetric multi-core processors and Linux OSs. Moreover, our evaluations have shown its effectiveness, demonstrating three fundamental features: a successful example on ARM MPCore, high performance, low state transition time, and small code size.

# CHAPTER 7
# CONCLUSION

This dissertation has presented a bold new paradigm, known as open embedded systems. While traditional embedded systems provide only closed base applications to users, open embedded systems allow the users to use open applications as well as base applications. Platforms used for open embedded systems require the achievement of two major design objectives: the scalable extension of base applications and the secure execution of open applications.

The primary contributions of this dissertation are the attainment of a multi-core processor platform for open embedded systems. Four innovative techniques feature our multi-core platform: (1) seamless communication, by which legacy base applications designed for a single-core processor make it possible to be executed over multiple processors without any software modifications; (2) secure processor partitioning, by which OSs are mutually protected on separate processors; (3) asymmetric virtualization, by which many OSs over the number of processors are securely executed under secure processor partitioning; and (4) secure dynamic partitioning as an extension of secure processor partitioning, by which the number of processors allocated to individual OSs makes it possible to be dynamically changed on SMP under secure processor partitioning.

Evaluations show the effectiveness of the four innovative techniques. Seamless communication achieves a successful example with actual mobile terminal software, high performance and small code size. Secure processor partitioning provides excellent hardware specifications and a high security level. Asymmetric virtualization achieves a successful example on three ARM processors, high performance and small code size. Secure dynamic partitioning demonstrates a successful example on an MPCore, high performance, low state transition time, and small code size. As a result, our multi-core processor platform is ideally suited to open embedded systems since our platform satisfies two important requirements for open embedded systems.

In future work, we would proceed with the research in three technology directions: the enhancement of data security, the support of many-core processors, and the application to reliable embedded systems. First, the secure integration of data security techniques, such as XOM [Lie 00], AEGIS [Suh 05], TPM [TCG 06], and SENSS [Zhang 05], with our platform would be a big challenge since keys need to be perfectly protected from open applications. Second, in many-core processors [Vangal 08], our centralized bus management unit needs to be extended to distributed network management units since each processor is connected with an inter-connection network in a chip. Specifically, the consistent setting of distributed network management units would be a difficult challenge since many processors that belong to open domains need to be independently checked. Finally, the concept of this work would seem to be extensible to reliable embedded systems, such as automotive systems In automotive systems enhancing real-time responses, however, would be an important challenge since open applications interfere with base applications through the shared bus [Abe 07].